\documentclass[twocolumn]{aastex63}
\pdfoutput=1 

\usepackage{amsfonts}
\usepackage{amsmath}
\usepackage{amssymb}
\usepackage{upgreek}
\usepackage{bm}
\usepackage{dcolumn}
\usepackage{epsfig}
\usepackage{graphicx}
\usepackage{placeins}
\usepackage{dsfont}
\usepackage{comment}
\usepackage{color}
\usepackage{booktabs}
\usepackage{braket}
\usepackage{mathtools}
\usepackage{multirow}
\usepackage{booktabs}

\usepackage{longtable}
\usepackage{supertabular}

\newcommand\msun{$M_\sun$}
\newcommand\at{AT\,2017gfo} 
\newcommand\zhoriz{typical redshift reach} 
\newcommand\Ninstr{13} 
\newcommand\Nfilt{44} 

\newcommand{\cieraaffil}{Center for Interdisciplinary Exploration and Research in Astrophysics (CIERA),
Evanston, IL, 60201,
USA}
\newcommand{\nuaffil}{
Department of Physics and Astronomy,
Northwestern University,
Evanston, IL, 60208,
USA}
\newcommand{\ctaaffil}{Center for Theoretical Astrophysics, Los Alamos National Laboratory,
Los Alamos, NM, 87545,
USA}
\newcommand{\xcpaffil}{Computational Physics Division, Los Alamos National Laboratory,
Los Alamos, NM, 87545,
USA}
\newcommand{\gwuphysaffil}{Department of Physics, The George Washington University, 
Washington, DC, 20052,
USA}
\newcommand{\gwuastroaffil}{Astronomy, Physics and Statistics Institute of Sciences (APSIS), The George Washington University,
Washington, DC, 20052,
USA}
\newcommand{\umdaffil}{Department of Astronomy, University of Maryland,
College Park, MD, 20742, 
USA}
\newcommand{\goddardaffil}{Astrophysics Science 
Division, NASA Goddard Space Flight Center,
Greenbelt, MD, 20771,
USA}
\newcommand{\ccsaffil}{Computer, Computational, and Statistical Sciences Division, Los Alamos National Laboratory, 
Los Alamos, NM, 87545, 
USA}
\newcommand{\arizaffil}{The University of Arizona,
Tucson, AZ, 85721, 
USA}
\newcommand{\unmaffil}{Department of Physics and Astronomy, The University of New Mexico,
Albuquerque, NM, 87131,
USA}
\newcommand{\jinaaffil}{Joint Institute for Nuclear Astrophysics - Center for the Evolution of the Elements,
USA}
\newcommand{\ritaffil}{Center for Computational Relativity and Gravitation, Rochester Institute of Technology,
Rochester, NY, 14623,
USA}

\received{}
\submitjournal{ApJ}

\shorttitle{Kilonova Detectability}
\shortauthors{Chase et al.}

\begin{document}

\title{Kilonova Detectability with Wide-Field Instruments}

\correspondingauthor{Eve~A. Chase}
\email{evechase2021@u.northwestern.edu}

\author[0000-0003-1005-0792]{Eve~A. Chase}
\affiliation{\ctaaffil}
\affiliation{\xcpaffil}
\affiliation{\cieraaffil}
\affiliation{\nuaffil}

\author[0000-0002-9700-0036]{Brendan O'Connor}
\affiliation{\gwuphysaffil}
\affiliation{\gwuastroaffil}
\affiliation{\umdaffil}
\affiliation{\goddardaffil}

\author[0000-0003-2624-0056]{Christopher~L. Fryer}
\affiliation{\ctaaffil}
\affiliation{\ccsaffil}
\affiliation{\arizaffil}
\affiliation{\unmaffil}
\affiliation{\gwuphysaffil}

\author[0000-0002-1869-7817]{Eleonora Troja}
\affiliation{\umdaffil}
\affiliation{\goddardaffil}

\author[0000-0003-4156-5342]{Oleg Korobkin}
\affiliation{\ctaaffil}
\affiliation{\jinaaffil}
\affiliation{\ccsaffil}

\author[0000-0003-3265-4079]{Ryan~T. Wollaeger}
\affiliation{\ctaaffil}
\affiliation{\ccsaffil}

\author{Marko Ristic}
\affiliation{\ritaffil}

\author[0000-0003-1087-2964]{Christopher~J. Fontes}
\affiliation{\ctaaffil}
\affiliation{\xcpaffil}

\author[0000-0001-6893-0608]{Aimee~L. Hungerford}
\affiliation{\ctaaffil}
\affiliation{\jinaaffil}
\affiliation{\xcpaffil}

\author{Angela~M. Herring}
\affiliation{\xcpaffil}

\begin{abstract}
Kilonovae are ultraviolet, optical, and infrared transients powered by the radioactive decay of heavy elements following a neutron star merger.
Joint observations of kilonovae and gravitational waves can offer key constraints on the source of Galactic \textit{r}-process enrichment, among other astrophysical topics.
However, robust constraints on heavy element production requires rapid kilonova detection (within $\sim 1$ day of merger) as well as multi-wavelength observations across multiple epochs.
In this study, we quantify the ability of \Ninstr{} wide field-of-view instruments to detect kilonovae, leveraging a large grid of over 900 radiative transfer simulations with 54 viewing angles per simulation. 
We consider both current and upcoming instruments, collectively spanning the full kilonova spectrum. 
The Roman Space Telescope has the highest redshift reach of any instrument in the study, observing kilonovae out to $z \sim 1$ within the first day post-merger. 
We demonstrate that BlackGEM, DECam, GOTO, the Vera C. Rubin Observatory's LSST, ULTRASAT, and VISTA can observe some kilonovae out to $z \sim 0.1$ ($\sim$475~Mpc), while DDOTI, MeerLICHT, PRIME, \textit{Swift}/UVOT, and ZTF are confined to more nearby observations. 
Furthermore, we provide a framework to infer kilonova ejecta properties following non-detections and explore variation in detectability with these ejecta parameters.
\end{abstract}




\section{Introduction}

Neutron star mergers have long been invoked as a cosmic source of heavy elements, produced through rapid neutron capture (\textit{r}-process) nucleosynthesis \citep{lattimer1974,lattimer1977,symbalisty1982,eichler1989,freiburghaus1999,cote2018}. 
Heavy lanthanides and actinides fuse in material gravitationally unbound during the coalescence of either binary neutron stars (BNS) or neutron star--black hole (NSBH) binaries with near-equal mass ratios \citep{metzger2019}.
The residual radioactive decay of these \textit{r}-process elements spurs electromagnetic emission, called a kilonova  \citep{li1998,kulkarni2005,metzger2010}. 

Kilonova emission spans ultraviolet, optical, and near-infrared wavelengths (UVOIR), with lower wavelength ``blue'' emission fading within a day of merger, giving way to ``red'' emission persisting for over a week.
Observations of a kilonova's rapid evolution can reveal the role of neutron star mergers in Galactic \textit{r}-process enrichment, while  holding promise for additional discoveries in cosmology, nuclear physics, and stellar astrophysics.

Gamma-ray burst emission may also accompany a neutron star merger \citep{blinnikov1984,paczynski1986,eichler1989,narayan1992}, with later theories \citep{popham1999,fryer1999} linking neutron star mergers to short-duration gamma-ray bursts (sGRBs): transient signals with hard gamma-ray emission persisting for less than two seconds \citep{norris1984,kouveliotou1993}.
Numerous follow-up observations of sGRBs reveal viable kilonova candidates \citep{perley2009,tanvir2013,berger2013,yang2015,jin2016,troja2018,gompertz2018,ascenzi2019,lamb2019,troja2019,jin2020,rossi2020,fong2021,rastinejad2021,oconnor2020}, strengthening the theorized connection between sGRBs, kilonovae, and neutron star mergers.
Astronomers solidified this connection with the joint detection of gravitational-wave (GW) event GW170817 \citep{lvc170817}, GRB 170817A \citep{lvc_grb,goldstein2017,savchenko2017}, and kilonova \at~\citep{andreoni2017,arcavi2017,chornock2017,coulter2017,covino2017,cowperthwaite2017,drout2017,evans2017,kasliwal2017,lipunov2017,nicholl2017,pian2017,shappee2017,smartt2017,tanvir2017,troja2017,utsumi2017,valenti2017}.
GW170817 also serves as the first GW detection containing a neutron star \citep{lvc170817} and the first multi-messenger detection with GWs \citep{mma2017}. 

Several serendipitous circumstances allowed for detection of the kilonova associated to GW170817.
The source's comparatively tight sky localization \citep[31~deg$^2$ at 90\% credible level;][]{gcn2_gw170817,obs_scen2020} combined with its nearby distance estimate \citep[40$^{+8}_{-8}$~ Mpc;][]{gcn2_gw170817} promoted prompt detection of an UVOIR counterpart within 12 hours post-merger \citep{arcavi2017,coulter2017,cowperthwaite2017,lipunov2017,tanvir2017,valenti2017}.
Due to its nearby distance, GW170817 was confined to a localization volume of $\sim$\,520~Mpc$^3$ at the 90\% credible level \citep{gcn2_gw170817}, leaving only a few dozen plausible host galaxies, which were efficiently observed with galaxy-targeted observations. Additionally, GW170817's favorable viewing angle \citep[$\lesssim 20^\circ$;][]{ghirlanda2019,troja2019b} aided coincident detection of a sGRB without excessive afterglow contamination of the kilonova signal.

However, detecting a kilonova at further distances becomes increasingly difficult, often requiring the use of wide field-of-view (FoV) instruments.
For example, large sky localization, exacerbated by the lack of coincident sGRB detection, spurred fruitless counterpart searches \citep{hosseinzadeh2019,paterson2020} following the GW detection of GW190425 \citep{lvc190425,gcn_gw190425}.
This BNS merger's sky localization was constrained to a massive 10,183~deg$^2$ at the 90\% credible level \citep{gcn_gw190425}, requiring an area of the sky over 300 times the size of GW170817's localization to be rapidly searched for transient signals.
Additionally, the source was more distant than GW170817 \citep[155$^{+45}_{-45}$~Mpc;][]{gcn_gw190425} and confined to a larger localization volume \citep[$9.7\times 10^6$~Mpc$^3$;][]{gcn_gw190425}, making efficient galaxy-targeted kilonova searches unfeasible. 

As GW detector sensitivity increases, wide-field instruments will remain necessary tools for kilonova detection.
The advent of more sensitive GW detectors will lead to compact binary merger detection at farther distances, which jointly increases the sky volume required for counterpart searches in addition to reducing the effectiveness of galaxy-targeted searches.
In the third Observing Run (O3) of advanced LIGO \citep{aLIGO} and advanced Virgo \citep{adVirgo}, the average redshift horizon of a 1.4\msun{} + 1.4\msun{} BNS was 300~Mpc, 240~Mpc, and 100~Mpc for the LIGO Livingston, LIGO Hanford, and Virgo detectors, respectively.\footnote{Horizon redshifts computed by multiplying the BNS detectability ranges in \citealt{gwtc2} by the standard factor of 2.26 described in \citealt{finn1993} and \citealt{chen2021}.}
We note that mergers comprised of more massive objects (e.g., NSBHs) are detectable at even farther distances.

The fourth Observing Run (O4) will considerably increase the sensitivity, with BNS horizon redshifts anticipated to increase to 360--430~Mpc, 200--270~Mpc, and 70--290~Mpc for advanced LIGO, advanced Virgo, and KAGRA \citep{kagra}, respectively \citep{obs_scen2020}.
Furthermore, in the 2030s, proposed third-generation GW detectors including Einstein Telescope \citep{einstein_telescope} and Cosmic Explorer \citep{cosmic_explorer} will detect BNS mergers out to cosmological redshifts of $z\sim4$ and $z\sim10$, respectively \citep{hall2019}, far beyond the reaches of currently existing instruments.

Although sky localization is expected to significantly improve with larger networks of GW detectors \citep{schutz2011,nissanke2013,obs_scen2020}, localization estimates will still encompass significant areas of the sky.
In O4, the four-detector network of advanced LIGO, advanced Virgo, and KAGRA will observe BNS (NSBH) systems with a median sky localization area of 33~deg$^2$ (50~deg$^2$) and volume of 52,000~Mpc$^3$ (430,000~Mpc$^3$), encapsulating thousands of plausible host galaxies \citep{obs_scen2020}. Moreover, the addition of a fifth GW detector in India will improve sky localization areas by approximately a factor of two \citep{pankow2018}.
However, even in the era of third-generation GW detectors, sky localization areas will remain large for distant BNS mergers with more than 50\% of BNS mergers at $z \gtrsim 0.4$ constrained to localization areas in excess of 100~deg$^2$ with a network of three Cosmic Explorer instruments \citep{mills2018}.
Therefore, wide-field instruments will remain a necessary tool for kilonova detection into the 2030s.

In this paper, we assess kilonova detectability with \Ninstr{} wide-field instruments, quantifying the redshift out to which kilonovae are detectable for a variety of filters. 
This study builds upon previous detectability studies \citep{scolnic2018} by employing the Los Alamos National Laboratory (LANL) grid of radiative transfer kilonova simulations \citep{wollaeger2020} to explore detectability for a diverse range of kilonova ejecta masses, velocities, morphologies, compositions, and inclinations. 

We describe the set of wide-field instruments selected for the study in Section~\ref{sec: instruments} and then quantify the \zhoriz{} of each instrument in Section~\ref{sec: detectability}. In Section~\ref{sec: variation}, we explore how each instrument's ability to observe a kilonova varies with ejecta properties. We build upon these results in Section~\ref{sec: infer_params} to provide a framework for inferring kilonova properties from non-detections. We summarize each instrument's capacity for kilonova detection in Section~\ref{sec: instr_results} and offer suggestions for future kilonova searches in Section~\ref{sec: conclusion}. Throughout the study, we adopt a standard $\Lambda$-CDM cosmology with parameters $H_0=67.4$, $\Omega_M=0.315$, and $\Omega_\Lambda=0.685$ \citep{planck2020}. Data products and software produced through this study are available on GitHub.\footnote{\url{https://github.com/eachase/kilonova_detectability/}}

\section{Wide-Field Instruments} \label{sec: instruments}

The optimal observing strategy for multi-wavelength follow-up of LIGO/Virgo/KAGRA candidate events is an evolving science \citep[e.g.,][]{Kasliwal2014,Gehrels2016,Artale2020}, and varies greatly depending on the instrument FoV, wavelength range, and sensitivity \citep[see, e.g.,][]{Gehrels2015,Bartos2016,Cowperthwaite2019,Graham2019}. 
Wide-field instruments provide the ability for rapid searches of GW sky localization regions with high cadence.
In the next decade, a number of ground- and space-based facilities will be dedicated to these types of follow-up, enabling the ability to cover large sky regions in a short amount of time. 
We base our kilonova detectability study on a selection of current and future wide-field instruments with planned follow-up of LIGO/Virgo/KAGRA candidate events.
In addition, we include instruments that have the sensitivity and wavelength coverage to contribute significantly to the kilonova detection rate, but that lack a formal GW follow-up strategy \citep[i.e., LSST, Roman;][]{Cowperthwaite2019,Foley2019wfirst}. 

We note that this study is focused on the ability of wide-field instruments to detect kilonovae arising from poorly-localized mergers.
We primarily focus on kilonova searches following a LIGO/Virgo/KAGRA candidate BNS or NSBH event. 
However, the results of this study apply to kilonova searches following sGRB detections with large sky localization areas, such as some sGRBs detected with the \textit{Fermi Gamma-Ray Burst Monitor} \citep{goldstein2020}. 
The follow-up strategy for well-localized events is significantly different (as it does not require covering large sky areas), and is beyond the scope of this work.

All instruments in this study are either already active or plan to see first light within the 2020s, in the advanced ground-based GW detector era.
Our study includes several existing instruments such as the Deca-Degree Optical Transient Imager \citep[DDOTI;][]{watson2016}, the Dark Energy Camera \citep[DECam;][]{Flaugher2015}, the Gravitational-wave Optical Transient Observer \citep[GOTO;][]{Dyer2018,Dyer2020}, MeerLICHT \citep{blackgem}, \textit{Swift}'s Ultra-Violet Optical Telescope \citep[UVOT;][]{swift}, the Visible and Infrared Survey Telescope for Astronomy \citep[VISTA;][]{vista}, and the Zwicky Transient Facility \citep[ZTF;][]{ztf}.
In addition, we include numerous near-term instruments such as BlackGEM \citep{blackgem}, the Vera C.\ Rubin Observatory’s Legacy Survey of Space and Time \citep[LSST;][]{LSST}, the PRime-focus Infrared Mirolensing Experiment (PRIME\footnote{\url{http://www-ir.ess.sci.osaka-u.ac.jp/prime/index.html}}), the Nancy Grace Roman Space Telescope \citep[Roman;][]{roman}, the Ultraviolet Transient Astronomy Satellite \citep[ULTRASAT;][]{ultrasat}, and the Wide-Field Infrared Transient Explorer \citep[WINTER;][]{winter}.
This study is not a comprehensive selection of all wide-field instruments available in the coming decade. 

The selected wide-field instruments span UVOIR wavelengths ($\sim$2,000-22,000 $\mathrm{\AA}$) typical of kilonova emission. 
The 5$\sigma$ limiting magnitudes were compiled from the literature based either on the instrument design sensitivity (i.e., BlackGEM, LSST, PRIME, Roman, ULTRASAT, WINTER) or the instrument performance during previous LIGO/Virgo observing runs (i.e., DECam, DDOTI, GOTO, MeerLICHT,  UVOT, VISTA, ZTF). For instruments that have already demonstrated their capability to effectively cover GW localization regions, we focus our study on the filters used in those searches. 
The filter selection\footnote{When available, filter response functions were taken from the SVO Filter Profile Service \citep{rodrigo2020}. The filter response functions for BlackGEM, DDOTI, GOTO, PRIME, and WINTER were obtained through private communication with the instrument teams.} and limiting magnitudes are outlined here for each instrument (see also Table~\ref{tbl:kn}):

\startlongtable
\begin{deluxetable*}{cccccccccc}
\tablecaption{Kilonova detectability metrics for a wide-field instruments. \label{tbl:kn}  
}
\tablenum{1}
\tablehead{\colhead{Instrument} & \colhead{FoV (deg$^2$)} & \colhead{Exp. Time (s)} & \colhead{Filter} & \colhead{$\lambda_{\mathrm{eff}}$ ($\mathrm{\AA}$)} & \colhead{$m_{\mathrm{lim}}$ (AB)} & \colhead{$z_{50\%}$} & \colhead{$z_{95\%}$} & \colhead{$z_{5\%}$ } & \colhead{Ref.}} 
\startdata
         BlackGEM & 8.1 &  300 & \textit{u} & 3800 & 21.5 & 0.037 & 0.013 & 0.091 & 1  \\
         & & & \textit{g} & 4850 & 22.6 & 0.057 & 0.011 & 0.21  \\
         & & & \textit{q} & 5800 & 23.0 & 0.072 & 0.027 & 0.23 \\
         & & & \textit{r} & 6250 & 22.3 & 0.052 & 0.022 & 0.14 \\
         & & & \textit{i} & 7650 & 21.8 & 0.044 & 0.013 & 0.13 \\
         & & & \textit{z} & 9150 & 20.7 & 0.025 & 0.011 & 0.069 \\
         \hline
         DDOTI & 69 & 7200 &  \textit{w} & 6190 & 20.5\tablenotemark{a} & 0.023 & 0.0083 & 0.073 & 2,3 \\
         \hline
         DECam & 2.2 & 90 & \textit{i} & 7870  & 22.5 & 0.058 & 0.022 & 0.18 & 4 \\
         &  &  & \textit{z} & 9220  & 21.8 & 0.042 & 0.014 & 0.12 &  \\
         \hline
         GOTO & 40\tablenotemark{b} & 360 &\textit{L} & 5730 & 21.0 & 0.029 & 0.0097 & 0.097 & 5,6 \\
         \hline
         LSST & 9.6 & 30 & \textit{u} & 3690 & 23.6 & 0.078 & 0.012 & 0.28 & 7 \\
         & & & \textit{g} & 4830 & 24.7 & 0.14 & 0.035 & 0.48 \\
         & & & \textit{r} & 6220 & 24.2 & 0.12 & 0.043 & 0.43 \\
         & & & \textit{i} & 7570 & 23.8 & 0.099 & 0.038 & 0.31 \\
         & & & \textit{z} & 8700 & 23.2 & 0.079 & 0.029 & 0.24 \\
         & & & \textit{y} & 9700 & 22.3 & 0.052 & 0.021 & 0.14 \\ 
         \hline
         MeerLICHT & 2.7  & 60 & \textit{u} & 3800 & 19.1 & 0.013 & 0.0052 & 0.035 & 1,8 \\
         & & & \textit{g} & 4850 & 20.2 & 0.019 & 0.0032 & 0.073 \\
         & & & \textit{q} & 5800 & 20.6 & 0.024 & 0.010 & 0.074 \\
         & & & \textit{r} & 6250 & 19.9 & 0.019 & 0.0067 & 0.046 \\
         & & & \textit{i} & 7650 & 19.4 & 0.016 & 0.0053 & 0.047 \\
         & & & \textit{z} & 9150 & 18.3 & 0.0085 & 0.0023 & 0.024 \\ 
         \hline
          PRIME & 1.56 & 100 & \textit{Z}  & 9030  & 20.5 & 0.023 & 0.0099 & 0.067 & 9 \\
         & & &\textit{Y}  & 10200 & 20.0 & 0.019 & 0.0079 & 0.048 \\
         & & &\textit{J}  & 12400 & 19.6 & 0.017 & 0.0067 & 0.044 \\
         & & &\textit{H}  & 16300 & 18.4 & 0.0088 & 0.0023 & 0.023 \\ \tablebreak
         \hline
         Roman & 0.28 & 67 & \textit{R} & 6160 & 26.2 & 0.29 & 0.10 & 0.96 & 10,11\\
         & & & \textit{Z} & 8720 & 25.7 & 0.24 & 0.10 & 0.79 \\
         & & & \textit{Y} & 10600 & 25.6 & 0.23 & 0.10 & 0.79 \\
         & & & \textit{J} & 12900 & 25.5 & 0.22 & 0.10 & 0.65 \\
         & & & \textit{H} & 15800 & 25.4 & 0.22 & 0.095 & 0.48 \\
         & & & \textit{F} & 18400 & 24.9 & 0.17 & 0.037 & 0.38 \\ 
         \hline
         \textit{Swift}/UVOT & 0.08 & 80 & \textit{u} & 3500 & 19.9 & 0.015 & 0.0019 & 0.061 & 12 \\ 
         \hline
         ULTRASAT & 200 & 900 & \textit{NUV} & 2550 & 22.3 & 0.022 & 0.0022 & 0.10 & 13\tablenotemark{c} \\
         \hline
         VISTA  & 1.6 & 360 & \textit{Y} & 10200 & 21.5 & 0.038 & 0.012 & 0.094 & 14,15 \\
         & & & \textit{J} & 12600 & 21.0 & 0.032 & 0.011 & 0.071  \\
         & & & \textit{H} & 16500 & 21.0 & 0.031 & 0.011 & 0.068 \\
         & & & \textit{$K_s$} & 21400 & 20.0 & 0.018 & 0.007 & 0.045 \\ 
         \hline
         WINTER & 1.0 & 360 &\textit{Y} & 10200 & 21.5 & 0.038 & 0.012 & 0.093 & 16\\
         & & &\textit{J} & 12500 & 21.3 & 0.036 & 0.012 & 0.075 \\
         & & &\textit{$H_s$} & 15800 & 20.5 & 0.023 & 0.010 & 0.052 \\
         \hline
         ZTF & 47 & 30 & \textit{g} & 4770 & 20.8 & 0.025 & 0.0063 & 0.092 & 17 \\
         & & & \textit{r} & 6420 & 20.6 & 0.024 & 0.0092 & 0.074 \\
         & & & \textit{i} & 8320 & 19.9 & 0.019 & 0.0061 & 0.059 \\
         \hline
         \hline
\enddata

\tablecomments{The column labeled $z_{50\%}$ represents the maximum redshift at which 50\% of LANL kilonova models are observable at any one time in a given band. Columns labeled $z_{95\%}$ and $z_{5\%}$ enumerate similar redshifts for 95\% and 5\% of modeled kilonovae, respectively.}

\tablenotetext{a}{10$\sigma$ limiting magnitude.}
\tablenotetext{b}{FoV of one GOTO-8 system.}
\tablenotetext{c}{ULTRASAT filter modeled with a top-hat function between 2200 and 2800 $\mathrm{\AA}$. Limiting magnitudes were taken from the instrument website (see footnote \ref{ultraweb}).}

\tablerefs{(1) Paul Groot, 2021, Private Communication; (2) \citet{thakur2020}; (3) Becerra et al., in prep.; (4) \citet{SoaresSantos2016}; (5) Ben Gompertz \& Martin Dyer, 2021, Private Communication; (6) \citet{dyer_thesis}; (7) \citet{LSST}; (8) \citet{deWet2021}; (9) Takahiro Sumi, 2021, Private Communication; (10) \citet{scolnic2018}; (11) \citet{hounsell2018}; (12) Oates et al., in prep.; (13) \citet{ultrasat}; (14) \citet{McMahon2013}; (15) \citet{Banerji2015}; (16) Nathan Lourie \& Danielle Frostig, 2021, Private Communication; (17) \citet{ztf}}

\end{deluxetable*}

\begin{figure*}
\centering
\includegraphics[width=0.75\textwidth]{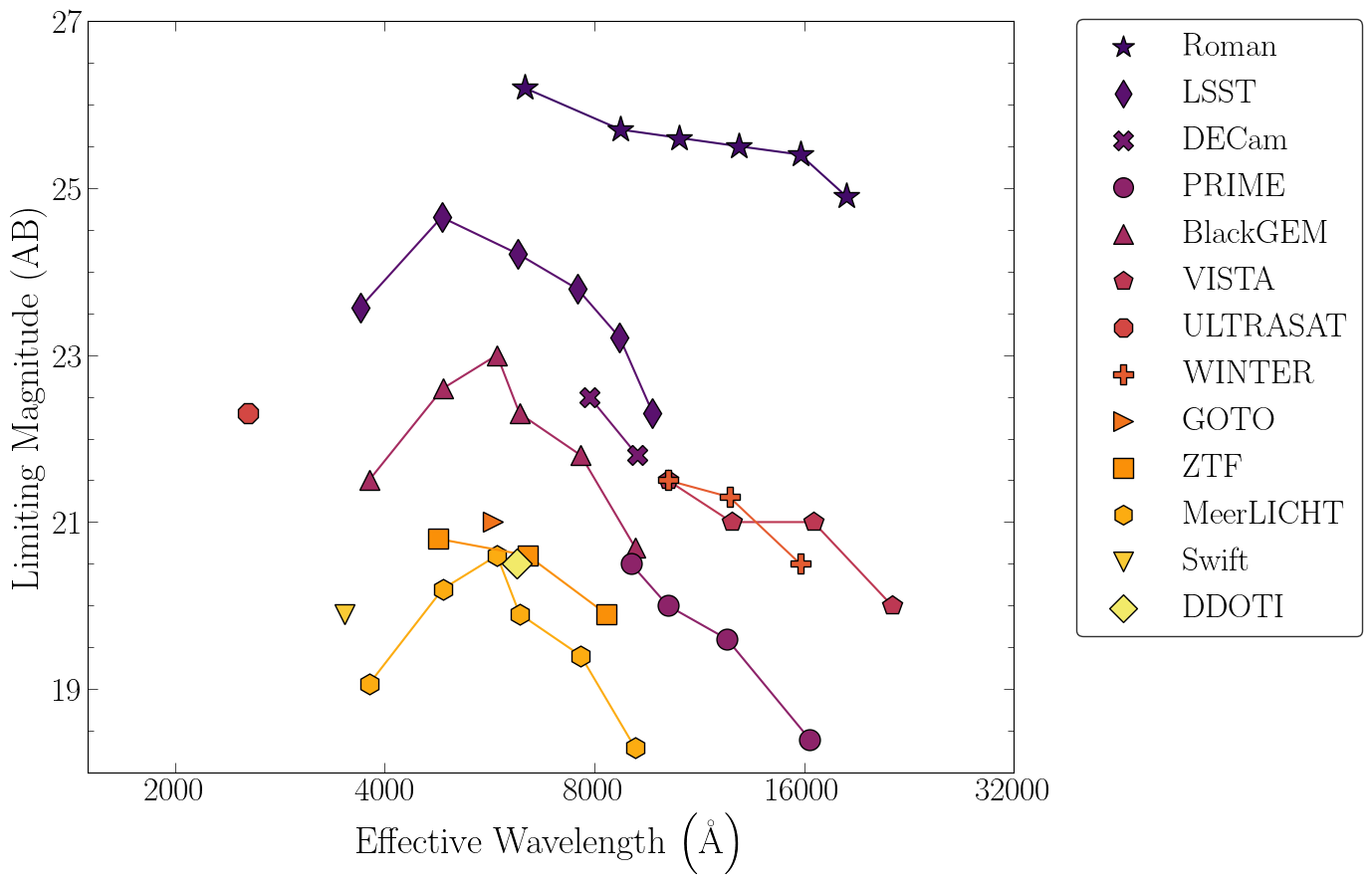}
\caption{Limiting magnitudes for a selection of wide-field instruments. Each symbol corresponds to one instrument, with an instrument's filters represented by its bandpass filter function's effective wavelength. Lines connect filters from the same instrument. All limiting magnitudes are presented at the 5$\sigma$ confidence level, unless otherwise indicated in Table~\ref{tbl:kn}.
}
\label{fig: lim_mag}
\end{figure*}

\par
\textit{BlackGEM} $-$ BlackGEM is a planned array of wide-field optical telescopes located at the La Silla Observatory in Chile. It will initially consist of three 0.65-m optical telescopes, each with a 2.7~deg$^2$ FoV \citep{blackgem,Groot2019}. Each telescope is equipped with a six-slot ($ugriz$ and broad $q$) filter wheel. The main focus of the BlackGEM mission is the follow-up of LIGO/Virgo/KAGRA candidate events, with a goal of achieving a cadence of two hours on low-latency sky localization using the $u$, $q$, and $i$ filters. In order to determine the most sensitive filters for KN detection, we also include $g$, $r$, and $z$ in our study. The $5\sigma$ limiting magnitude for a 300~s integration time under photometric observing conditions ($1\arcsec$ seeing) is $q\gtrsim 23$~mag (Paul Groot, 2021, Private Communication).

\par
\textit{DDOTI} $-$ DDOTI is a wide-field, robotic imager located at the Observatorio Astron\'omico Nacional (OAN) in Sierra San Pedro M\'artir, Mexico \citep{watson2016}. The instrument is comprised of six 28-cm telescopes with a combined 69~deg$^2$ FoV. DDOTI produces unfiltered images, referred to as the $w$-band. It has previously been using in the follow-up of GW190814 \citep{thakur2020}. Based on DDOTI follow-up during O3 (Becerra et al., in prep.), we adopt a median exposure time of $\sim$\,$2$~hr, yielding a median $10\sigma$ limiting magnitude $w\gtrsim20.5$~mag.

\par
\textit{DECam} $-$ DECam is mounted on the  4-m Victor M. Blanco telescope at the Cerro Tololo Inter-American Observatory (CTIO) in Chile. The instrument has a 2.2~deg$^2$ FoV, and was designed for the purpose of wide-field optical ($ugrizY$) surveys \citep{Flaugher2015}. Electromagnetic follow-up of LIGO/Virgo candidate events has been carried out in previous observing runs by the Dark Energy Survey GW (DESGW) Collaboration \citep[e.g.,][]{SoaresSantos2016,Soares-Santos2017,andreoni2020,Herner2020,Morgan2020} using the $i$- and $z$-bands. In this work, we focus on kilonova detection by DECam at these wavelengths.

\par
\textit{GOTO} $-$ 
At design specifications, GOTO will include an array of $16\times$ 40-cm telescopes on two robotic mounts (producing a 160~deg$^2$ FoV) at two separate locations in Spain and Australia, allowing for coverage of both hemispheres \citep{Dyer2018,Dyer2020,dyer_thesis}. A prototype, referred to as GOTO-4 (18~deg$^2$ FoV), was instituted in La Palma, Spain in 2017 with an initial array of 4 telescopes. This was upgraded to an 8 telescope array in 2020 (GOTO-8), yielding a 40~deg$^2$ FoV \citep{Dyer2020}.
The prototype mission, GOTO-4 has demonstrated its capability to cover large GW localization regions during O3 with good sensitivity to optical transients \citep{Gompertz2020}. 
In a 360~s integration time, GOTO can reach depths of $L\gtrsim 21.0$~mag (Ben Gompertz \& Martin Dyer, 2021, Private Communication), where $L$ is a wide-band filter\footnote{\url{https://github.com/GOTO-OBS/public_resources/tree/main/throughput}}, approximately equivalent to $g+r$.

\par
\textit{LSST} $-$ The Vera C.\ Rubin Observatory, currently under construction on Cerro Pachon in Chile, has a planned first light in 2023. The observatory consists of an 8.4-m wide-field, optical telescope covering a 9.6~deg$^2$ FoV.
The Rubin Observatory's LSST will survey half the sky every three nights in the $ugrizy$ filters. The planned Wide-Fast-Deep (WFD) survey will reach $r\gtrsim 24.2$~mag (assuming airmass $\sim$\,$1.2$) for a 30~s integration time \citep{LSST}. 
Despite the lack of a formal GW follow-up strategy, we have included LSST within this study to highlight the sensitivity of the WFD survey \citep[see also, e.g.,][]{scolnic2018,Cowperthwaite2019} in order to encourage GW follow-up, and demonstrate the prospect of serendipitous detection.

\par
\textit{MeerLICHT} $-$ MeerLICHT is a fully robotic 0.65-m optical telescope located at the South African Astronomical Observatory (SAAO) in Sutherland, South Africa \citep{blackgem}. MeerLICHT was designed as the prototype instrument for the BlackGEM array \citep{Groot2019}. The telescope has a 2.7~deg$^2$ FoV, and covers the same $ugr(q)iz$ wavelengths as BlackGEM. MeerLICHT has been used for GW follow-up during past LIGO/Virgo observing runs \citep[e.g., GW190814;][]{deWet2021} with observations in the $u$, $q$, and $i$ filters. For comparison to BlackGEM, we likewise include the $g$, $r$, and $z$ filters in our study. MeerLICHT is sky background limited in 60~s exposures, yielding a limiting magnitude $q\gtrsim20.6$~mag.

\par
\textit{PRIME} $-$ PRIME is a 1.8-m wide-field infrared telescope under construction at SAAO. The telescope has a 1.56~deg$^2$ FoV, and covers wavelengths $ZYJH$. In a 100~s integration time PRIME reaches depths of $z\gtrsim 20.5$~mag and $Y\gtrsim 20.0$~mag (Takahiro Sumi, 2021, Private Communication).

\par
\textit{Roman} $-$ The 2.4-m Roman Space Telescope (formerly WFIRST), with  planned launch in 2025, will cover a 0.28~deg$^2$ FoV \citep[$\sim$\,$200$ times larger than \textit{HST};][]{roman}. The Wide Field Instrument (WFI) is sensitive to optical/infrared wavelengths between 5000--20000~$\mathrm{\AA}$. In this study, we focus on the $RZYJHF$ filters. We have chosen to include Roman within this study in order to encourage GW follow-up \citep[see also][]{Foley2019wfirst}, and demonstrate its sensitivity to kilonovae out to cosmological distances. The instrument sensitivity is adopted following \citet{hounsell2018} and \citet{scolnic2018}.

\par
\textit{UVOT} $-$ The UVOT onboard the \textit{Neil Gehrels Swift Observatory} has a wavelength coverage of 1600--8000~$\mathrm{\AA}$, and a 0.08~deg$^2$ FoV \citep{swift}. Despite its smaller FoV, \textit{Swift} is able to cover large regions of the sky in $<24$~hr with a rapid response (a few hours) to LIGO/Virgo alerts \citep{Evans2016,evans2017,Klingler2019,Page2020,Klingler2021}. In addition, \textit{Swift} has the added benefit of simultaneous X-ray coverage from the X-ray Telescope (XRT). UVOT follow-up has been optimized for its smaller FoV by targeting galaxies with high probabilities of being the host, ensuring they are completely within the FoV \citep{Klingler2019}. The large majority of UVOT tiles are observed with the $u$-band, and, therefore, we focus our analysis on this filter. Based on GW follow-up during O3, we adopt a median $5\sigma$ limiting magnitude $u\gtrsim 19.9$ mag (Oates et al., in prep.).

\par
\textit{ULTRASAT} $-$ ULTRASAT\footnote{\url{https://www.weizmann.ac.il/ultrasat/mission/mission-design-overview}\label{ultraweb}} is an ultraviolet telescope with planned launch to geostationary orbit in 2024. The instrument will have a 200~deg$^2$ FoV, and cover wavelengths between 2200--2800~$\mathrm{\AA}$, referred to as $NUV$. The estimated limiting magnitude in 900~s integration time is $NUV\gtrsim22.3$~mag. Since ULTRASAT does not have a publicly available filter function, we approximated the filter response with a top-hat function between 2200 and 2800~$\mathrm{\AA}$.

\par
\textit{VISTA} $-$ VISTA is a 4-m wide-field survey telescope equipped with the VISTA InfraRed CAMera (VIRCAM) covering wavelengths $ZYJHK_s$ with a $1.6$~deg$^2$ FoV \citep{vista}. VISTA is located at the Cerro Paranal Observatory in Chile, and operated by the European Southern Observatory (ESO). Follow-up of LIGO/Virgo candidate events has largely occurred through the Vista Near infra-Red Observations Uncovering Gravitational wave Events (VINROUGE) project in the $Y$, $J$, and $K_s$ filters \citep[e.g,][]{tanvir2017,ackley2020}. The $5\sigma$ limiting magnitudes were taken from the Vista Hemisphere Survey \citep{McMahon2013,Banerji2015} and re-scaled to a 360~s exposure time, yielding $Y\gtrsim 21.5$~mag, $J\gtrsim 21.0$~mag, $H\gtrsim 21.0$~mag, and $K_s \gtrsim 20.0$~mag. 

\par
\textit{WINTER} $-$ WINTER is a new infrared instrument, with planned first light in mid-2021 \citep{winter,Frostig2020}. WINTER will use a 1-m robotic telescope located at the Palomar Observatory in California, United States. The instrument has a $\sim$\,1~deg$^2$ FoV, and covers infrared wavelengths $YJH_s$. In a 360~s integration time, WINTER reaches $5\sigma$ depth $Y\gtrsim 21.5$~mag, $J\gtrsim 21.3$~mag, and $H_s\gtrsim 20.5$~mag (Nathan Lourie \& Danielle Frostig, 2021, Private Communication). 

\par
\textit{ZTF} $-$ ZTF employs a wide-field camera (47~deg$^2$ FoV) on the Palomar 48-in (P48) Oschin (Schmidt) telescope in San Diego County, California, United States \citep{ztf}. Using 30~s exposures, it can cover 3760~deg$^2$ hr$^{-1}$ to limiting magnitude $r\gtrsim 20.6$~mag \citep{Graham2019}. This capability makes ZTF an effective tool for kilonova searches, both for GW candidate events \citep{coughlin2019,Kasliwal2020} and serendipitous discovery \citep{Andreoni2021}.

We note that the follow-up strategy and sensitivity of these instruments varies greatly depending on a number of factors, such as the observing conditions, sky location, and the size of the GW localization region. The sensitivity assumed in this work is considered to be an approximate limiting magnitude for GW follow-up, while noting that this sensitivity is variable day-to-day for each ground-based instrument. We also note that the limiting magnitudes are computed for ``snapshot'' exposures, and more sensitive images can be obtained by stacking multiple exposures over the course of a night.
Figure~\ref{fig: lim_mag} displays the assumed limiting magnitudes, indicating separate filters by their effective wavelength. 
For consistency, we compute effective wavelengths from all filter response functions using Eq. 1 of \citet{king1952}.
These instruments provide an excellent coverage of the range of wavelengths expected for kilonovae. In Table~\ref{tbl:kn}, we present the FoV, typical exposure time of GW tiling, filter effective wavelengths, and limiting magnitudes for each instrument.

\section{Assessing Kilonova Detectability} \label{sec: detectability}

\begin{deluxetable*}{cc}
\tablecaption{Properties of LANL kilonova simulations (adapted from \citealt{wollaeger2020}). \label{tbl:nome}}
\tablenum{2}
\tablehead{\colhead{Property} & \colhead{Values}}
\startdata
\hline
    Dyn. ejecta mass & $\{0.001,0.003,0.01,0.03,0.1\}$ \msun{} \\
    Wind ejecta mass & $\{0.001,0.003,0.01,0.03,0.1\}$ \msun{} \\
    Dyn. ejecta velocity & $\{0.05,0.15,0.3\}$ $c$ \\
    Wind ejecta velocity & $\{0.05,0.15,0.3\}$ $c$ \\
    Dyn. ejecta morphology & Toroidal (T; Cassini oval family;~\citealt{korobkin2020}) \\
    Wind ejecta morphology & Spherical (S) or ``Peanut'' (P; Cassini oval family;~\citealt{korobkin2020}) \\
    Dyn. ejecta composition & initial $Y_e = 0.04$  (see Table 2 in \citealt{wollaeger2020}) \\
    Wind ejecta composition & initial $Y_e = 0.27$ or $0.37$  (see Table 2 in \citealt{wollaeger2020}) \\
    \hline
\enddata 
\end{deluxetable*}

We employ the Los Alamos National Laboratory (LANL) grid of kilonova simulations \citep{wollaeger2020} to assess each instrument's ability to observe a kilonova.
Rather than limiting our study to \at-like kilonovae, these radiative transfer simulations span a wide range of ejecta parameters, including expected values from both BNS and NSBH progenitors.
The LANL simulations are two-component, multi-dimensional, axisymmetric kilonova models, generated with the Monte Carlo radiative transfer code {\tt SuperNu}~\citep{wollaeger2013,wollaeger2014}. 
These simulations rely on nucleosynthesis results from the {\tt WinNet} code \citep{winteler2012, korobkin2012} in addition to a set of tabulated binned opacities \citep{fontes2020} from the LANL suite of atomic physics codes \citep{fontes2015}.
The most recent LANL grid of kilonova simulations \citep{wollaeger2020} includes a full set of lanthanides and fourth row elements, some of which were not included in previous datasets \citep{wollaeger2018}.

\begin{figure}
  \centering
  \includegraphics[width=0.60\columnwidth]{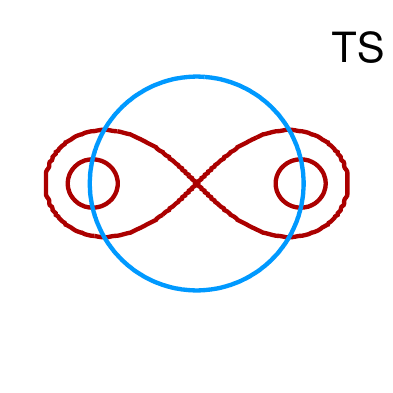}\\
  \includegraphics[width=0.60\columnwidth]{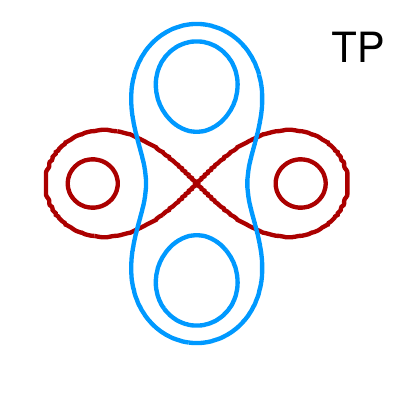}
  \caption{
    Schematics of the two combined morphologies used in the simulation grid \citep{wollaeger2020}.
    All models have a toroidal (T, red) dynamical ejecta, 450 models are simulated
    with a spherical wind (S, blue), and 450 models are simulated with a peanut-shaped
    wind (P, blue).
    Each component is varied over the mass-velocity grid in Table~\ref{tbl:nome},
    and hence is not necessarily drawn to scale here (adapted from \citealt{korobkin2020} and \citealt{wollaeger2020}).
  }
  \label{fig:morph}
\end{figure}

These radiative transfer simulations simultaneously evolve two ejecta components: a dynamical ejecta and wind ejecta component. 
The dynamical ejecta component is initiated with a low electron fraction ($Y_e = 0.04$), and corresponds to the lanthanide-rich or ``red'' emission.
This component has composition consistent with the robust ``strong'' \textit{r}-process pattern such as the one repeatedly found in metal-poor \textit{r}-process enriched stars and in the Solar \textit{r}-process residuals \citep{sneden09,holmbeck2020b}. 
The second, lanthanide-poor wind ejecta component corresponds to the ``blue'' kilonova emission and is primarily composed of wind-driven ejecta from the post-merger accretion disk. 
The wind ejecta is initiated with two separate wind compositions, representing either high- ($Y_e = 0.37$) or mid- ($Y_e = 0.27$) latitude wind composition, consistent with winds induced by a wide range of post-merger remnants \citep{lippuner2017,wollaeger2020}.
Dynamical ejecta is initiated with a toroidal morphology, constrained near the binary's orbital plane, while the wind ejecta is modeled by either a spherical or ``peanut-shaped'' geometry \citep{korobkin2020}.
Figure~\ref{fig:morph} presents the two morphological configurations used in the simulation set.
Models span five ejecta masses (0.001, 0.003, 0.01, 0.03, and 0.1\msun{}) and three ejecta velocities (0.05, 0.15, and 0.3$c$) for each component, covering the full range of kilonova properties anticipated from numerical simulations (see \citealt{wollaeger2020} and references therein).
By considering all possible combinations of ejecta properties, the LANL grid includes 900 kilonova simulations.
The full set of simulation parameters are compiled in Table~\ref{tbl:nome}.

\begin{figure}
\centering
\includegraphics[width=\columnwidth, trim=20 0 100 00, clip]{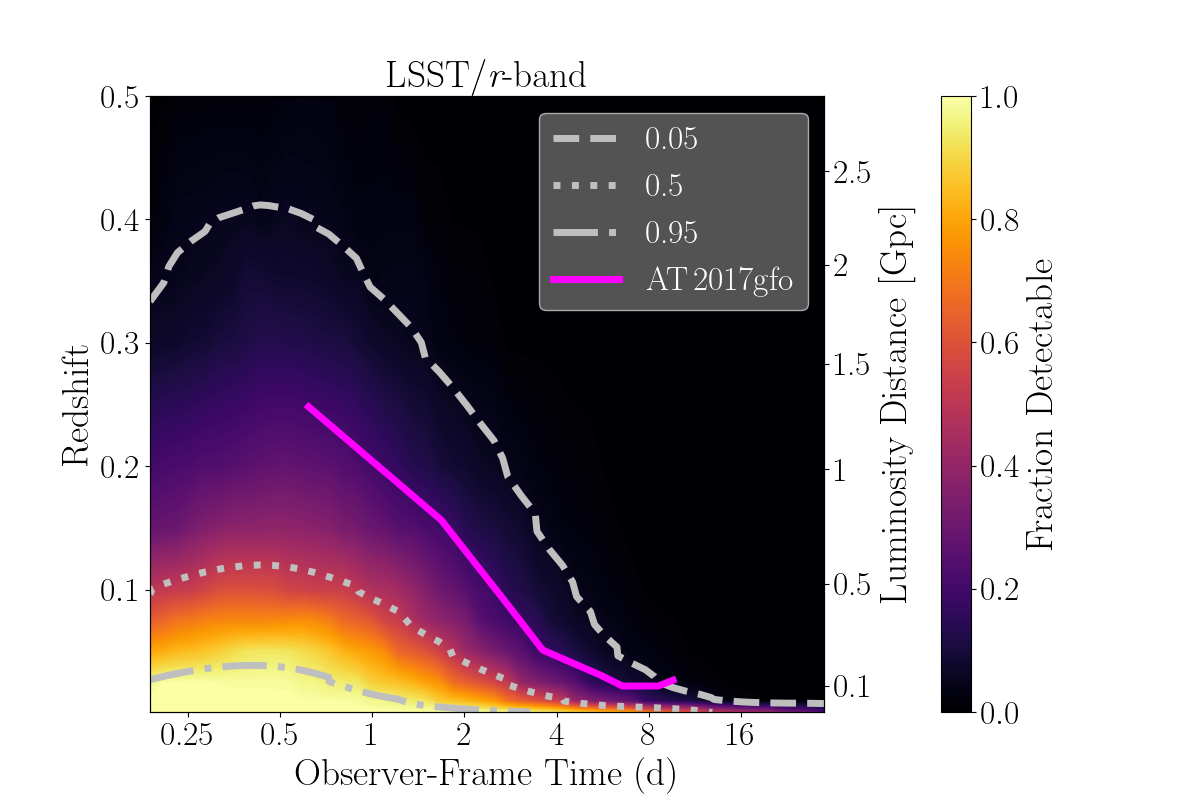}
\includegraphics[width=\columnwidth, trim=20 0 100 0, clip]{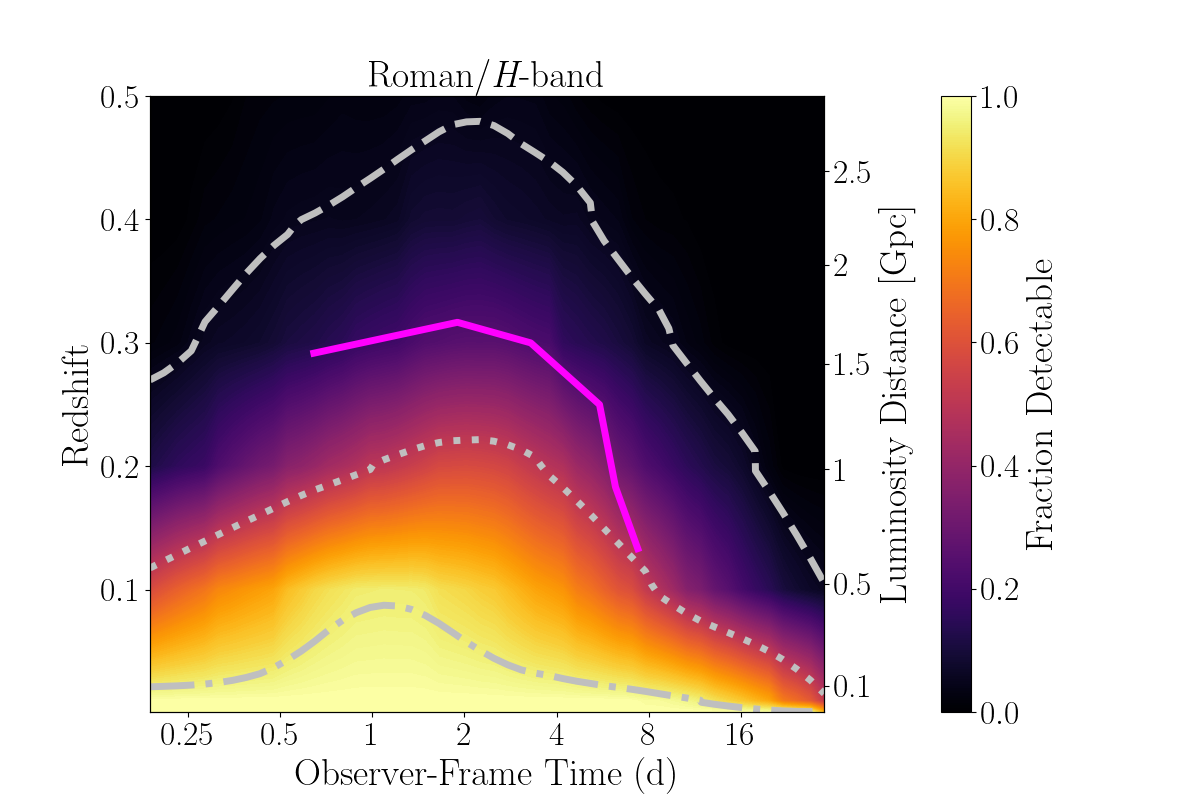}
\caption{Detectability constraints for two filters: LSST/\textit{r}-band (\textit{top}) and Roman/\textit{H}-band (\textit{bottom}). Contours indicate the fraction of 48,600 simulated kilonovae (900 simulations each rendered at 54 viewing angles) with apparent magnitudes brighter than the limiting magnitude in each filter, for a given redshift and observer-frame time. The three white contours demarcate regions where 5\%, 50\%, and 95\% of simulated kilonovae are detectable. The magenta curve represents each filter's ability to detect \at-like kilonovae.
}
\label{fig: contours}
\end{figure}

These 900 simulations have previously been used to estimate kilonova properties associated with follow-up of GW190814 \citep{thakur2020} and targeted observations of two cosmological sGRBs \citep{oconnor2020,bruni2021}.
Additionally, the LANL simulation grid is the basis for an active learning-based approach to a kilonova lightcurve surrogate modeling and parameter estimation framework \citep{ristic2021}.

All 900 multi-dimensional simulation are rendered in 54 viewing angles, each subtending an equal solid angle of $4\pi/54$~sr.
In this study, we use spectra from all 54 viewing angle renditions to quantify detectability, making the generous assumption that a GW event has equal probability of detection with all viewing angles.\footnote{Gravitational waves from compact-binary coalescences are preferentially detected near face-on and face-off inclinations \citep{finn1993, nissanke2010, schutz2011}, although a wide range of viewing angles are detectable at small distances \citep{gwtc1,lvc190412,lvc190814}. We can further constrain kilonova detectability by including low-latency estimates on inclination.}
This differs from kilonova searches aimed at sGRBs, such as in \citealt{oconnor2020}, where we limit our simulation grid to face-on viewing angles ($\theta_v\lesssim15.64^\circ$) as sGRBs are typically detected on-axis. 
By including all 54 viewing angles of the 900 LANL kilonova simulations, our detectability study incorporates a full suite of 48,600 kilonova models.

Each model includes a set of time-dependent spectra, which are converted to lightcurves by computing a magnitude associated with each spectrum. 
Spectra are not simulated for the first three hours post-merger (rest frame). 
We render each simulated spectra set into a lightcurve for various redshifts, $z$, and for a broad range of filters with corresponding wavelength-dependent bandpass filter functions $R\left(\lambda_0\right)$. 
Magnitudes are computed according to 
\begin{equation}
m_{\mathrm{AB}} = -2.41 -\frac{\int_{0}^{\infty} f\left( \lambda_0 \left(1+z\right)^{-1}\right)  R\left( \lambda_0\right) \lambda_0  d\lambda_0}{ \left(1+z\right) \int_{0}^{\infty} R\left( \lambda_0\right) \lambda_0^{-1} d\lambda_0},
\label{eq:m_AB}
\end{equation}
where $\lambda_0$ is the observed wavelength and $f\left(\lambda_0\right)$ is the wavelength-dependent spectral flux density  \citep{hogg2002, blanton2003}.
In Equation~\ref{eq:m_AB}, we account for cosmological K-corrections \citep{humason1956, oke1968} by converting rest-frame spectral emission to observer-frame wavelengths. 

We define a kilonova as detectable in a given filter if it outshines the limiting magnitude of the filter, as listed in Table~\ref{tbl:kn}.
Then, we trace the detectability's variation with redshift, by considering lightcurves at various cosmological distances. 
Figure~\ref{fig: contours} displays detectability constraints for two representative filters: the LSST \textit{r}-band (optical) and the Roman Space Telescope's  \textit{H}-band (near-infrared).
We present detectability, defined as the fraction of detectable kilonova simulations, as a function of both redshift and observer-frame time.
Detectability in the LSST/\textit{r}-band decreases over time, as kilonovae optical emission fades within $\sim$1~day post-merger.
However, the higher wavelength Roman/\textit{H}-band reaches peak detectability two days post-merger, with 5\% of nearby ($<$ 1~Gpc) kilonovae detectable over two weeks after merger.
For comparison, Figure~\ref{fig: contours} includes detectability constraints for \at-like kilonovae, computed from spectroscopic data \citep{chornock2017,cowperthwaite2017,nicholl2017,pian2017,shappee2017,smartt2017}.\footnote{Compiled on \url{kilonova.space} \citep{guillochon2017}}
We do not present \at-like detectability within 12 hours of merger, as no spectra were collected at this point in the kilonova evolution. 
Our \at-like detectability metrics are broadly consistent with previous studies \citep{scolnic2018,rastinejad2021}.
Similar figures for all filters in Table~\ref{tbl:kn} are available in Appendix~\ref{sec: app}.

\begin{figure*}
\centering
\includegraphics[width=0.75\textwidth]{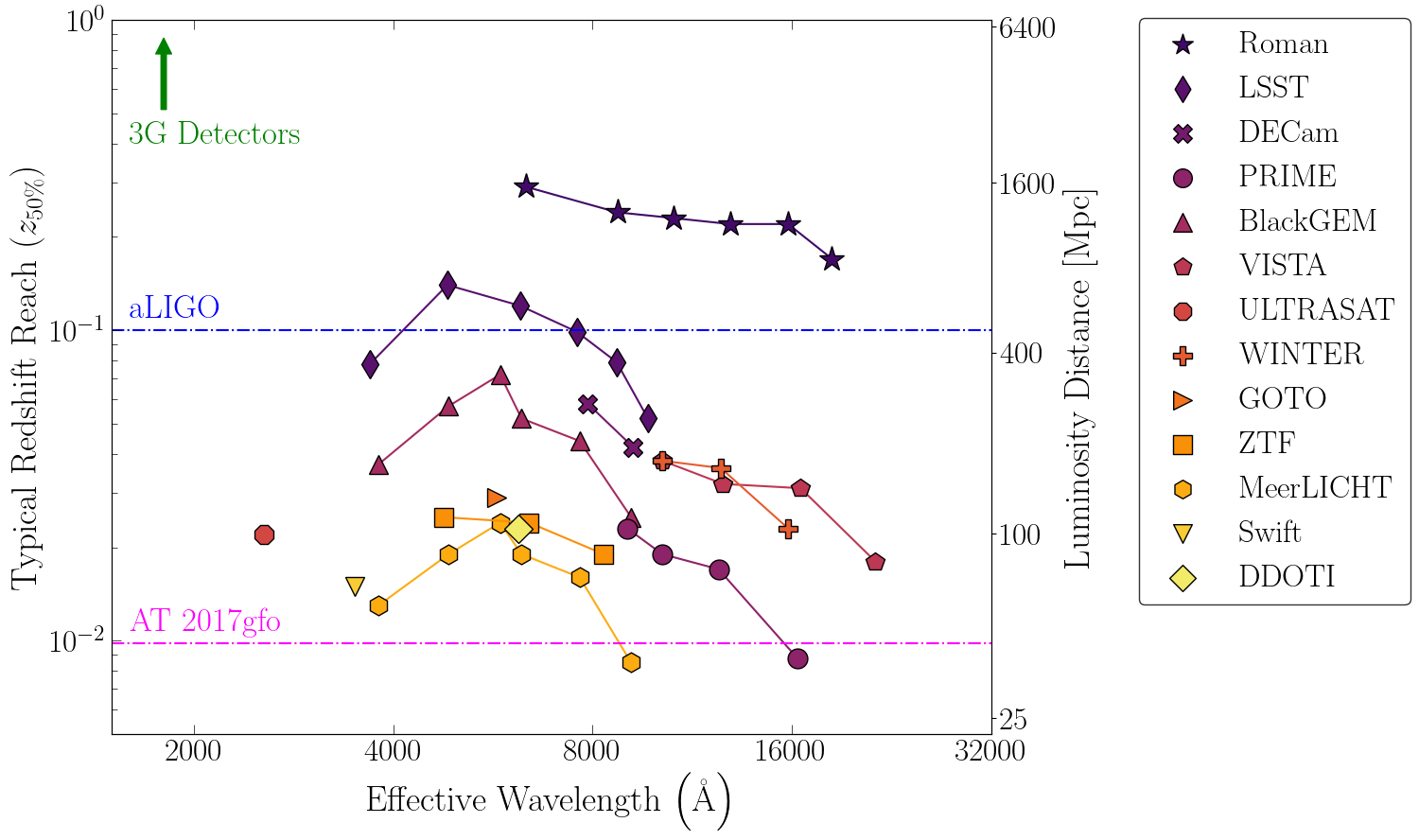}
\caption{
Typical redshift reach for a selection of instruments. The vertical axis indicates the maximum redshift at which 50\% of the LANL simulated lightcurves are detectable in a given filter at any one time. Similarly to Figure~\ref{fig: summary}, each symbol corresponds to one instrument, with an instrument's filter represented by their bandpass filter function's effective wavelength. Lines connect filters from the same instrument. The magenta horizontal line indicates the redshift of \at's host galaxy \citep{kourkchi2017}, while the blue line indicates the redshift horizon for a 1.4\msun{} + 1.4\msun{} BNS at advanced LIGO's design sensitivity \citep{hall2019}. Horizon redshifts for third-generation GW detectors, including Cosmic Explorer and Einstein Telescope, exceed the maximum redshift in the figure.  
}
\label{fig: summary}
\end{figure*}

For each filter, we compute the maximum redshift at which 50\% of simulated kilonovae are detectable at any one time, $z_{50\%}$, called the ``\zhoriz{}.''  
With this definition, we ascribe a \zhoriz{} of $z_{50\%}=0.12$ and $z_{50\%}=0.22$ to the LSST/\textit{r}-band and Roman/\textit{H}-band in Figure~\ref{fig: contours}, respectively. 
Table~\ref{tbl:kn} lists the \zhoriz{}es for all instruments and filters, in addition to compiling maximum detectable redshifts for both 95\% and 5\% of simulated kilonovae as $z_{95\%}$ and $z_{5\%}$, respectively.

Additionally, Figure~\ref{fig: summary} compares the \zhoriz{} for all instruments in this study, highlighting the differences as a function of their effective wavelengths (see Table~\ref{tbl:kn}). Typical redshift reaches vary from $z_{50\%}=0.0085$ ($\sim$38~Mpc) for the MeerLICHT/\textit{z}-band up to $z_{50\%}=0.29$ ($\sim$1.5~Gpc) for the Roman/\textit{R}-band.  
As anticipated, the \zhoriz{} for each instrument correlates significantly with the limiting magnitude distribution in Figure~\ref{fig: lim_mag}.
We remind the reader that simulated lightcurves are not available earlier than three hours post-merger (rest frame), resulting in no detectability predictions in this time range. 
This omission may bias the detectability estimates of low-wavelength (ultraviolet) filters, such as the ULTRASAT/$NUV$-band, LSST/\textit{u}-band, and UVOT/\textit{u}-band.

\section{Detectability Variations with Kilonova Properties} \label{sec: variation}

The \zhoriz{} often varies significantly with kilonova properties, such as the ejecta mass or velocity. 
Larger ejecta masses generally result in more luminous kilonova emission, allowing for detection at higher redshifts, while kilonovae containing lower ejecta masses are only detectable at nearby distances.
This mass dependency makes it difficult to determine whether a GW candidate event will produce an observable kilonova for a given instrument.
Robust detectability metrics are further muddled by compounding degeneracies with other parameters such as velocity of the expanding ejecta, viewing angle, and composition.

Differing kilonova properties induce variations in kilonova detectability.
For example, a kilonova with large wind ejecta mass and small dynamical ejecta mass may be easily detectable in the ultraviolet but difficult to observe in near-infrared filters, while other kilonova parameters may produce emission that is primarily detectable in the infrared. 
Lower-wavelength filters probe the wind ejecta, with peak emission in optical wavelengths at early times. 
At the low-wavelength extreme, ultraviolet instruments capture the early structure of the outermost wind-driven ejecta \citep{arcavi2018,banerjee2020}. 
However, these low-wavelength filters offer little insight into the dynamical ejecta, which peaks at redder wavelengths.
Additionally, a filter's variation with kilonova parameters depends on the source redshift: high-redshift kilonova emission is shifted to higher wavelengths, requiring subsequently redder filters to capture variability in dynamical ejecta mass.
If kilonovae properties are unknown, multi-band observations across UVOIR wavelengths are necessary to maximize the probability of kilonova detection.

\begin{figure*}
\centering
\includegraphics[width=\columnwidth, trim=20 0 100 00, clip]{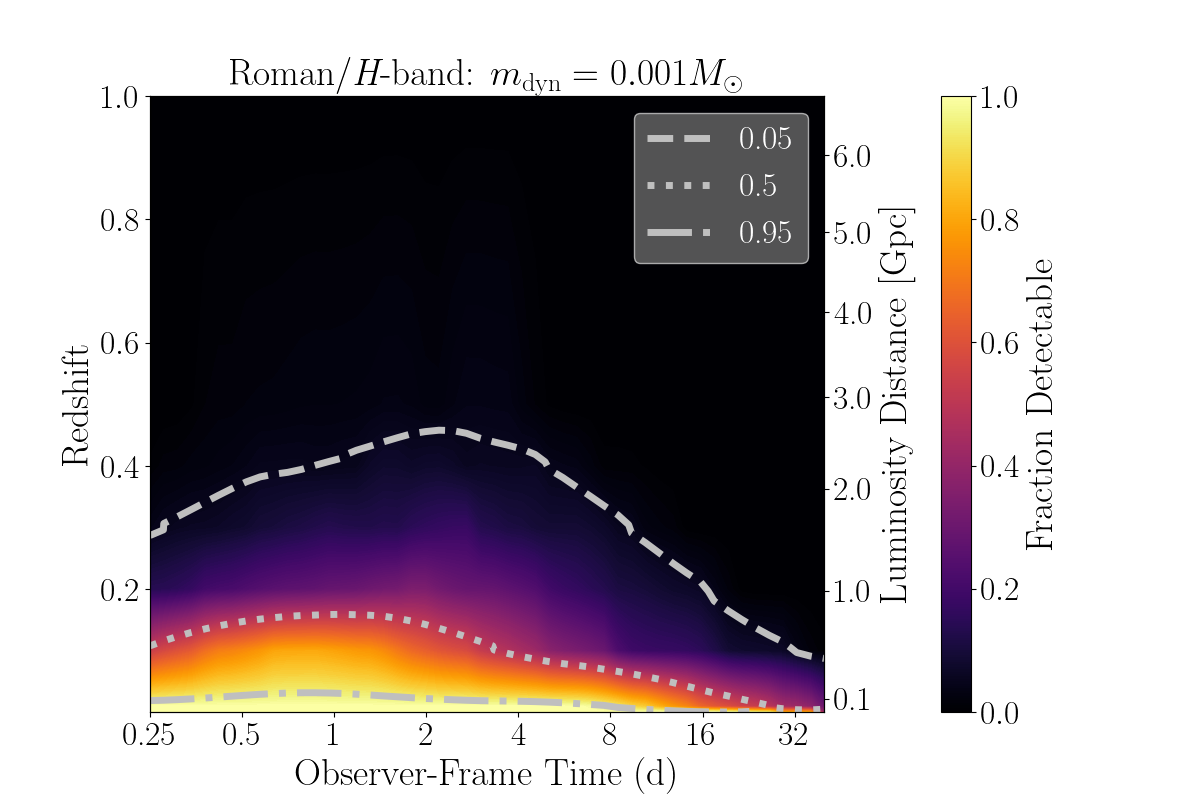}
\includegraphics[width=\columnwidth, trim=20 0 100 0, clip]{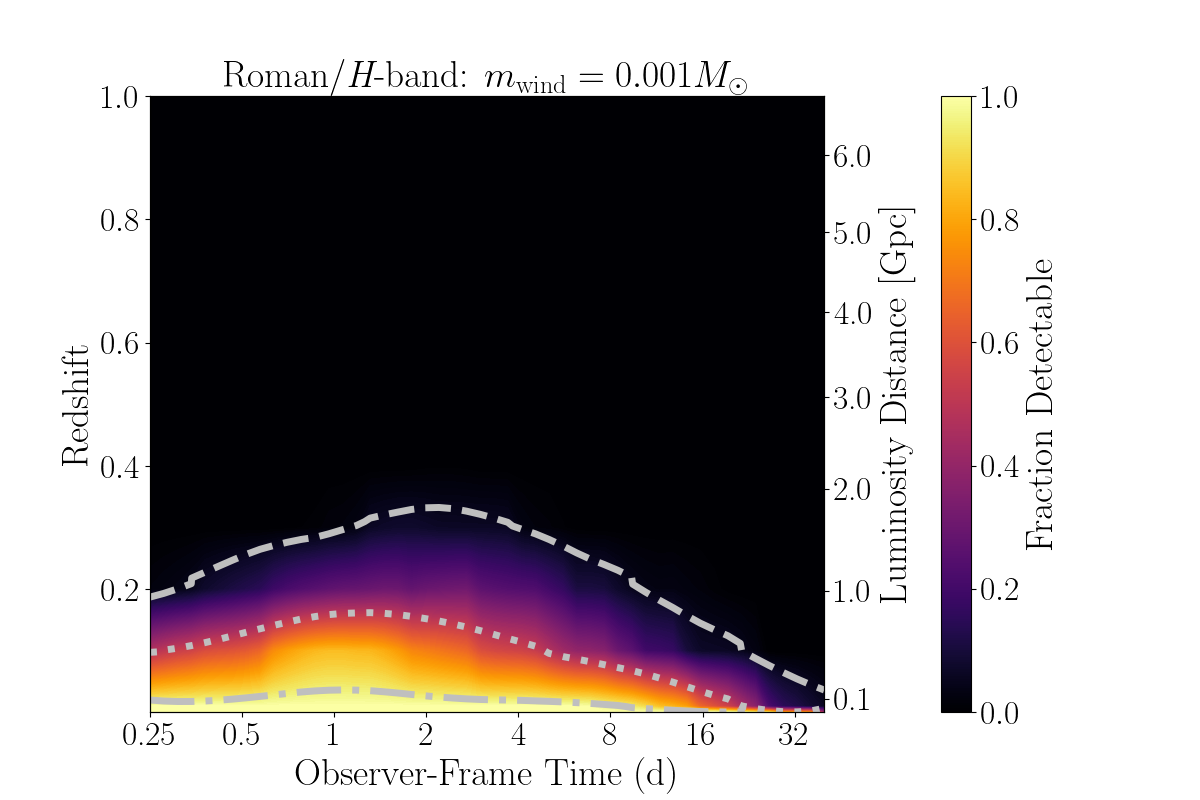}
\includegraphics[width=\columnwidth, trim=20 0 100 00, clip]{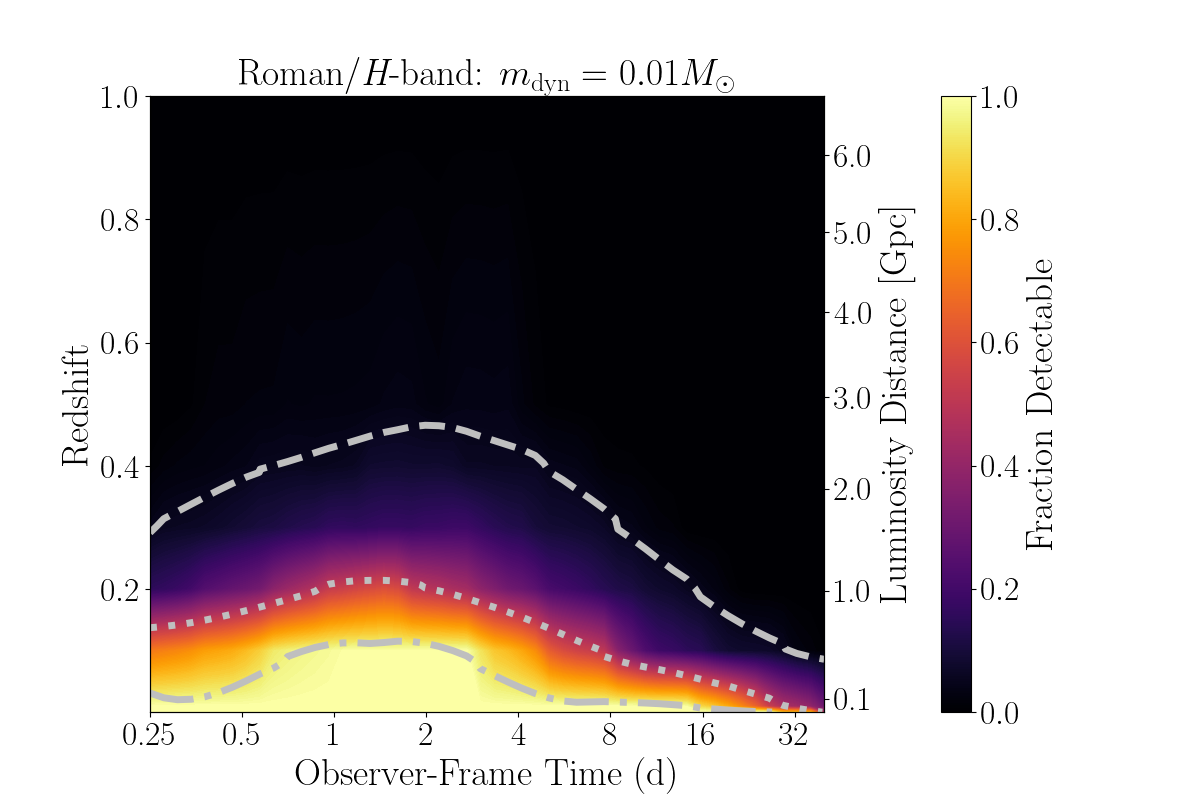}
\includegraphics[width=\columnwidth, trim=20 0 100 0, clip]{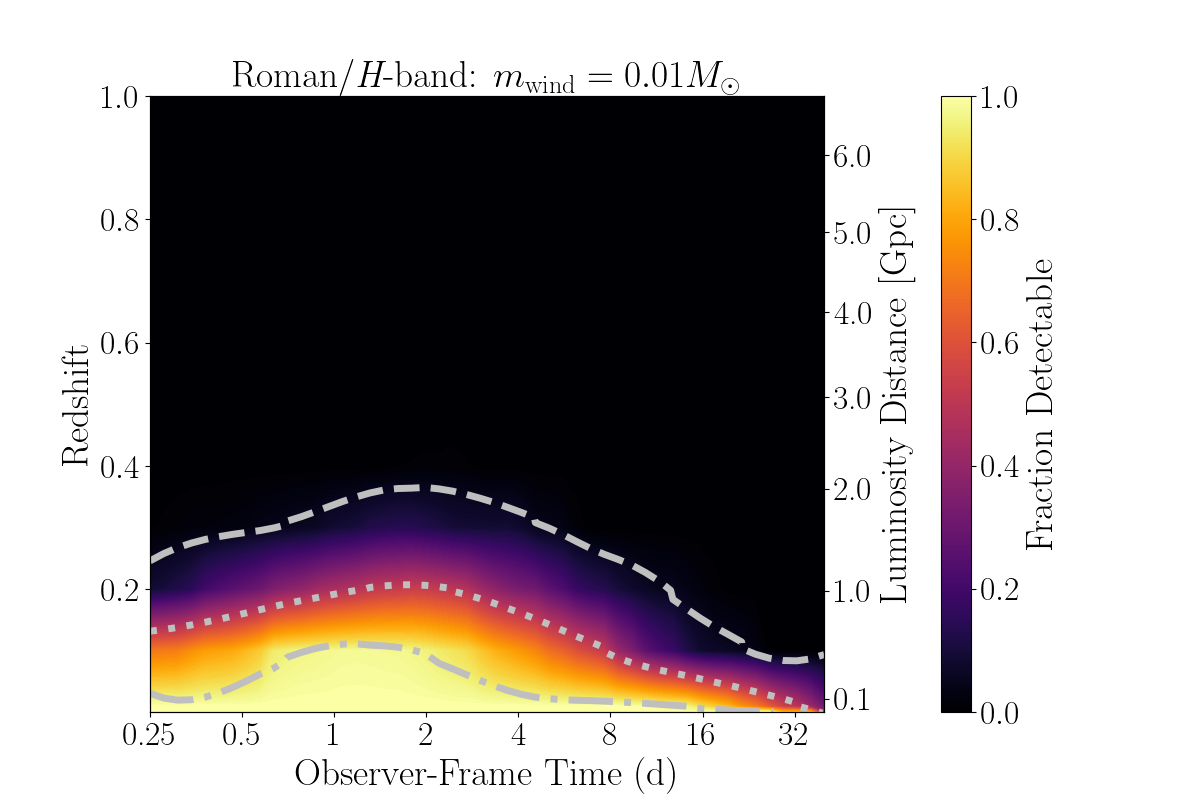}
\includegraphics[width=\columnwidth, trim=20 0 100 0, clip]{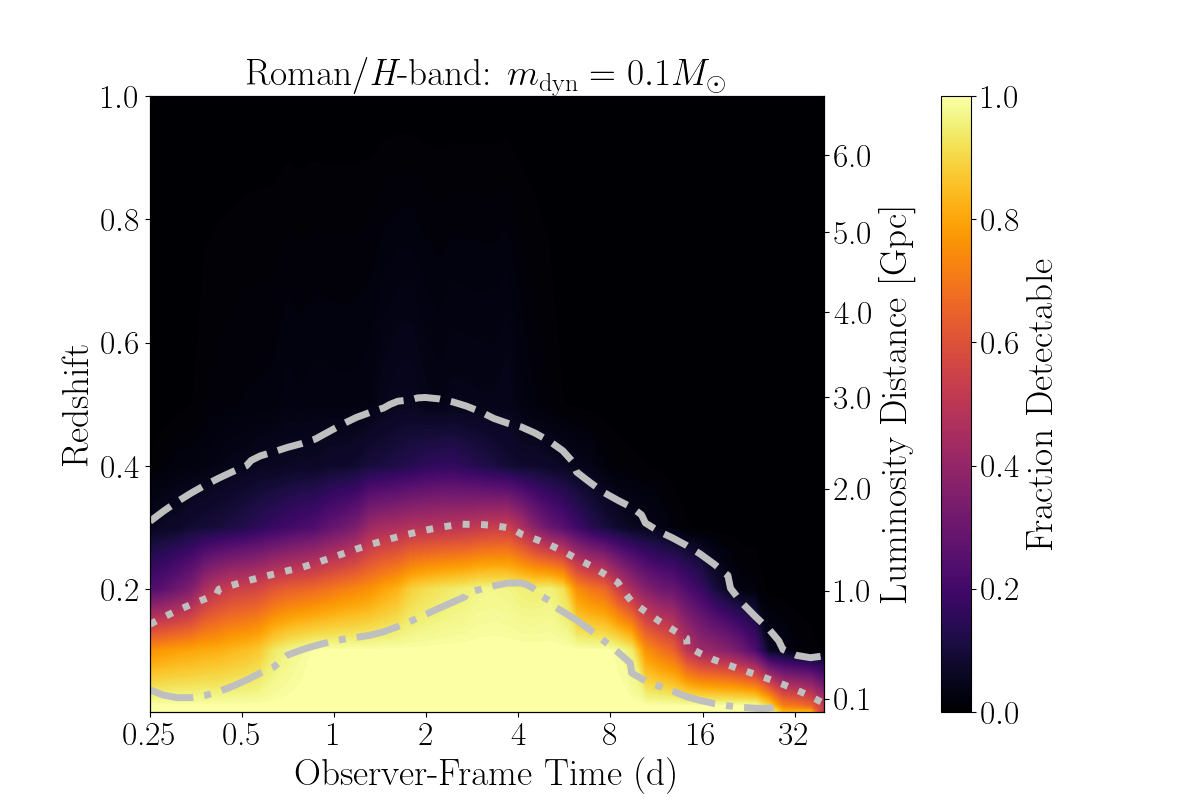}
\includegraphics[width=\columnwidth, trim=20 0 100 0, clip]{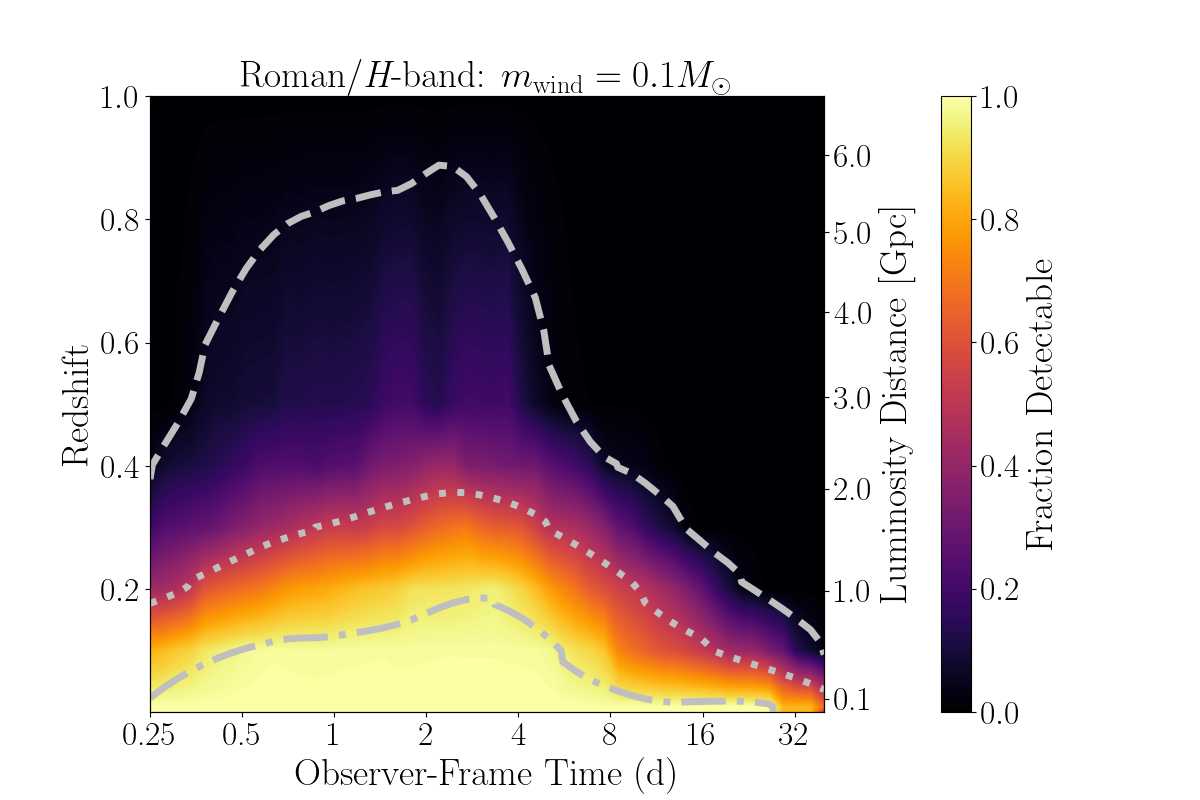}
\caption{Detectability constraints for the Roman/\textit{H}-band for six sets of kilonova simulations. The left column represents variation with dynamical ejecta mass, while the right presents variation with wind ejecta mass. The top row corresponds to lower masses (0.001\msun{}), middle row corresponds to an intermediate mass (0.01\msun{}), while the bottom row present high ejecta masses (0.1\msun{}). We place no restrictions on ejecta velocities, morphology, composition, or viewing angle, such that each panel includes 9720 simulations (180 simulated kilonovae each with 54 viewing angles) of the full set of 48,600 LANL kilonovae.
}
\label{fig: Hband_variability}
\end{figure*}

Figure~\ref{fig: Hband_variability} demonstrates the variation in kilonova detectability in the Roman/\textit{H}-band for different ejecta masses. 
The left (right) column restricts the simulation set to one dynamical (wind) ejecta mass, while allowing all other parameters to vary, resulting in 9720 simulations per panel (180 simulations each with 54 viewing angles). 
In the left column, dynamical ejecta mass varies from the lowest (0.001\msun{}) to the largest (0.1\msun{}) values in the LANL simulation grid.
Larger ejecta masses enhance detectability, as 50\% of simulations with dynamical ejecta masses of 0.1\msun{} are detectable out to $z=0.31$, with peak emission three days post-merger. 
Lower ejecta masses induce both dimmer emission and an earlier peak timescale \citep[e.g., ][]{kasen2017}, as shown by the diminished late-time detectability and smaller \zhoriz{} ($z=0.16$) for dynamical ejecta masses of 0.001\msun{}. 
However, a small subset of kilonovae with low dynamical ejecta masses remain detectable at redshifts $z > 0.3$, consistent with the addition of a large wind ejecta mass. 
The variation is a bit more pronounced for wind ejecta: 50\% of simulated kilonovae with large wind ejecta masses (0.1\msun{}) are detectable out to $z=0.37$, while less than 5\% of low wind ejecta (0.001\msun{}) kilonovae are detectable at such high redshifts. 
Additionally, as higher redshift ($z \gtrsim 0.4$) kilonova emission is proportionally shifted to higher wavelengths, the Roman/\textit{H}-band becomes less effective at detecting emission from high dynamical ejecta mass mergers. 
As a result, variation with dynamical ejecta mass decreases with redshift.

Different wavelength bands exhibit varying dependency on ejecta mass.
Similarly to in the near-infrared (Roman/\textit{H}-band), detectability at optical and ultraviolet wavelengths varies significantly with wind ejecta masses.
However, optical and ultraviolet bands show little variability with dynamical ejecta masses. 
We define the variable $v$ to quantify a given filter's sensitivity to a kilonova property such as ejecta mass. 
As an example, we can quantify the Roman/\textit{H}-band's dependence on wind ejecta mass by comparing the 50\% detectability contours (dot--dashed lines) in Figure~\ref{fig: Hband_variability}.
We label the 50\% detectability contour for the lowest ejecta mass (top panel) and highest ejecta mass (bottom panel) as $g(t)$ and $f(t)$, respectively.
We then quantify variability with mass as:
\begin{equation} \label{eq: var}
v = \frac{\int_{t_{\mathrm{min}}}^{t_{\mathrm{max}}} \left|f\left(t\right) - g\left(t\right)\right| dt}{\frac{1}{2} \int_{t_{\mathrm{min}}}^{t_{\mathrm{max}}} \left[f\left(t\right) + g\left(t\right)\right] dt},
\end{equation} 
integrating from the smallest rest-frame time, $t_{\mathrm{min}} = 0.125$~d, to a maximum time of $t_{\mathrm{max}} = 20$~d. 
Values of $v$ close to zero indicate negligible variation with a given parameter, while higher values of $v$ indicate a significant dependence. 
There is no upper limit on $v$, although we note that a value of $v=1$ corresponds to a three-fold enhancement in $z_{50\%}$ between two subsets of parameters (i.e. $f\left(t\right) = 3g\left(t\right)$).
Based on the 50\% contours in Figure~\ref{fig: Hband_variability}, the Roman/\textit{H}-band produces $v=0.91$ for dynamical ejecta mass (left column) and $v=0.94$ for wind ejecta mass (right column), suggesting that the Roman/\textit{H}-band's detectability is slightly more dependent on wind ejecta mass than dynamical ejecta mass.

\begin{figure*}
\centering
\includegraphics[width=0.75\textwidth]{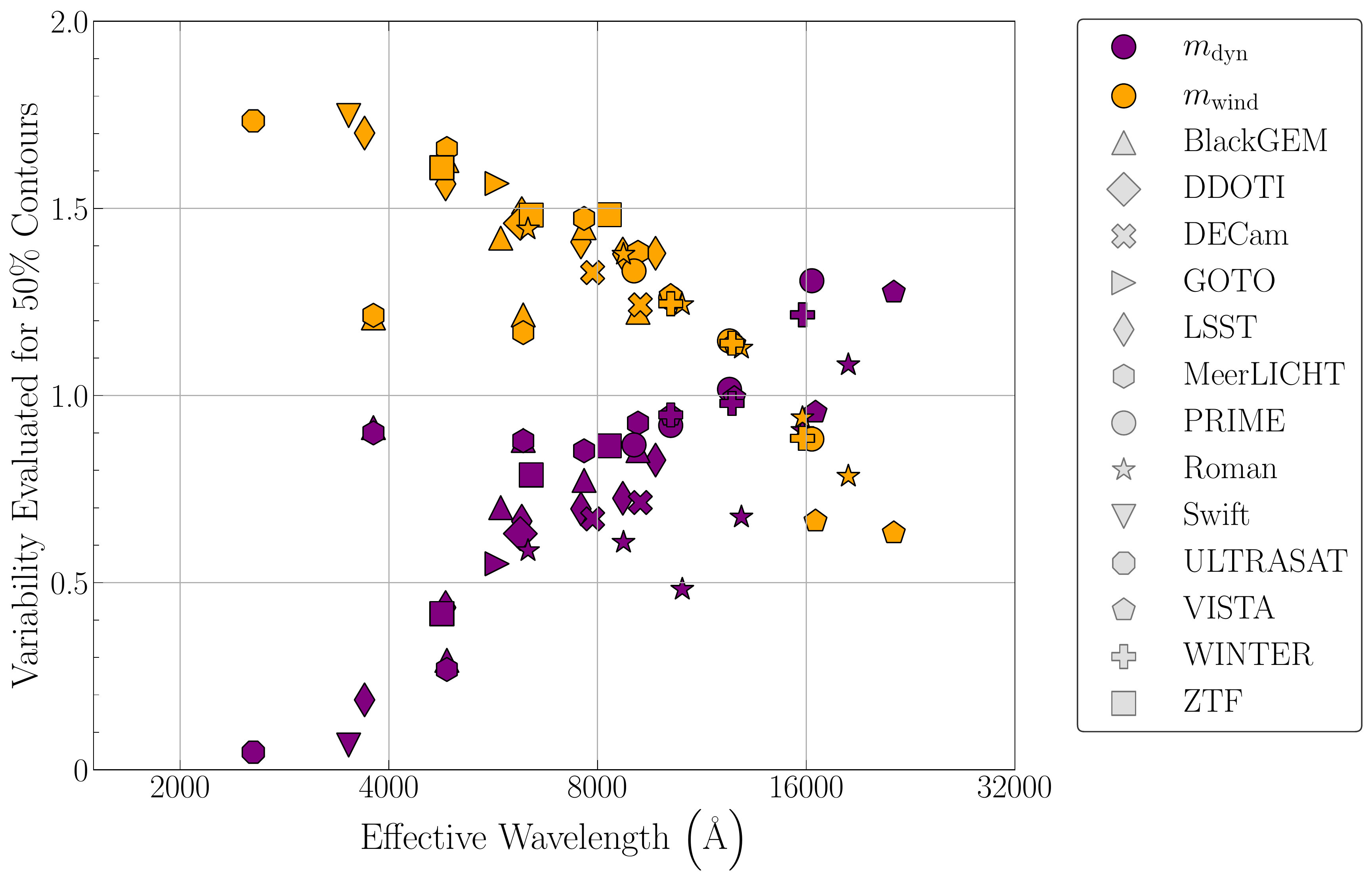}
\caption{Mass-dependent variability for a selection of filters. Each symbol corresponds to one instrument, with an instrument's filters represented by their bandpass filter function's effective wavelength. The vertical axis represents each filter's dependence to either dynamical ejecta mass (purple) or wind ejecta mass (orange), as defined in Equation~\ref{eq: var}. Higher-wavelength filters are increasingly sensitive to variations in dymamical ejecta mass, while dependence on wind ejecta mass decreases with increasing filter wavelength.
}
\label{fig: mass_summary}
\end{figure*}

Figure~\ref{fig: mass_summary} presents the variability scores for all filters in Figure~\ref{fig: summary} as a function of both dynamical ejecta mass (purple) and wind ejecta mass (orange). 
As anticipated, variability with dynamical ejecta mass increases with filter wavelength, while wind ejecta mass variability decreases with wavelength.
Ultraviolet filters exhibit the largest dependence on wind ejecta mass, with the UVOT/\textit{u}-band and ULTRASAT/\textit{NUV}-band both yielding $v=1.7$.
The PRIME/\textit{$H$}-band demonstrates the largest variability with dynamical ejecta mass, with $v=1.3$.
Variability scores directly relate to a filter's ability to constrain a given parameter from photometric observations (see Section~\ref{sec: infer_params}).

\begin{figure}
\centering
\includegraphics[width=\columnwidth, trim=20 0 100 00, clip]{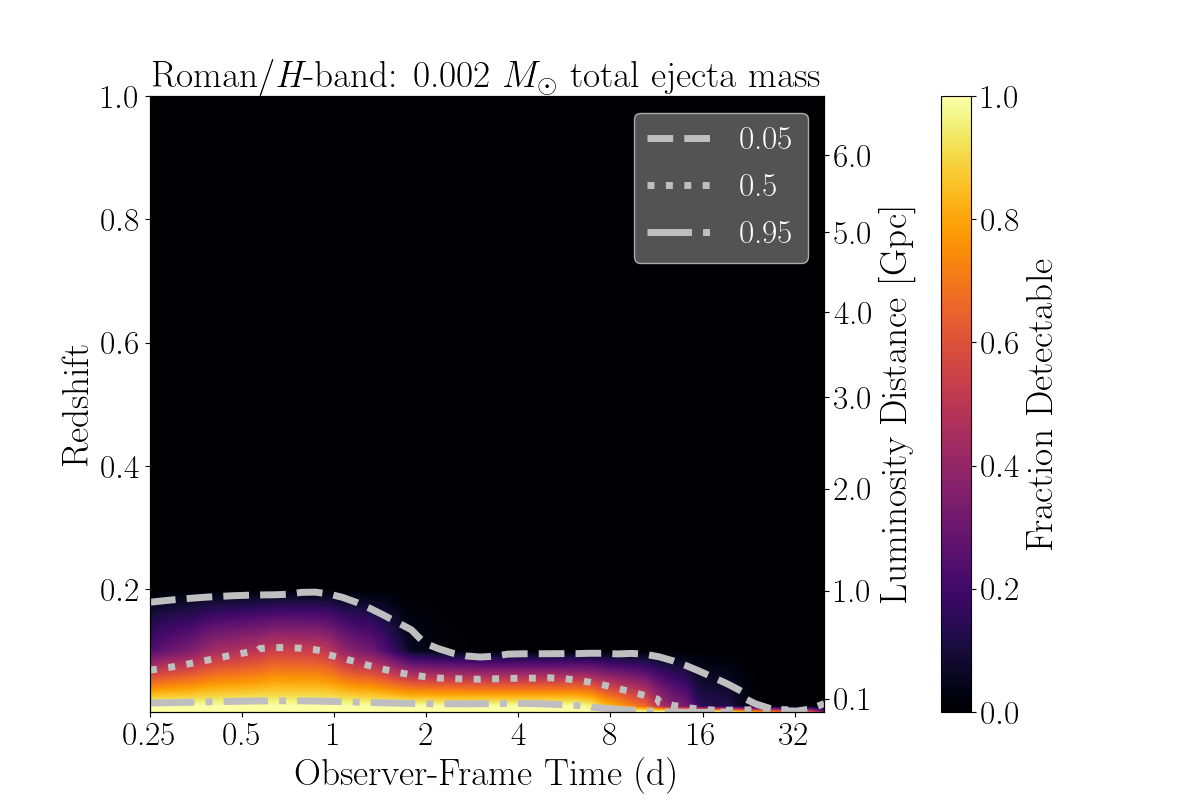}
\includegraphics[width=\columnwidth, trim=20 0 100 0, clip]{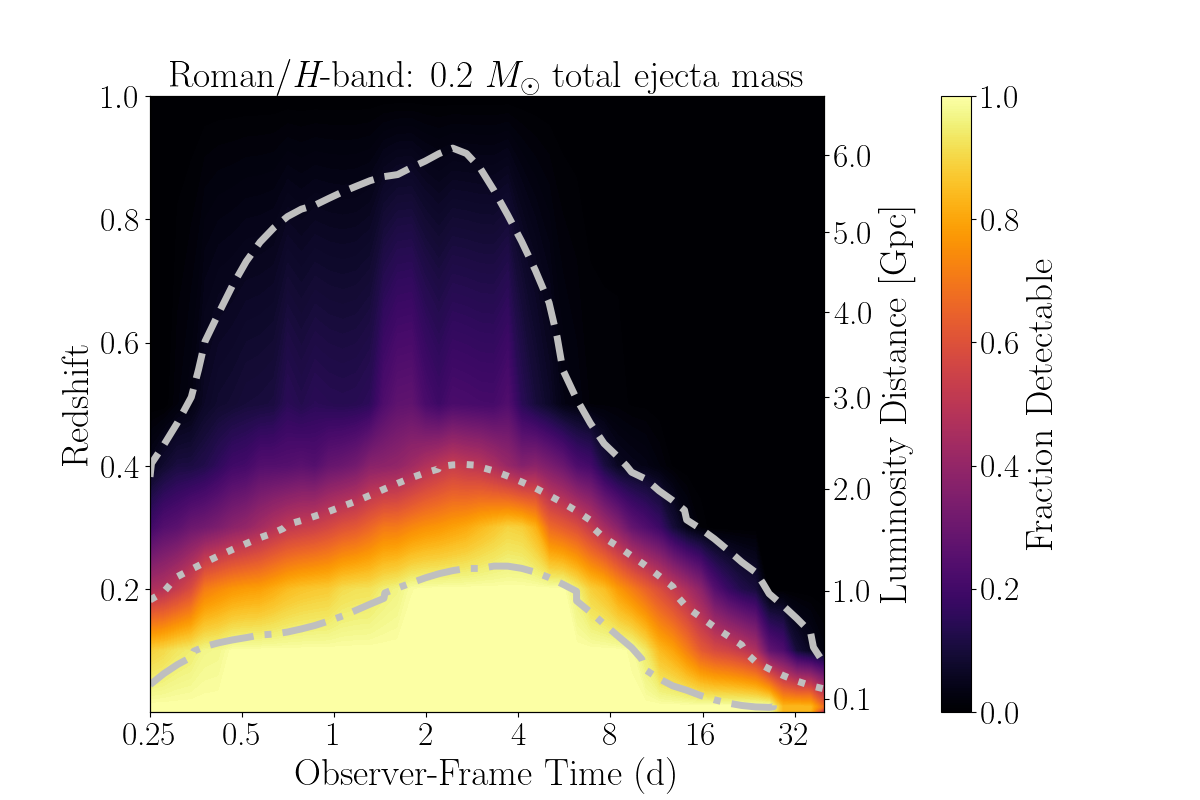}
\caption{Detectability constraints for the Roman/\textit{H}-band for two total ejecta masses, corresponding to the sum of dynamical and wind ejecta masses. The top panel displays the smallest total ejecta mass (0.002\msun{}), while the bottom panel corresponds to the largest total ejecta mass (0.2\msun{}). We place no restrictions on morphology, composition, viewing angle, or ejecta velocity, such that each panel includes 1944 simulations (36 simulations each with 54 viewing angles) of the full set of 48,600 LANL kilonova simulations.
}
\label{fig: total_mass}
\end{figure}

The interplay between various kilonova parameters must be considered to fully capture variations in detectability.
For example, masses for both dynamical and wind ejecta alter kilonova emission, and thus affect detectability.
We explore the interrelation of dynamical and wind ejecta masses in Figure~\ref{fig: total_mass}, using the Roman/\textit{H}-band as an example.
The top (bottom) panel restricts total ejecta mass to the lowest (highest) simulated mass, where total mass is the sum of both dynamical and wind ejecta masses.
In both panels, the total mass is fixed while composition, morphology, viewing angles, and ejecta velocities vary, resulting in 1944 simulations in each panel (36 simulations each rendered at 54 viewing angles).
Total mass significantly alters kilonova detectability in the Roman/\textit{H}-band, with a variability $v = 1.6$ between the total masses of 0.002\msun{} (top panel) and 0.2\msun{} (bottom panel). 
Kilonovae with higher total ejecta masses are detectable at significantly higher redshifts than their low mass counterparts.

In addition to mass, other properties of neutron star mergers affect kilonova detectability.
Luminosity may vary with viewing angle due to morphological effects \citep{korobkin2020} and lanthanide curtaining \citep{kasen2015}.
Additionally, ejecta composition significantly alters detectability in some filters, as lanthanide-poor ejecta yields significantly less luminous emission in the redder optical and infrared filters than mergers with lanthanide-rich ejecta.
Ejecta velocity also has a pronounced effect on detectability, with higher ejecta velocities leading to earlier peak emission and subsequently more luminous kilonovae \citep[e.g.,][]{kasen2017}.
Numerous other factors alter simulated lightcurves and kilonova detectability including decay product thermalization and nuclear mass models \citep{lippuner2015,hotokezaka2020}. 
Considerably more uncertain nuclear physics may be propagated by allowing synthesized \textit{r}-process abundance patterns to differ from the robust ``strong'' pattern, as done in \citet{barnes2020} and~\citet{zhu2021}.

\section{Inferring Kilonova Properties with Wide-Field Observations} \label{sec: infer_params}

We can leverage parameter-dependent variations in kilonova detectability to infer ejecta properties and guide future observations.
By coupling non-detections in wide-field transient searches with the kilonova detectability metrics presented in this work, constraints can be placed on kilonova ejecta properties.
Non-detections are particularly constraining when a very large fraction of the GW sky localization is covered by multiple instruments or filters in a short period of time. To explore this, we consider observations in the Roman/\textit{H}-band taking place two days post-trigger for a merger with a predicted redshift of $z = 0.2$ (an ambitious $\sim$1~Gpc for current GW detectors).
If these observations yield no transient detection, the total ejecta mass of an associated kilonova can be constrained to lower masses.
This relationship is supported by Figure~\ref{fig: total_mass}, as nearly 100\% of simulated kilonovae with total ejecta masses of 0.2\msun{} are detectable two days post-merger at $z=0.2$, while no kilonovae with total ejecta masses of 0.002\msun{} are detectable.
Thus, a non-detection rules out the presence of a high-ejecta mass kilonova at $z=0.2$.

This method has previously been employed by \citealt{thakur2020} to constrain parameters for a plausible kilonova associated with GW190814 \citep{lvc190814}. 
The authors compare upper limits from wide-field searches (including DDOTI) to the LANL kilonova grid, ruling out the presence of a high-mass ($>$ 0.1\msun{}) and fast-moving (mean velocity $\geq 0.3c$) wind ejecta. 
Total ejecta masses $\geq$ 0.2\msun{} are also strongly disfavored by the upper limits.

\begin{figure*}
\centering
\includegraphics[width=\columnwidth]{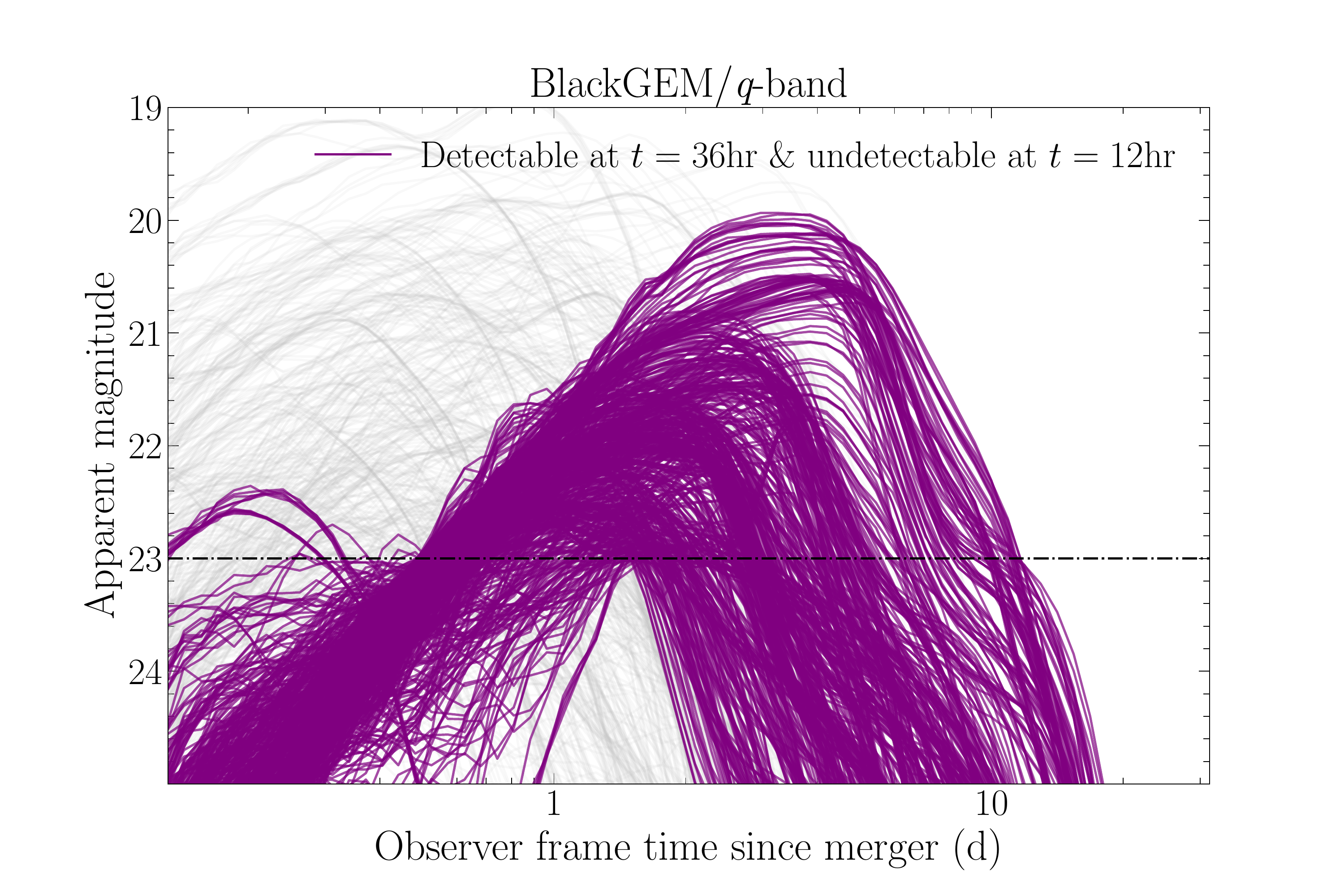}
\includegraphics[width=\columnwidth]{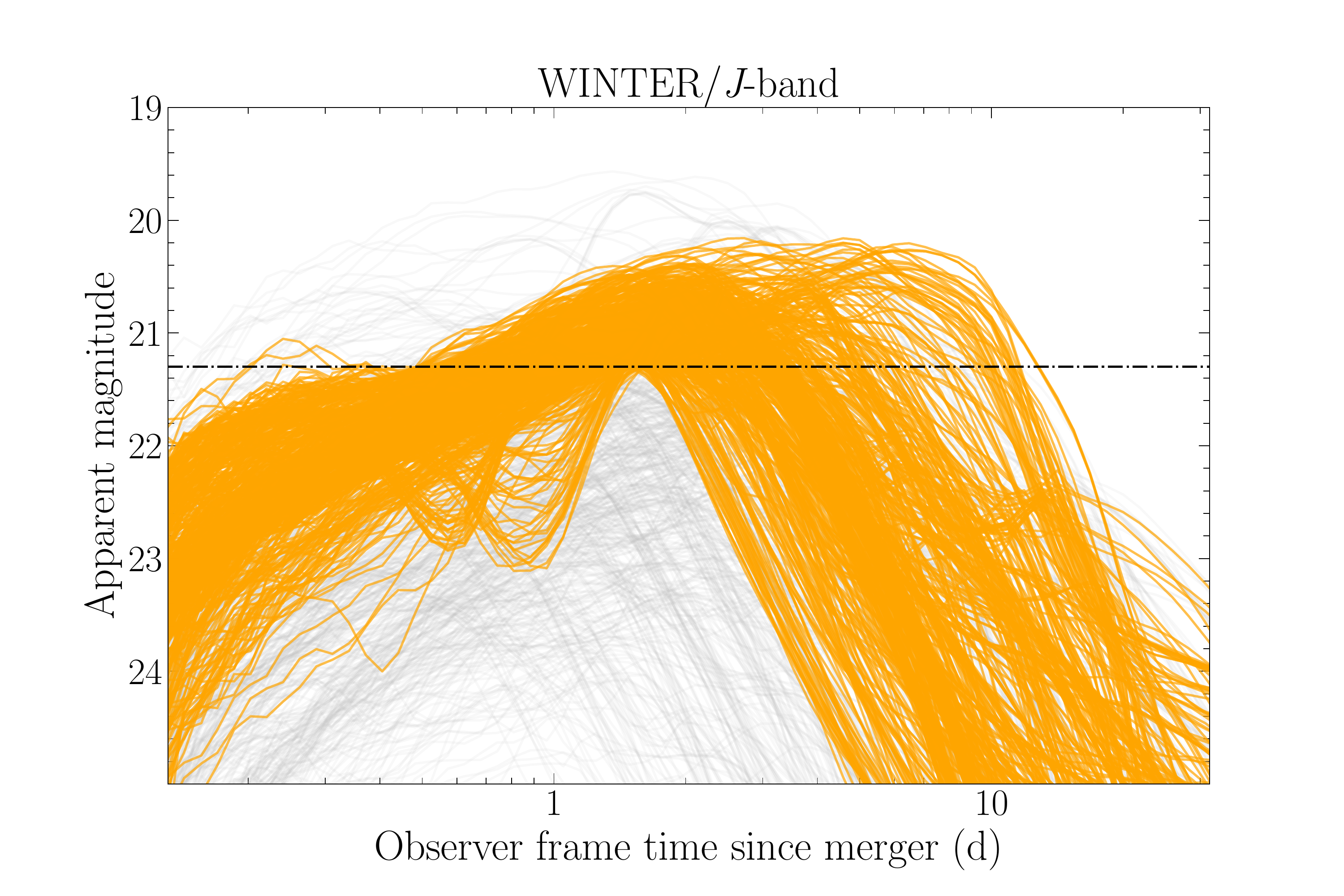}
\caption{Simulated kilonova lightcurves, in the observer frame, in the BlackGem/\textit{q}-band (\textit{left}) and WINTER/\textit{J}-band (\textit{right}) for a merger at $z=0.05$ ($\sim$230~Mpc). Purple and orange lightcurves are detectable at 36 hours but undetectable at 12 hours in the selected filter. Gray lightcurves do not meet this  criterion. Horizontal dot--dashed lines correspond to limiting magnitudes in each filter. Only a subset of the full LANL kilonova grid is displayed, for clarity. These panels also highlight the variation in lightcurve behavior between optical (BlackGEM) and near-infrared (WINTER) filters.
}
\label{fig: nondetect}
\end{figure*}

Non-detections in wide-field searches may also guide subsequent observing schedules. 
For example, assume that you are planning additional follow-up observations after a non-detection in the BlackGEM/\textit{q}-band at 12 hours post-merger for an event at $z=0.05$ ($\sim$230~Mpc) that has not been detected by any other instrument. 
Of all the instruments in this study intended to follow-up GW candidate events (i.e. not including LSST or Roman), an additional BlackGEM observation is best suited for follow-up observations. This conclusion was reached by searching the 48,600 LANL kilonova models to identify kilonovae that are undetectable with BlackGEM at this redshift and time, but detectable with other instruments between 12 and 36 hours post-merger.
Based on the LANL kilonova grid, 8\% of simulated kilonovae are detectable 36 hours post-merger in the BlackGEM/\textit{q}-band, but not detectable 12 hours post-merger.
A selection of detectable lightcurves is highlighted in the left panel of Figure~\ref{fig: nondetect}, suggesting the presence of high mass ($\geq$ 0.03\msun{}) and slow-moving (mean velocity $< 0.15c$) wind ejecta.
BlackGEM's \textit{q}-band does not offer strong constraints on dynamical ejecta properties, consistent with the discussion in Section~\ref{sec: variation} and Figure~\ref{fig: mass_summary}.
DECam can also provide useful follow-up observations after a non-detection at 12 hours, with 6\% of simulated kilonovae detectable in DECam but not BlackGEM at the aforementioned times.
Several instruments, including DDOTI, GOTO, VISTA, and PRIME, have negligible probability of detecting a kilonova at $z=0.05$ following a BlackGEM non-detection.
However, we note that both LSST and Roman have the highest probability of detecting a kilonova in this scenario, further highlighting their utility in kilonova searches. 
 
We performed a similar analysis assuming a non-detection with WINTER (\textit{J}-band) 12 hours post-merger for a GW event at $z=0.05$ ($\sim$230~Mpc). 
Again, we searched the 48,600 LANL kilonova models for kilonovae that are undetectable with the WINTER/\textit{J}-band at this redshift and time, but detectable with other instruments between 12 and 36 hours post-merger.
Based on the LANL kilonova grid, 16\% of simulated kilonovae are detectable with the WINTER/\textit{J}-band 36 hours post-merger but not detectable 12 hours post-merger.
These lightcurves are highlighted in the right panel of Figure~\ref{fig: nondetect}.
The WINTER non-detection and subsequent detection are consistent with large ejecta masses, with over 70\% of orange lightcurves in Figure~\ref{fig: nondetect} corresponding to total ejecta masses $\geq$ 0.1\msun{}.
WINTER's \textit{J}-band offers constraints on both dynamical and wind ejecta parameters, as expected from Figure~\ref{fig: mass_summary}.
In addition to WINTER, several instruments in this study are capable of observing kilonovae following a WINTER non-detection.
For example, over 40\% of simulated lightcurves are observable in the BlackGEM/\textit{q}-band at 36 hours post-merger following a WINTER non-detection 24 hours earlier.
Additional instruments have non-zero probabilities of detecting a kilonova under this scenario: GOTO can observe 6\% of simulated kilonovae following a WINTER non-detection, DDOTI can observe 12\%, VISTA can observe 15\%, ZTF can observe 3\%, and ULTRASAT can observe 1\%.

These examples demonstrate how non-detections offer powerful constrains on kilonova ejecta properties. 
However, targeted observations with large aperture telescopes are better suited to sample kilonova emission after an initial detection in a wide-field search, providing multi-band observations over the duration of kilonova evolution. The LANL set of kilonova models provides a rich data set for thorough statistical analysis and optimization of kilonova observing strategies, beyond the study of non-detections in wide-field surveys, which we reserve for future work. In addition, we refer the reader to several other studies on the efficacy of Bayesian parameter estimation to infer kilonova parameters \citep[i.e. ][]{coughlin2018,barbieri2019,nicholl2021,heinzel2021, ristic2021}. 

\section{Instrument Results} \label{sec: instr_results}

In this section, we briefly summarize each instrument's capacity for kilonova detection.
We describe the accessible redshifts for each filter and the most probable times for kilonova detection.
Additional figures for each instrument are presented in Appendix~\ref{sec: app}. 

\par
\textit{BlackGEM} $-$ BlackGEM's optical and near-infrared filters are well-suited to detect kilonova emission, especially within two days post-merger. The broad \textit{q}-band provides the best opportunities for detection, with 50\% of simulated kilonovae detectable at $z=0.072$ ($\sim$340~Mpc) and high wind ejecta mass kilonovae detectable out to $z \sim 0.2$. Our results largely support BlackGEM's plan to follow-up with the $\textit{u}$, $\textit{q}$, and $\textit{i}$ filters. However, we note that the $\textit{z}$-band increases the detectable kilonova parameter space at later times ($t \geq 2$~d). As an example, nearly 10\% of modeled kilonovae at $z=0.01$ ($\sim$45~Mpc) are only observable in the $\textit{z}$-band at $t = 4$~d (observer frame) and not observable in any other BlackGEM filters.

\par
\textit{DDOTI} $-$ DDOTI's $w$ filter can detect 50\% of simulated kilonovae out to $z=0.023$ ($\sim$100~Mpc), with a peak detectability 12 hours post-merger. DDOTI can follow-up some LIGO/Virgo/KAGRA candidate events with inferred redshifts out to $z=0.073$ ($\sim$340~Mpc).

\par
\textit{DECam} $-$ DECam's \textit{i}- and \textit{z}-bands are sensitive to kilonova detection, reaching $z_{50\%}$ of 0.058 ($\sim$270~Mpc) and 0.042 ($\sim$190~Mpc), respectively. Despite its lesser sensitivity the \textit{z}-band may be better suited to detect kilonovae than the \textit{i}-band, as nearly all simulated kilonovae detectable in the \textit{i}-band are also detectable by the \textit{z}-band.
DECam is able to observe high ejecta mass kilonovae out to $z=0.18$ ($\sim$900~Mpc).

\par
\textit{GOTO} $-$ GOTO's \textit{L}-band achieves a $z_{50\%} = 0.029$ ($\sim$130~Mpc), with peak detectability $\sim$8~hr post-merger. GOTO is able to follow-up high wind ejecta mass LIGO/Virgo/KAGRA candidate events with inferred redshifts out to $z=0.097$ ($\sim$460~Mpc).

\par
\textit{LSST} $-$ LSST can observe the most distant kilonovae out of all other ground-based instruments in the study. The \textit{g}-band reaches the highest \zhoriz{}, with a $z_{50\%} = 0.14$ peaking $\sim$8~hr post-merger. We especially encourage the use of LSST's \textit{g}, \textit{r}, and \textit{i} filters for follow-up of relatively distant LIGO/Virgo/KAGRA candidate events ($z=0.15-0.5$) that may not be observable with other ground-based, wide-field instruments. 

\par
\textit{MeerLICHT} $-$ Similarly to BlackGEM, MeerLICHT's broad \textit{q} filter is the most sensitive at detecting kilonovae.  MeerLICHT has better chances to observe kilonovae located at $z<0.074$ ($\sim$345~Mpc). For comparison, BlackGEM can detect kilonovae at over twice the distance of MeerLICHT.

\par
\textit{PRIME} $-$ This ground-based infrared instrument is useful for late-time kilonova searches, with peak detectability between one and eight days post-merger. The most sensitive filter reaches a $z_{50\%} = 0.023$ ($\sim$100~Mpc), with some high ejecta mass kilonovae detectable out to $z=0.067$ ($\sim$300~Mpc). Additionally, the PRIME/\textit{H}-band may offer strong constraints on dynamical ejecta masses. 

\par
\textit{Roman} $-$ Roman provides an excellent tool for kilonova detection and GW follow-up. The Roman/\textit{R}-band is the most sensitive filter in our study, reaching $z_{50\%} = 0.29$ with some kilonovae detectable out to $z \sim 1$. However, Roman's relatively small FoV precludes follow-up of GW candidate events with large localization areas. We encourage the use of the Roman Space Telescope for GW follow-up \citep[see also][]{Foley2019wfirst}, especially for distant and/or well-localized candidate events.

\par
\textit{UVOT} $-$ \textit{Swift}/UVOT's \textit{u}-band reaches a $z_{50\%}=0.015$ ($\sim$70~Mpc), with a peak detectability around four hours post-merger. The \textit{u}-band is well-suited to follow-up LIGO/Virgo/KAGRA candidate out to $z=0.061$ ($\sim$280~Mpc). UVOT's detectability estimates may be biased by the lack of simulated lightcurves within three hours post-merger.

\par
\textit{ULTRASAT} $-$ ULTRASAT provides an excellent tool for early kilonova detection. It's 200~deg$^2$ \textit{NUV} filter can detect 50\% of simuilated kilonovae out to $z=0.022$ ($\sim$100~Mpc), with high wind ejecta mass mergers detectable out to $z=0.1$ within the first 12 hours post-merger. We caution the reader in their interpretation of ULTRASAT results. This study is based on an approximate ULTRASAT filter, as no finalized filter is publicly available. We anticipate that ULTRASAT's detectability estimate would increase with the availability of simulated lightcurves before three hours.

\par
\textit{VISTA} $-$ VISTA is one of the most sensitive ground-based infrared facilities in the study, with $z_{50\%} = 0.038$ ($\sim$170~Mpc) for the \textit{Y}-band. VISTA is able to follow-up high ejecta mass LIGO/Virgo/KAGRA candidate events out to $z=0.094$ ($\sim$450~Mpc) one day post-merger. Additionally, VISTA observations may provide strong constraints on dynamical ejecta masses. 

\par
\textit{WINTER} $-$ Similarly to VISTA, WINTER is one of the most sensitive ground-based infrared facilities in the study, with $z_{50\%} = 0.038$ ($\sim$170~Mpc) for the \textit{Y}-band. WINTER is capable of follow-up some nearby (within $\sim$ 100~Mpc) mergers over two weeks post-merger. WINTER can observe high ejecta mass kilonovae out to $z=0.093$ ($\sim$440~Mpc) one day post-merger, in addition to offering tight constraints on dynamical ejecta mass.

\par
\textit{ZTF} $-$ ZTF is well-suited to search large sky localization regions for relatively nearby kilonovae ($z \lesssim 0.1$), with a \zhoriz{} of $z \sim 0.02$ ($\sim$100~Mpc) in the \textit{g}-, \textit{r}-, and \textit{i}-bands. ZTF is able to detect a subset of high ejecta mass kilonovae out to $z=0.094$ ($\sim$445~Mpc).

\section{Discussion \& Conclusions} \label{sec: conclusion}

This study explores kilonova searches with current and upcoming wide-field instruments.
This analysis relies on the LANL grid of kilonova simulations \citep{wollaeger2020}, a set of 48,600 radiative transfer models which span a variety of ejecta parameters.
Based on these simulations, we quantify each instrument's ability to detect a kilonova, recording the fraction of observable kilonovae from the simulation grid for various times and redshifts.
The \Nfilt{} filters in this study have significant variations in their \zhoriz{}; ultraviolet filters are restricted to observing nearby kilonovae ($z < 0.3$), while the Roman Space Telescope can observe a subset of kilonovae out to $z \sim 1$. 
We concentrate on wide-field kilonova searches following a GW candidate event, although the conclusions of this study are equally relevant to kilonova searches following poorly-localized short GRBs.
Additionally, the framework presented in this study is easily adaptable, and can be applied to searches performed by more sensitive, smaller FoV instruments (e.g., Gemini, Keck, GTC, VLT, SALT, \textit{HST}, \textit{JWST}) following the detection of a kilonova.
Additionally, we remind the reader that the facilities studied in this work do not constitute a comprehensive selection of wide-field instruments. 

Detecting, analyzing, and modeling kilonovae is a new and evolving science; many uncertainties remain after only a single confident multi-messenger observation with GWs \citep{mma2017}.
As a result, the models that lay the foundation of this study rely on numerous physical assumptions, including the use of a specific nuclear mass model \citep[FRDM, ][]{moeller1995} and decay product thermalization \citep{barnes2016, rosswog2017}. 
While the LANL grid of kilonova models covers a wide range of kilonova parameters (see Section~\ref{sec: detectability} and Table~\ref{tbl:nome}), this simulation grid does not span all possible ejecta scenarios. 
For instance, the simulations do not encompass all possible variations in ejecta morphology \citep{korobkin2020} or composition \citep{even2019}.
We also ignore changes in the detected UVOIR emission due to extinction or the presence of contamination from either a sGRB afterglow or host galaxy, which will be  expanded upon in a future work. 

Our work produces results that are consistent with the detectability study presented by \citet{scolnic2018}, which determined kilonova detectability using an \at-like model.
\citet{scolnic2018} predict a redshift range of $z=0.02-0.25$ for LSST, consistent with our redshift range of $z=0.035-0.48$ for the most sensitive LSST filter. 
Our results are also broadly consistent with the detectability estimates presented by \citet{rastinejad2021}, which leveraged three kilonova models (in addition to \at) to infer the maximum redshift of kilonova detection.

Similarly to \citet{scolnic2018}, we focus only on the ability of these instruments to detect a kilonova, and not its unambiguous identification (as this outcome depends on multi-epoch observations across a range of filters). 
Wide-field searches are likely to uncover numerous other transient signals (e.g., supernovae), potentially masquerading as a kilonova signal (see \citealt{cowperthwaite2015} for a discussion of kilonova contaminants).
LIGO/Virgo/KAGRA candidate events constrained to large localization volumes are increasingly likely to reveal false positive kilonova candidates, further necessitating the need for repeated observations of each field to rapidly distinguish a kilonova candidate.
In future work, we intend to provide methods to disentangle kilonova candidates from other transients with a limited number of photometric observations.

We caution the reader in the interpretation of our results and emphasize that the \zhoriz{} values presented in Table~\ref{tbl:kn} and Figure~\ref{fig: summary} refer to the redshifts at which 50\% of \textit{simulated kilonovae} are detectable. 
This is not equivalent to the probability of detection for the full population of kilonovae.
For example, our results claim that 50\% of simulated kilonovae are observable in the LSST/\textit{r}-band at $z=0.12$.
This does not imply that the LSST/\textit{r}-band has a 50\% probability of observing a kilonova at $z=0.12$.
Such a statement relies on a realistic astrophysical distribution of kilonova ejecta properties, while our study is based on a uniformly sampled grid of models.
This study could be expanded to draw from a distribution of kilonova models with the availability of more robust models for binary neutron star populations, neutron star equation of state, and mappings between compact object progenitors and ejecta. Additionally, this work could be further expanded to include redshift-dependent rates of neutron star mergers and associated GW detection.

Multi-band observations capture variability in kilonova emission and provide tighter constraints on kilonova properties, such as the composition and ejecta mass.
Once a kilonova is identified, targeted follow-up observations at several wavelengths (and times) are necessary to constrain ejecta properties.
Additionally, instruments must be distributed in geographically disparate locations, similarly to the Las Cumbres Observatory Global Telescope Network \citep{lcogtn}, to maximize the probability of kilonova detection. 
Instrument response time may also vary significantly based on the location of a neutron star merger on the sky.
We stress the utility of both ground-based and space-based detectors to allow for rapid follow-up post-trigger, aiding in kilonova detection.

Kilonova searches also benefit from a variety of fields-of-view and exposure times, such as the collection of instruments in this study.
Wide FoV instruments with short exposure times are best suited to rapidly scan the sky for GW candidate events localized to large regions of the sky, while smaller FoV instruments with longer exposure times are better suited for well-localized ($\lesssim$ 10~deg$^2$) events.
For example, while the Roman Space Telescope has the largest \zhoriz{} in this study, less sensitive, wider FoV instruments are better able to follow-up GW candidate events with large localization areas.

Based on the results of our study, we present the following findings to guide observing strategies and commissioning of future instruments: 
\begin{enumerate}
    \item This study demonstrates the utility of a diversity of wide-field instruments to search for kilonovae following GW detections. This includes a variety in instrument location, sensitivity, FoV, cadence, exposure time, and wavelength coverage. Additionally, rapid dissemination of both GW and electromagnetic search results is critical to optimize counterpart searches among this diverse array of instruments.
    \item Early observations increase the probability of observing a kilonova, especially at ultraviolet and optical wavelengths. Based on the results in this study, we stress the importance of early-time observations for kilonova detection. Although not a focus of this study, we  note that early observations offer the most stringent constraints on wind ejecta properties.
    \item More sensitive wide-field ultraviolet instruments are needed for kilonova detection as LIGO/Virgo/KAGRA reach design sensitivity. The ultraviolet instruments in this study can observe only a small fraction of kilonovae beyond $z \sim 0.1$, which is equivalent to the BNS horizon redshift of advanced LIGO at design sensitivity \citep{obs_scen2020}. 
    \item We promote the construction of additional wide-field near-infrared instruments. Compared to optical instruments, few ground-based near-infrared instruments are available for GW follow-up, potentially reducing the probability of kilonova detection. Near-infrared instruments hold immense promise for kilonova detection, as they are able to detect kilonovae within several days post-merger as opposed to the limited timescale accessible to ultraviolet and optical instruments.
    \item We recommend using this work as a guide for scheduling future kilonova observations. Based on the results of this study, we encourage observations if more than 5\% of simulated kilonovae are detectable at a given time and redshift based on the figures in Appendix~\ref{sec: app}. 
    \item Low-latency GW products, such as sky localization and distance estimates, should be used to alter observing strategies. We recommend comparing low-latency distance estimates to the detectability estimates provided in this work.
\end{enumerate}

The current and upcoming wide-field instruments in this study are well-poised to observe kilonovae coincident with GW events in the advanced detector era.
However, the landscape of multi-messenger astronomy with GWs will change through the construction of new detectors.
Sky localization will improve as additional GW detectors are commissioned \citep{obs_scen2020}, allowing larger aperture instruments to cover higher percentages of the localization region.
Simultaneously, enhanced GW detector sensitivity will reveal neutron star mergers at cosmic distances previously inaccessible through GWs alone, with the Einstein Telescope and Cosmic Explorer detecting BNS mergers at redshifts of out to $z\sim4$ and $z \sim 10$, respectively \citep{hall2019}.
To match the increase in GW detector sensitivity, innovative UVOIR instruments must be constructed to observe kilonovae in the era of third-generation GW detectors.

\acknowledgements

This work was supported by the US Department of Energy through the Los Alamos National Laboratory.
Los Alamos National Laboratory is operated by Triad National Security, LLC, for the National Nuclear Security Administration of U.S.\ Department of Energy (Contract No.\ 89233218CNA000001). 
Research presented in this article was supported by the Laboratory Directed Research and Development program of Los Alamos National Laboratory under project number 20190021DR.
This research used resources provided by the Los Alamos National Laboratory Institutional Computing Program, which is supported by the U.S.\ Department of Energy National Nuclear Security Administration under Contract No.\ 89233218CNA000001. EAC acknowledges financial support from the IDEAS Fellowship, a research traineeship program funded by the National Science Foundation under grant DGE-1450006. 
EAC also acknowledges her time spent as an LSSTC Data Science Fellow; the skills she gained through the fellowship greatly enhanced this study.
MR acknowledges support from NSF AST 1909534. BO and ET were supported in part by the National Aeronautics and Space Administration through grants NNX16AB66G, NNX17AB18G, and 80NSSC20K0389. 

We thank the following individuals for useful conversations during the production of this work: Christopher Berry, Brad Cenko, Derek Davis, Maya Fishbach, Wen-fai Fong, Erika Holmbeck, Griffin Hosseinzadeh, Vicky Kalogera, Carlos Moreno, Richard O'Shaughnessy, Jillian Rastinejad, and Michael Zevin. 

We are especially grateful for the many instrument team leaders and members who provided us with limiting magnitudes, exposure times, and filter functions. This includes Simone Dichiara (DDOTI), Martin Dyer (GOTO), Danielle Frostig (WINTER), Ben Gompertz (GOTO), Paul Groot (BlackGEM/MeerLICHT), Nathan Lourie (WINTER), Frank Marshall (UVOT), Takahiro Sumi (PRIME), and Alan Watson (DDOTI).

This research has made use of the SVO Filter Profile Service (\url{http://svo2.cab.inta-csic.es/theory/fps/}) supported from the Spanish MINECO through grant AYA2017-84089.

\software{\texttt{astropy} \citep{astropy2013,astropy2018},  \texttt{ligo.skymap} \citep{singer2016},  \texttt{matplotlib} \citep{matplotlib}, \texttt{numpy} \citep{numpy}, \texttt{scipy} \citep{scipy2020}}\\

\bibliography{biblio}{}

\appendix

\section{Supplementary Figures} \label{sec: app}

\begin{figure*}[h]
\centering
\includegraphics[width=0.4\columnwidth, trim=20 0 100 00, clip]{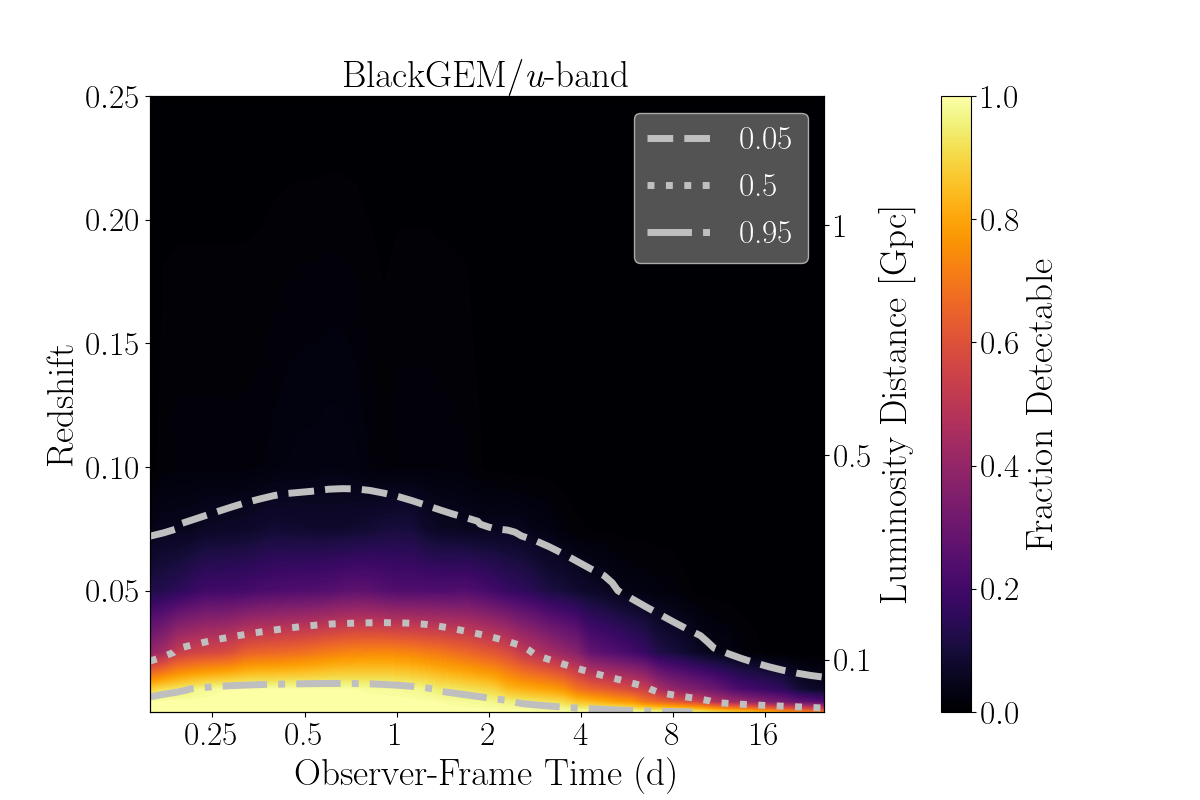}
\includegraphics[width=0.4\columnwidth, trim=20 0 100 0, clip]{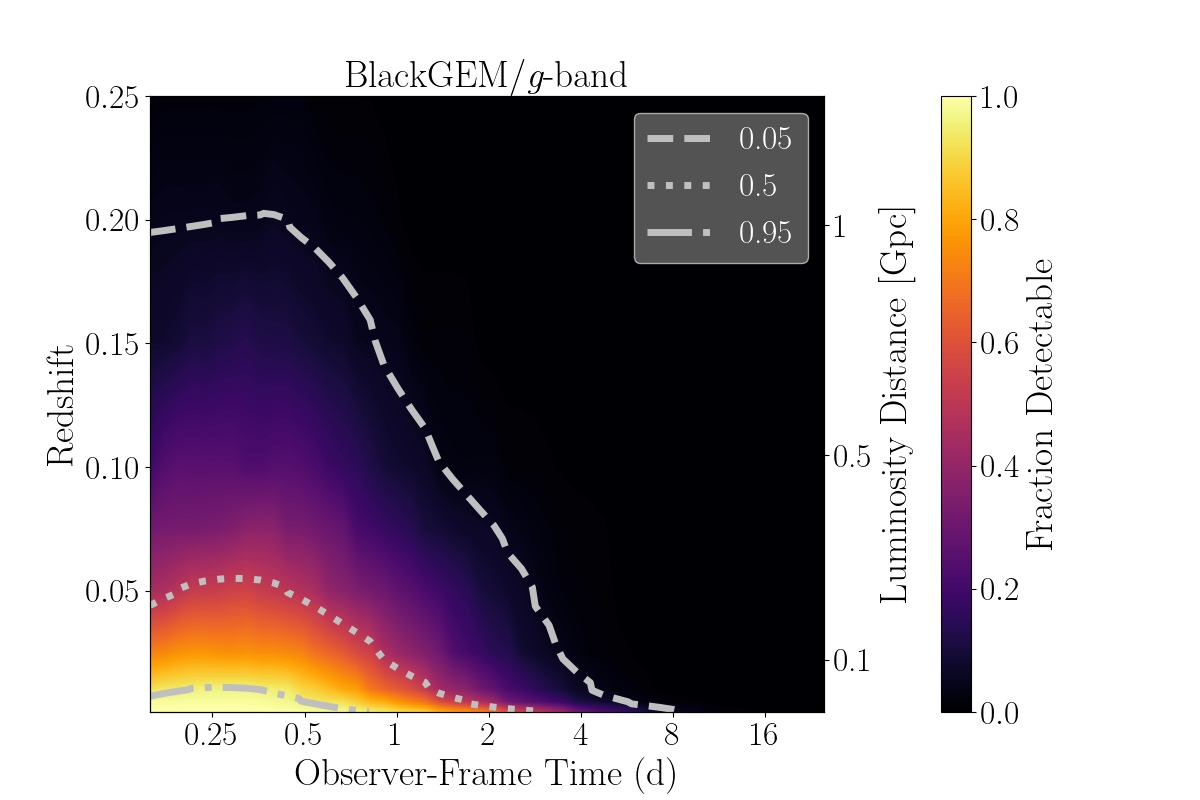}
\includegraphics[width=0.4\columnwidth, trim=20 0 100 0, clip]{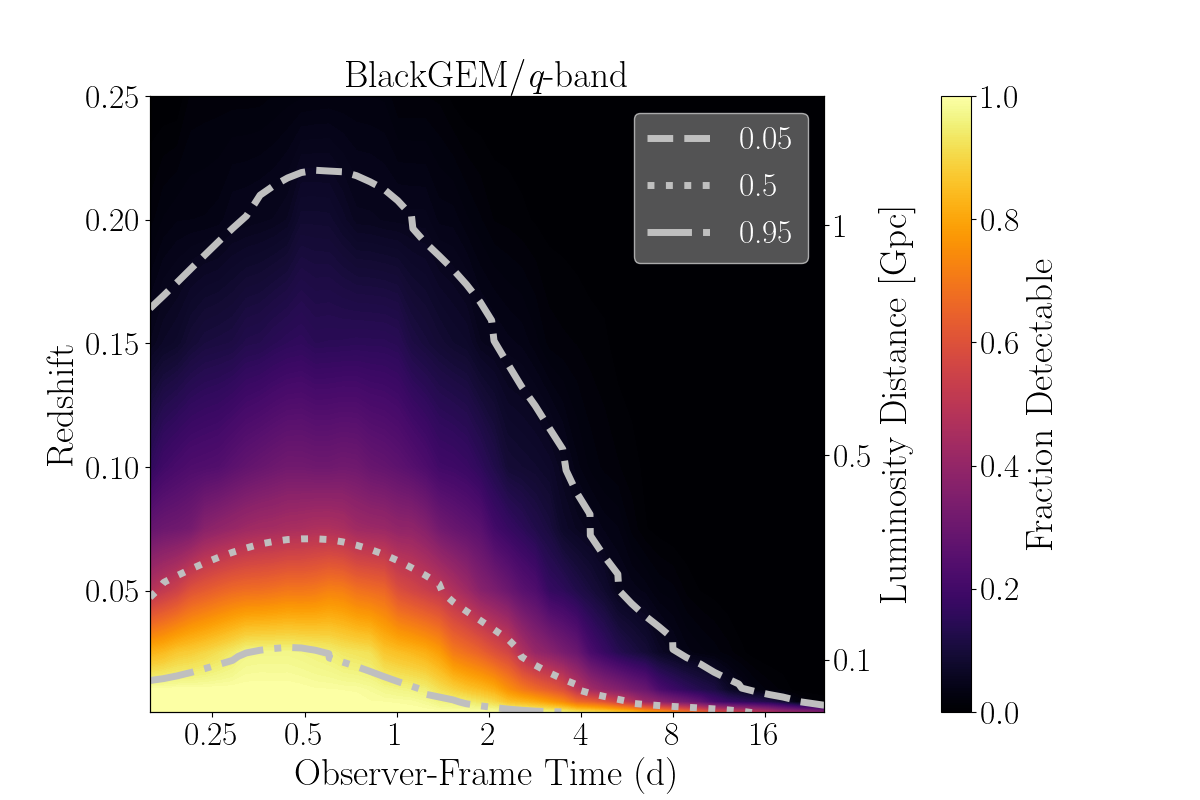}
\includegraphics[width=0.4\columnwidth, trim=20 0 100 0, clip]{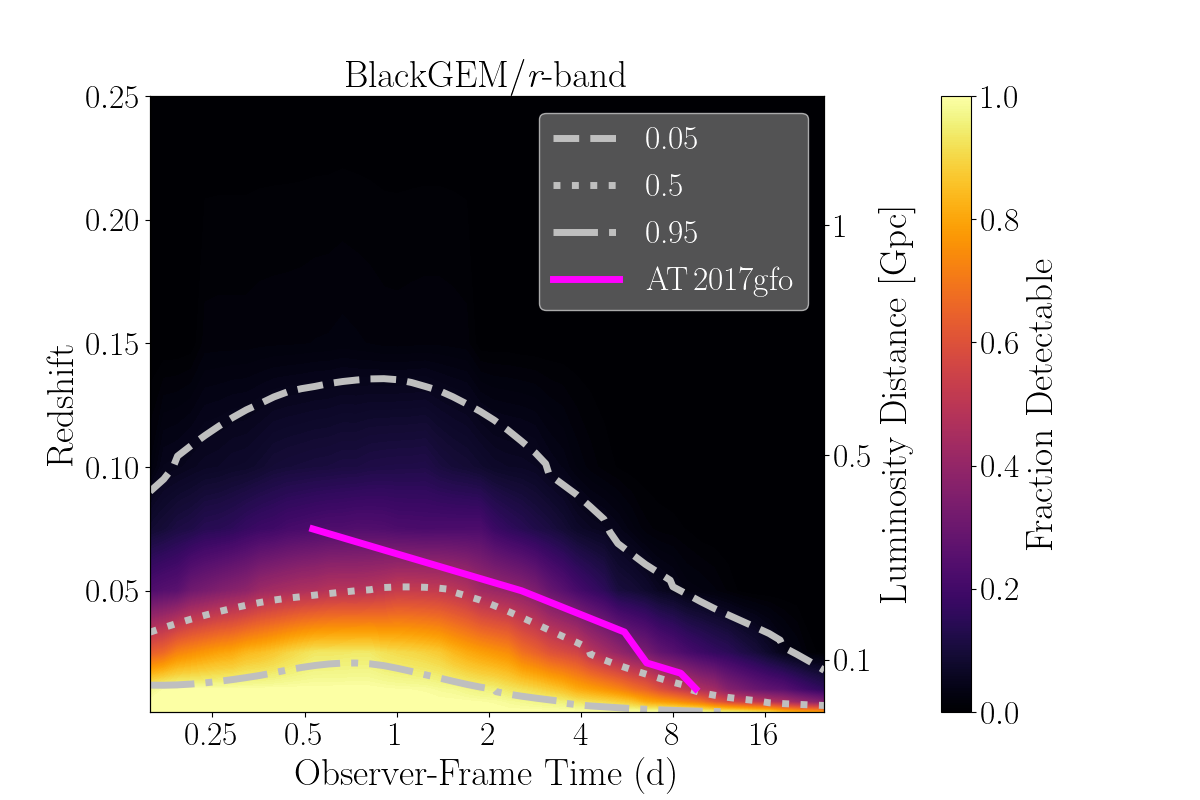}
\includegraphics[width=0.4\columnwidth, trim=20 0 100 0, clip]{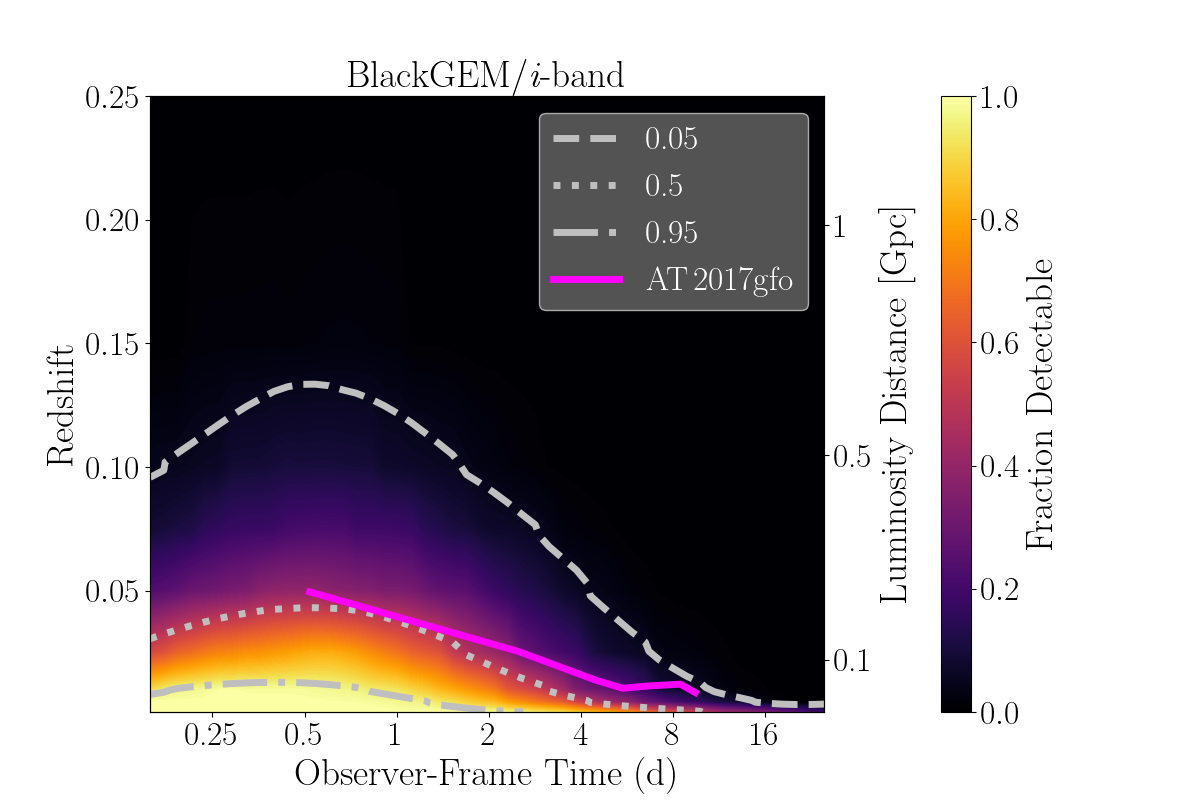}
\includegraphics[width=0.4\columnwidth, trim=20 0 100 0, clip]{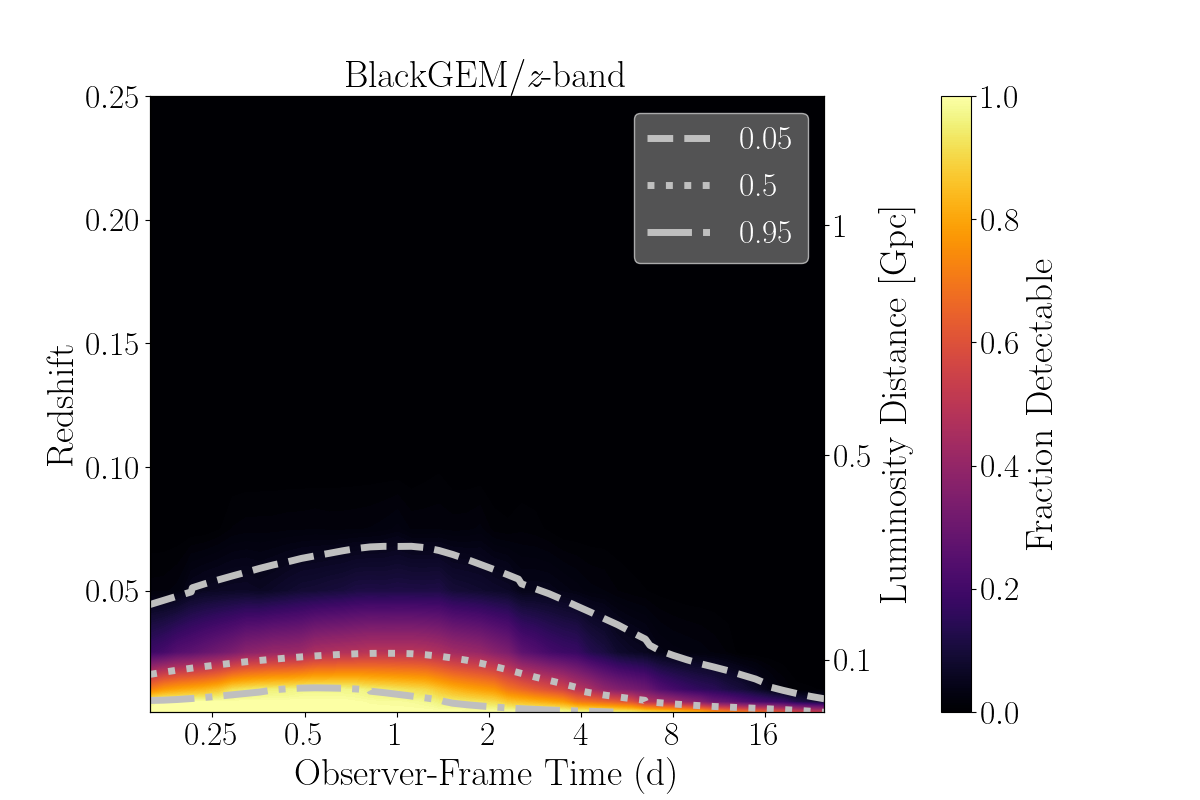}
\caption{Detectability constraints for six BlackGEM filters. Contours indicate the fraction of 48,600 simulated kilonovae (900 simulations each rendered at 54 viewing angles) with apparent magnitudes brighter than the limiting magnitude in each filter, for a given redshift and observer-frame time. The three white contours demarcate regions where 5\%, 50\%, and 95\% of simulated kilonovae are detectable. The magenta curve represents each filter's ability to detect \at-like kilonovae; we only present \at-like detectability for filters that are fully spanned by \at~spectral observations.
}
\label{fig: BlackGEM}
\end{figure*}

\begin{figure*}
\centering
\includegraphics[width=0.4\columnwidth, trim=20 0 100 00, clip]{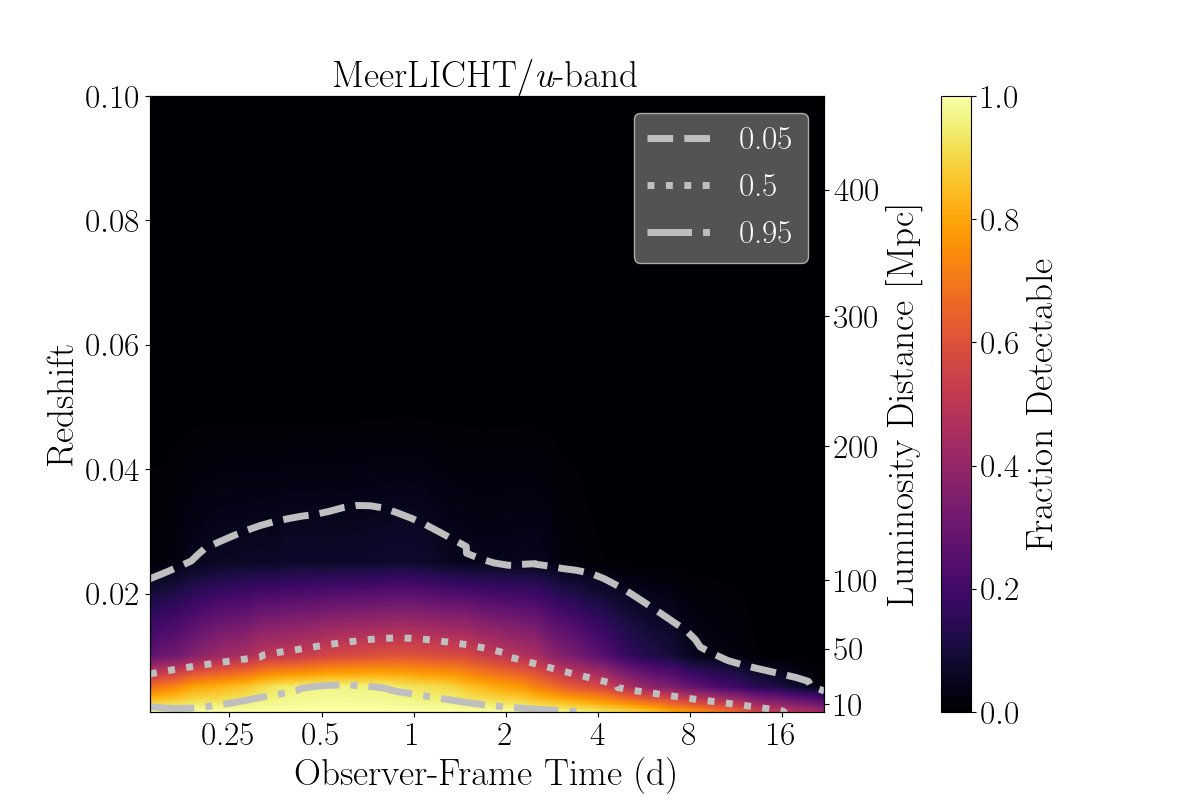}
\includegraphics[width=0.4\columnwidth, trim=20 0 100 0, clip]{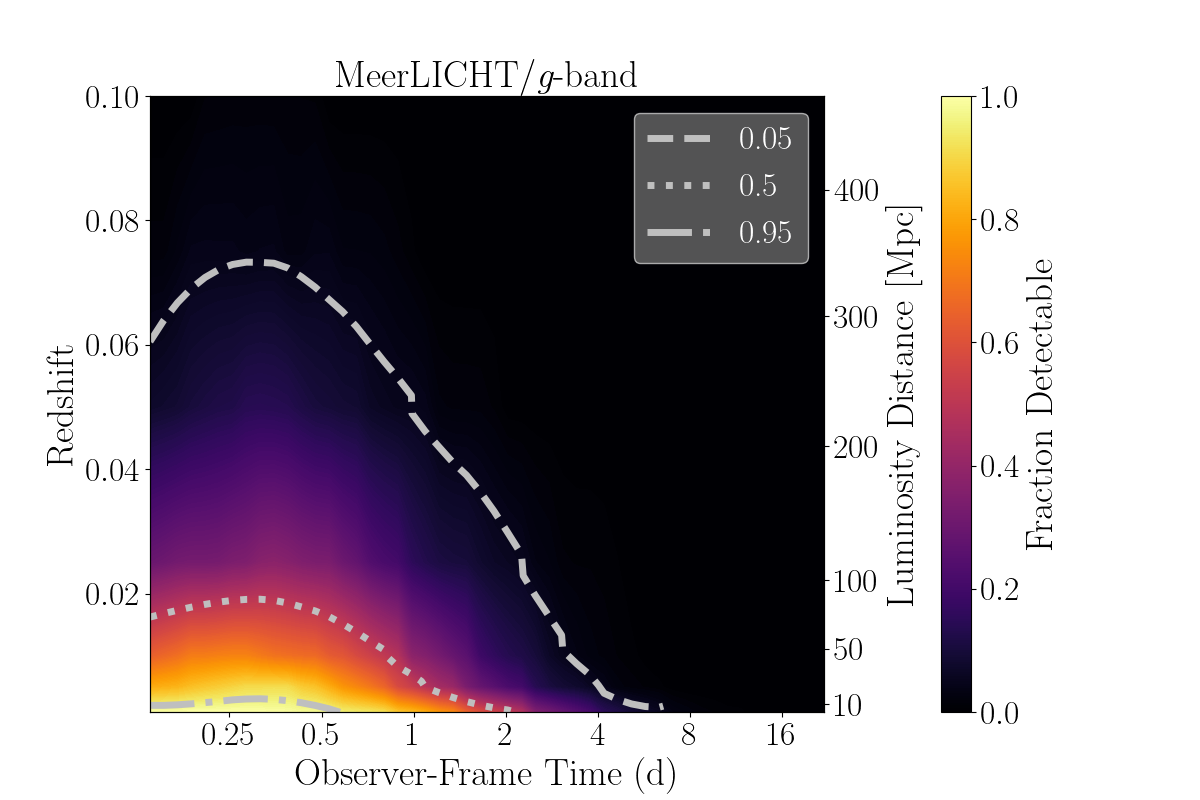}
\includegraphics[width=0.4\columnwidth, trim=20 0 100 0, clip]{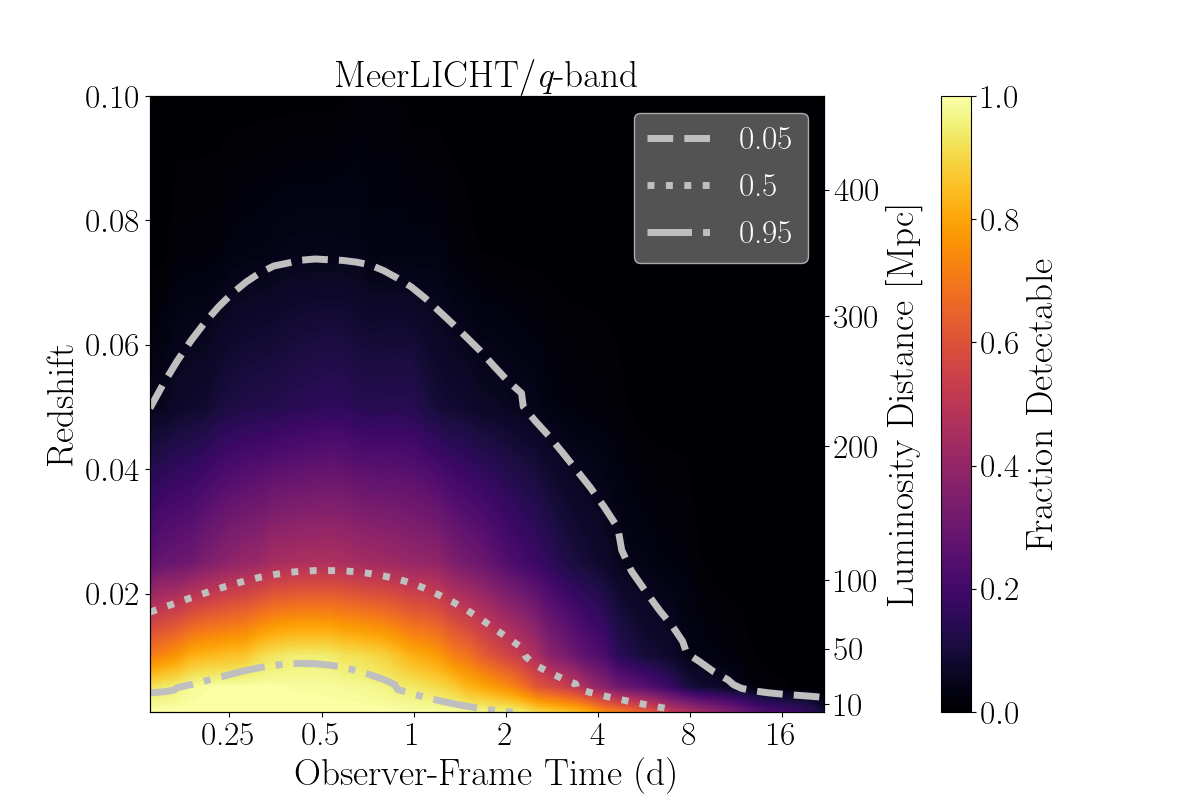}
\includegraphics[width=0.4\columnwidth, trim=20 0 100 0, clip]{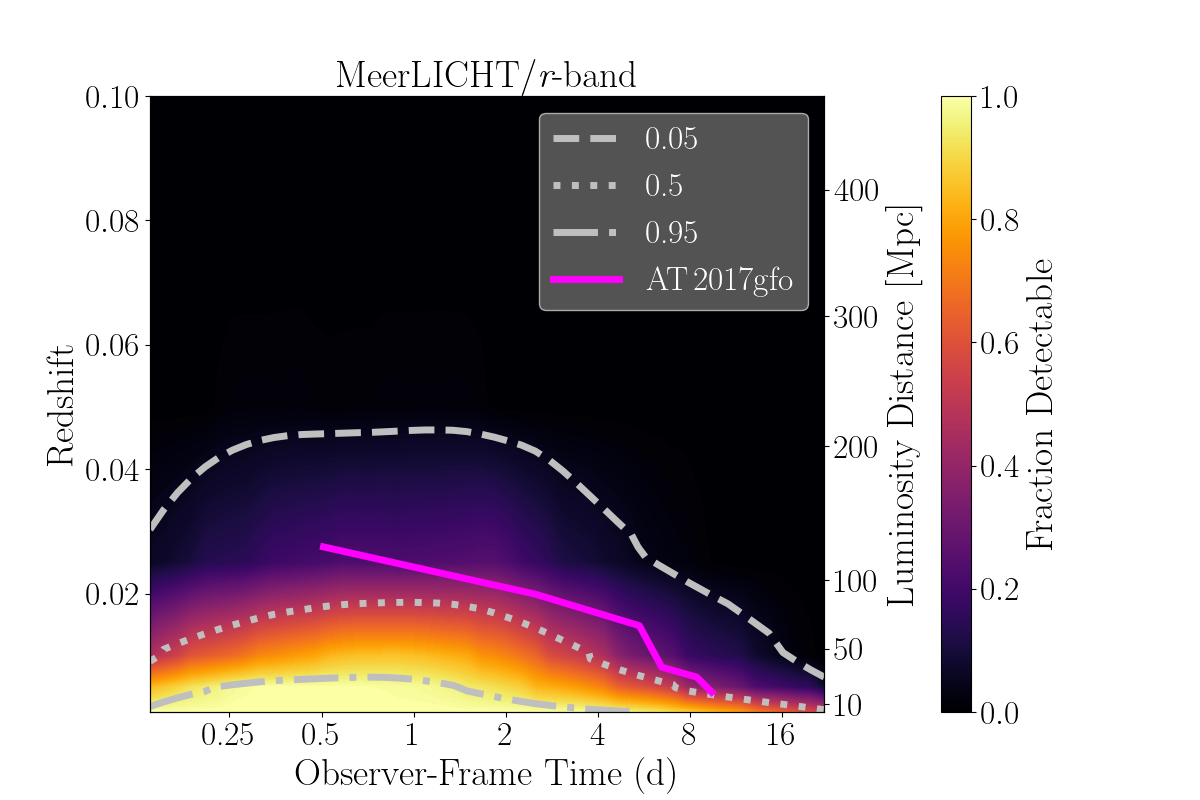}
\includegraphics[width=0.4\columnwidth, trim=20 0 100 0, clip]{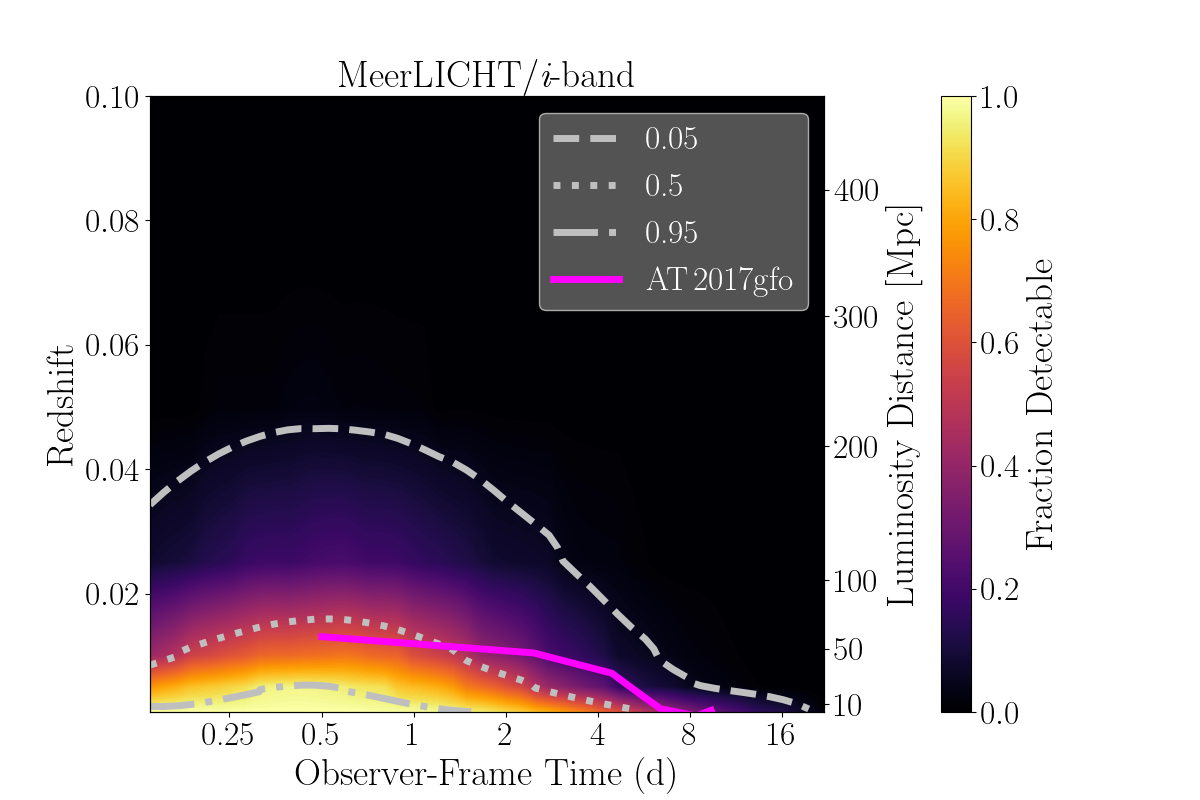}
\includegraphics[width=0.4\columnwidth, trim=20 0 100 0, clip]{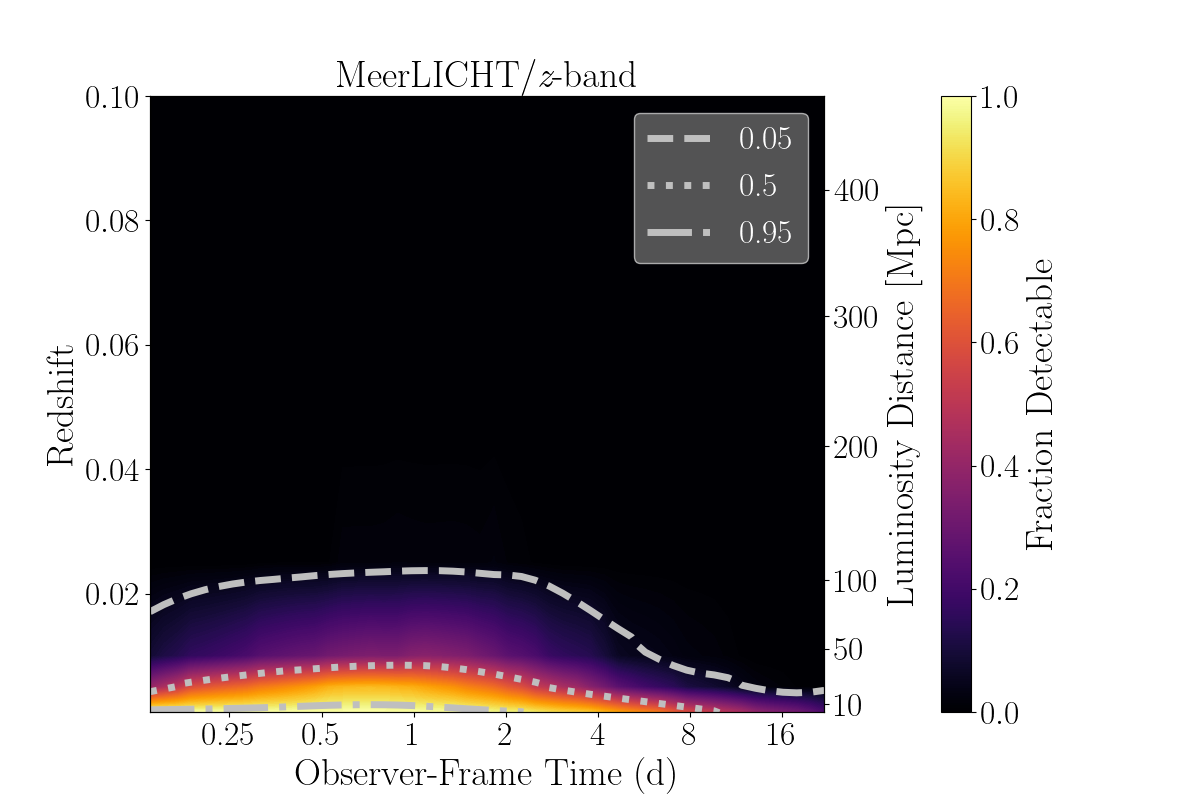}
\caption{Detectability contours for MeerLICHT (see Figure~\ref{fig: BlackGEM} caption).
}
\label{fig: MeerLICHT}
\end{figure*}

\begin{figure*}
\centering
\includegraphics[width=0.4\columnwidth, trim=20 0 100 00, clip]{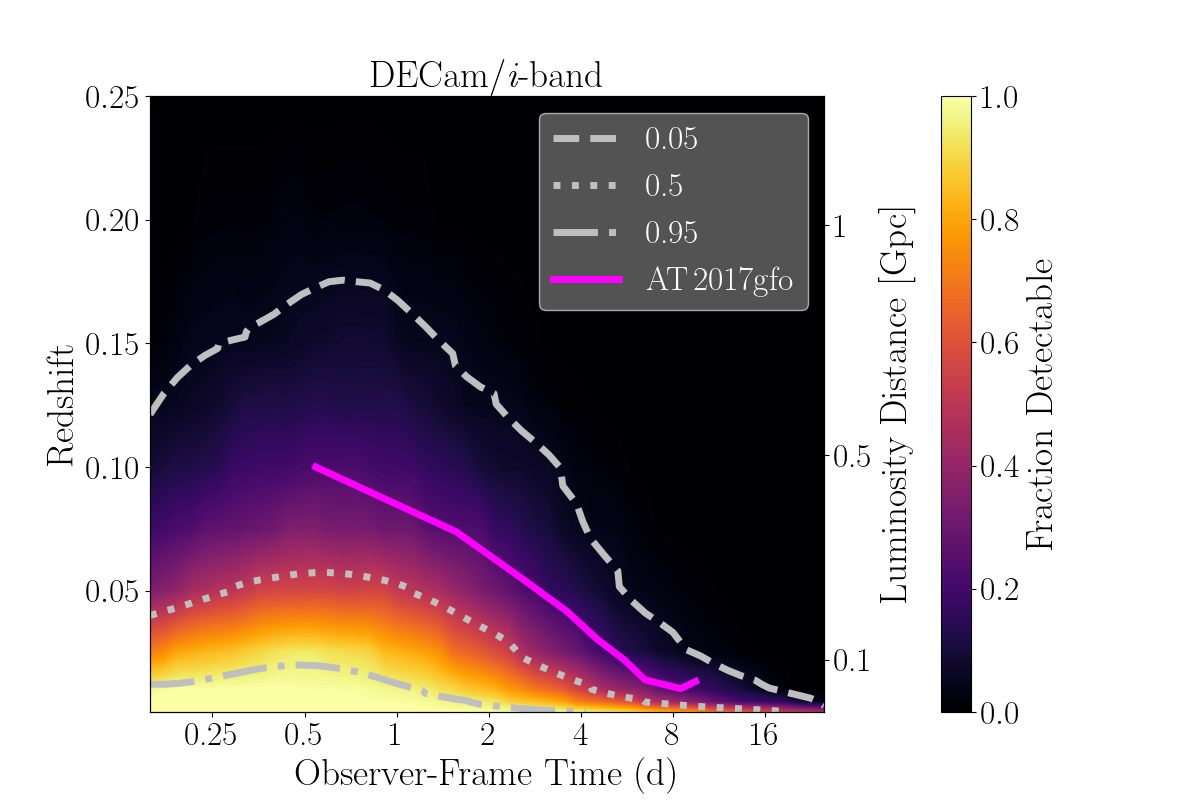}
\includegraphics[width=0.4\columnwidth, trim=20 0 100 00, clip]{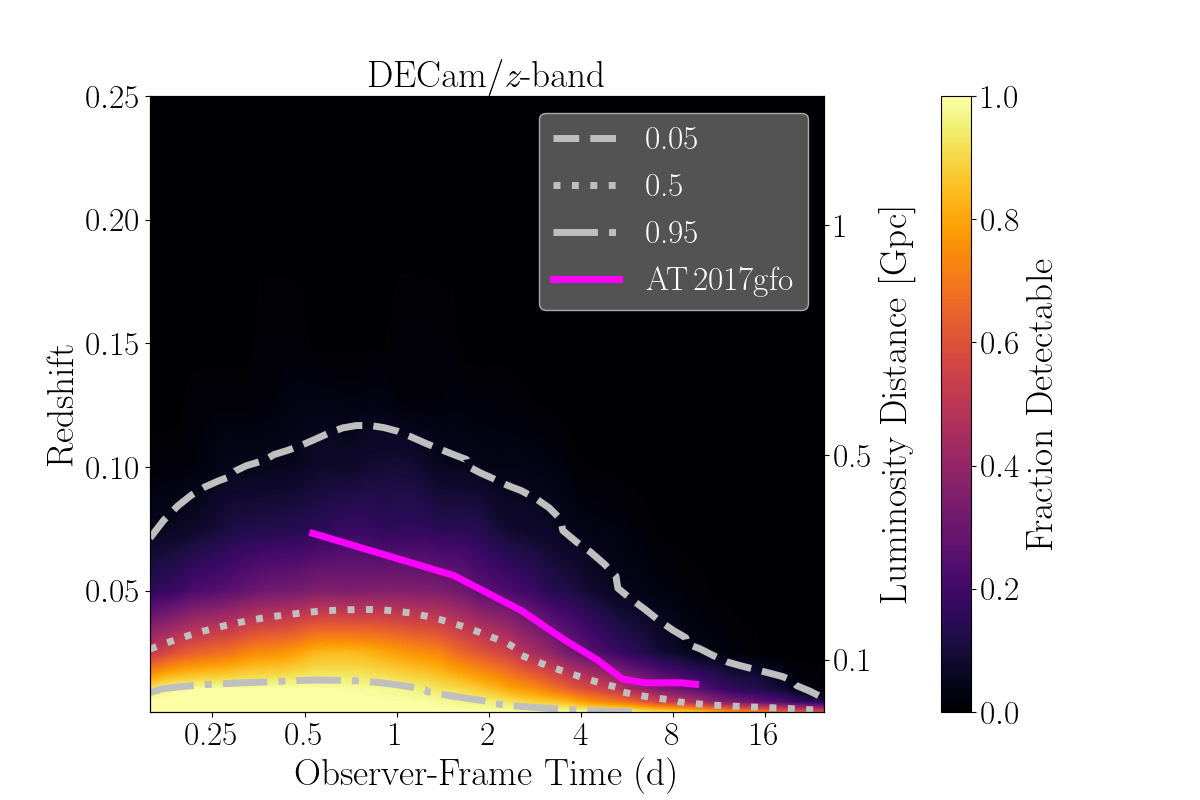}
\caption{Detectability contours for DECam (see Figure~\ref{fig: BlackGEM} caption).
}
\label{fig: DECam}
\end{figure*}

\begin{figure*}
\centering
\includegraphics[width=0.4\columnwidth, trim=20 0 100 00, clip]{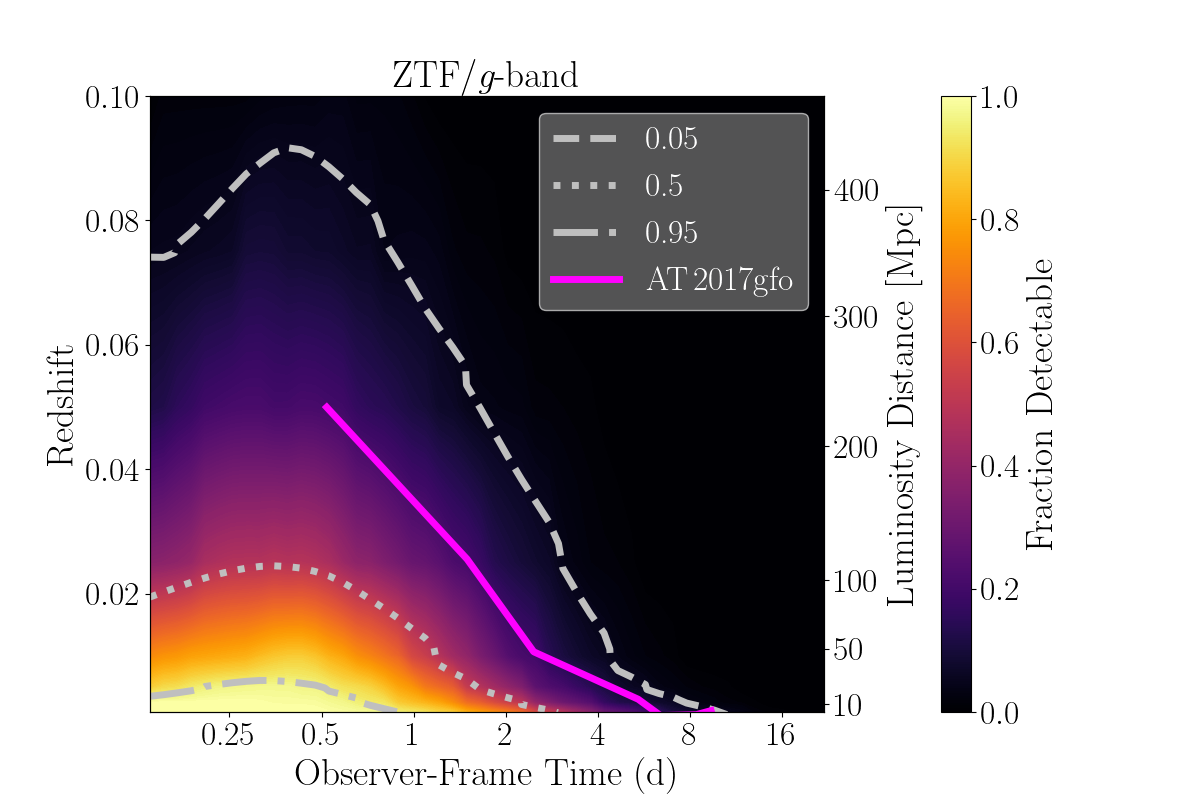}
\includegraphics[width=0.4\columnwidth, trim=20 0 100 00, clip]{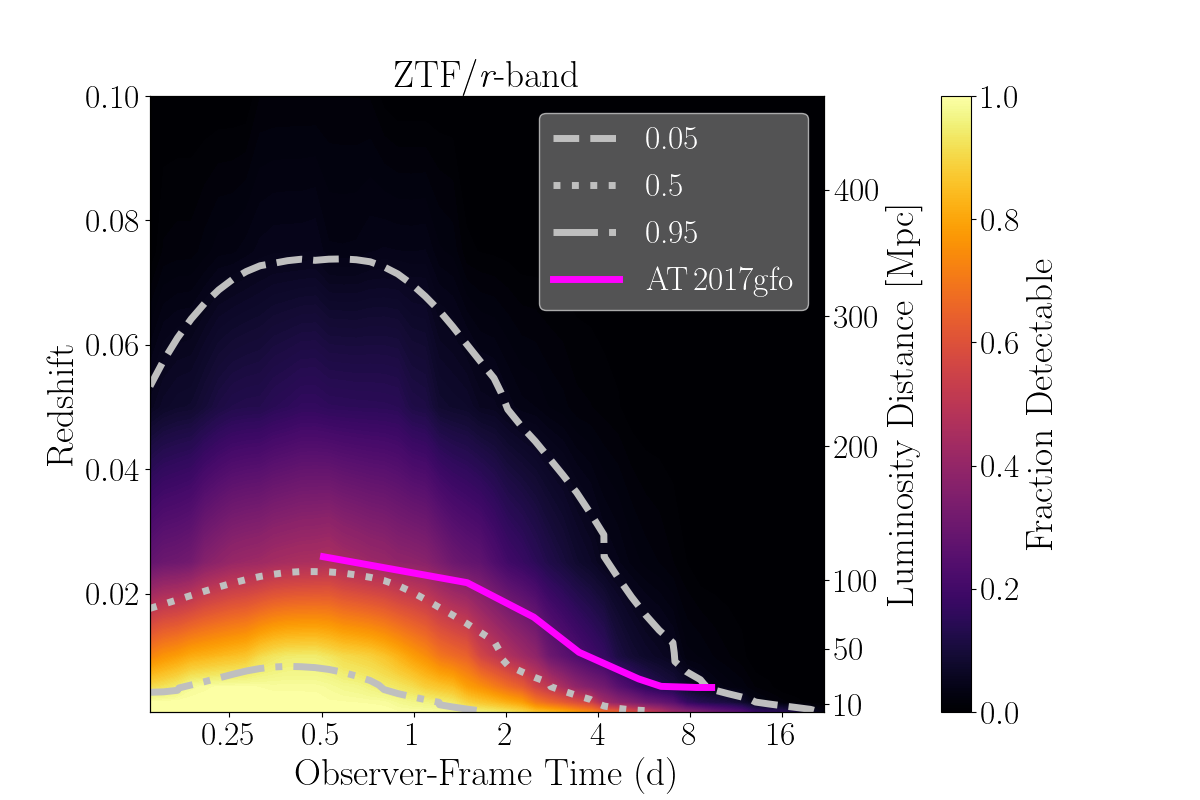}
\includegraphics[width=0.4\columnwidth, trim=20 0 100 00, clip]{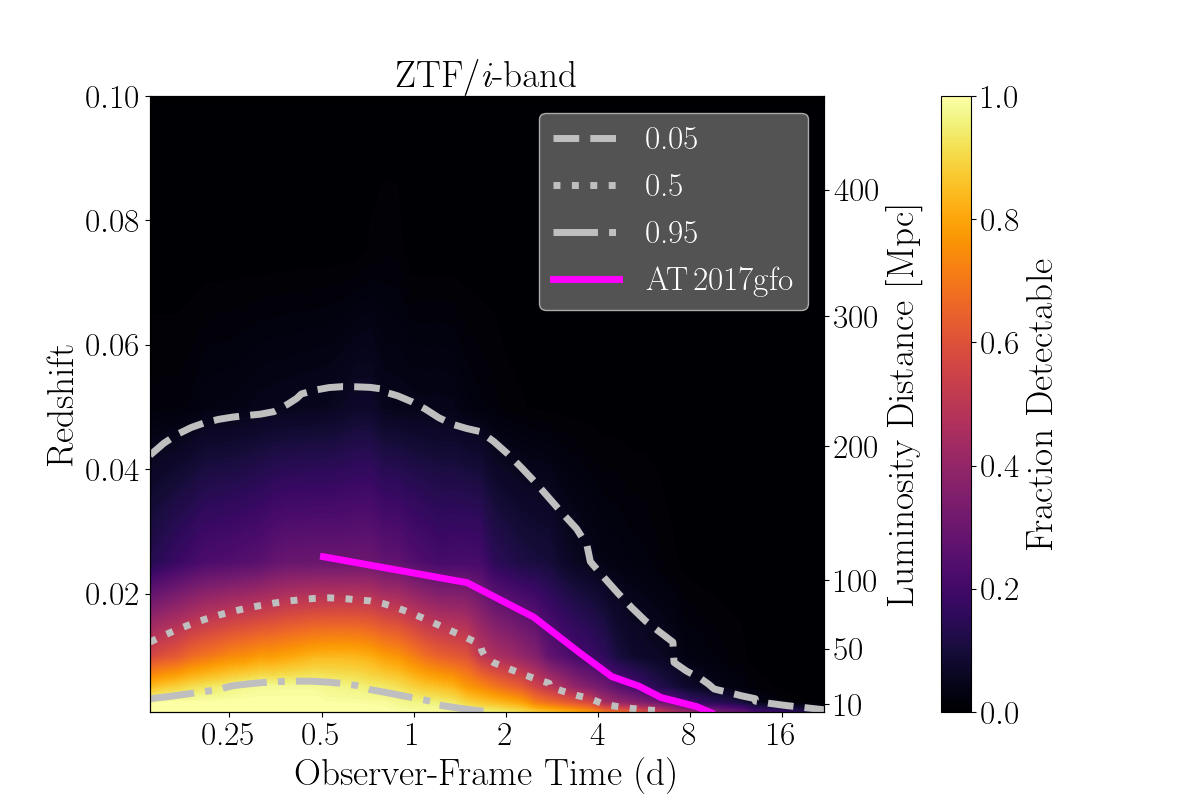}
\caption{Detectability contours for ZTF (see Figure~\ref{fig: BlackGEM} caption).
}
\label{fig: ZTF}
\end{figure*}

\begin{figure*}
\centering
\includegraphics[width=0.4\columnwidth, trim=20 0 100 00, clip]{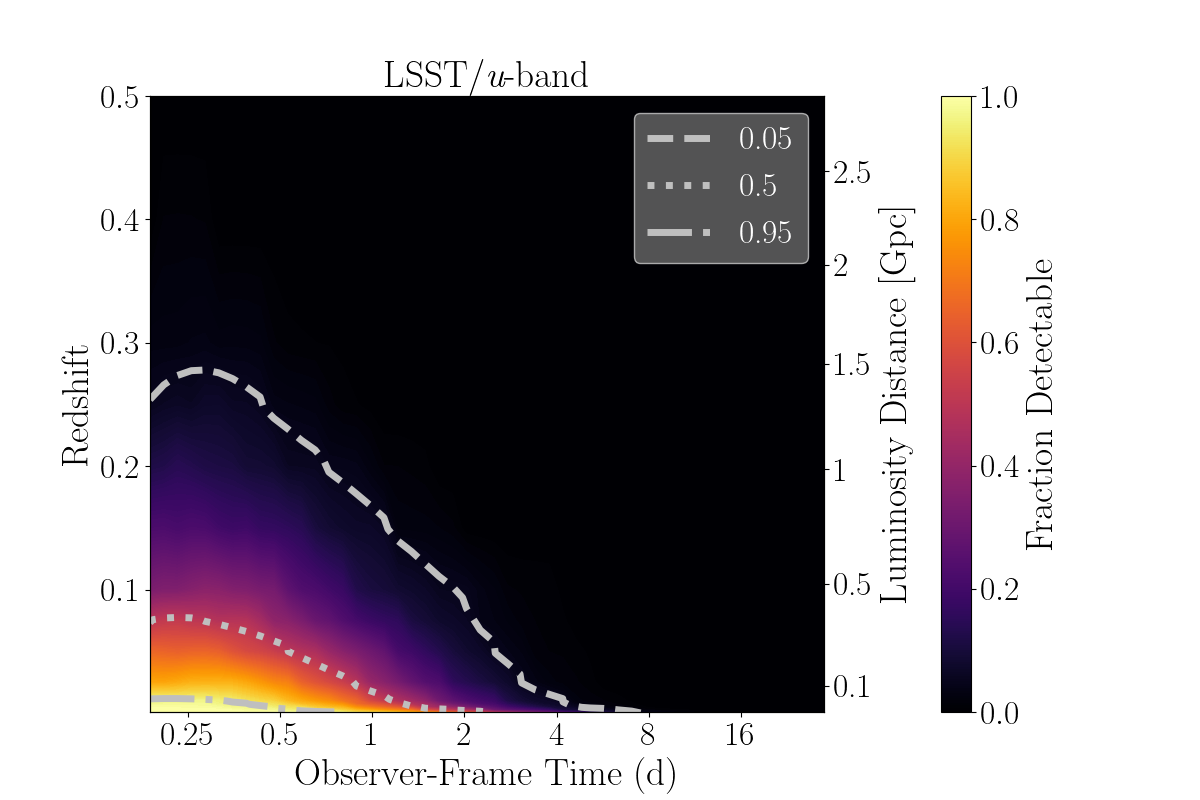}
\includegraphics[width=0.4\columnwidth, trim=20 0 100 0, clip]{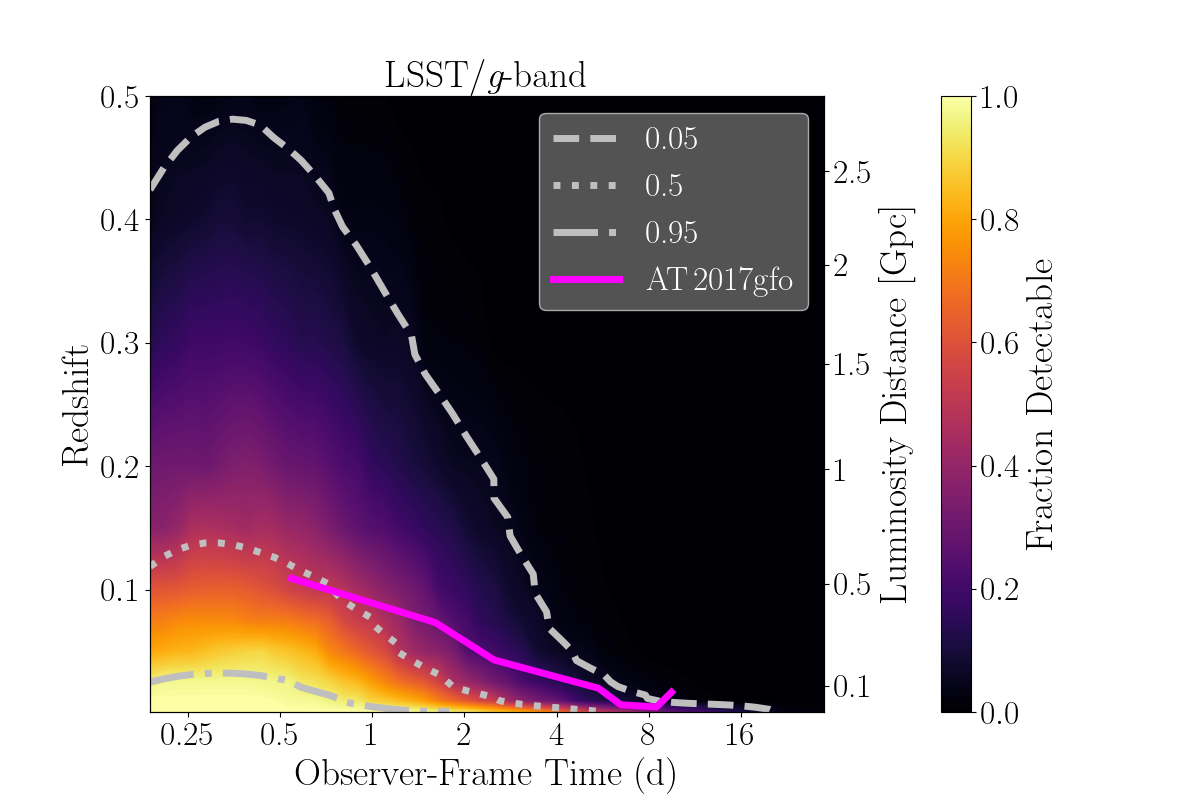}
\includegraphics[width=0.4\columnwidth, trim=20 0 100 0, clip]{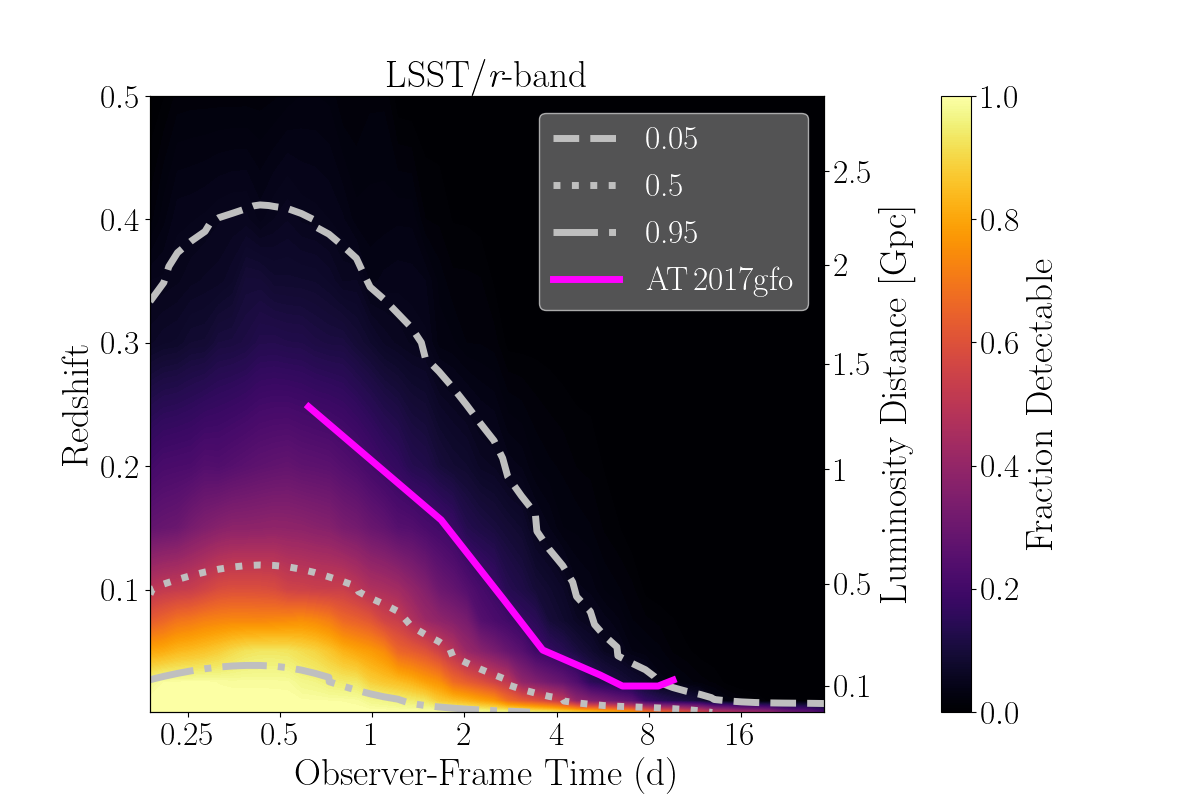}
\includegraphics[width=0.4\columnwidth, trim=20 0 100 0, clip]{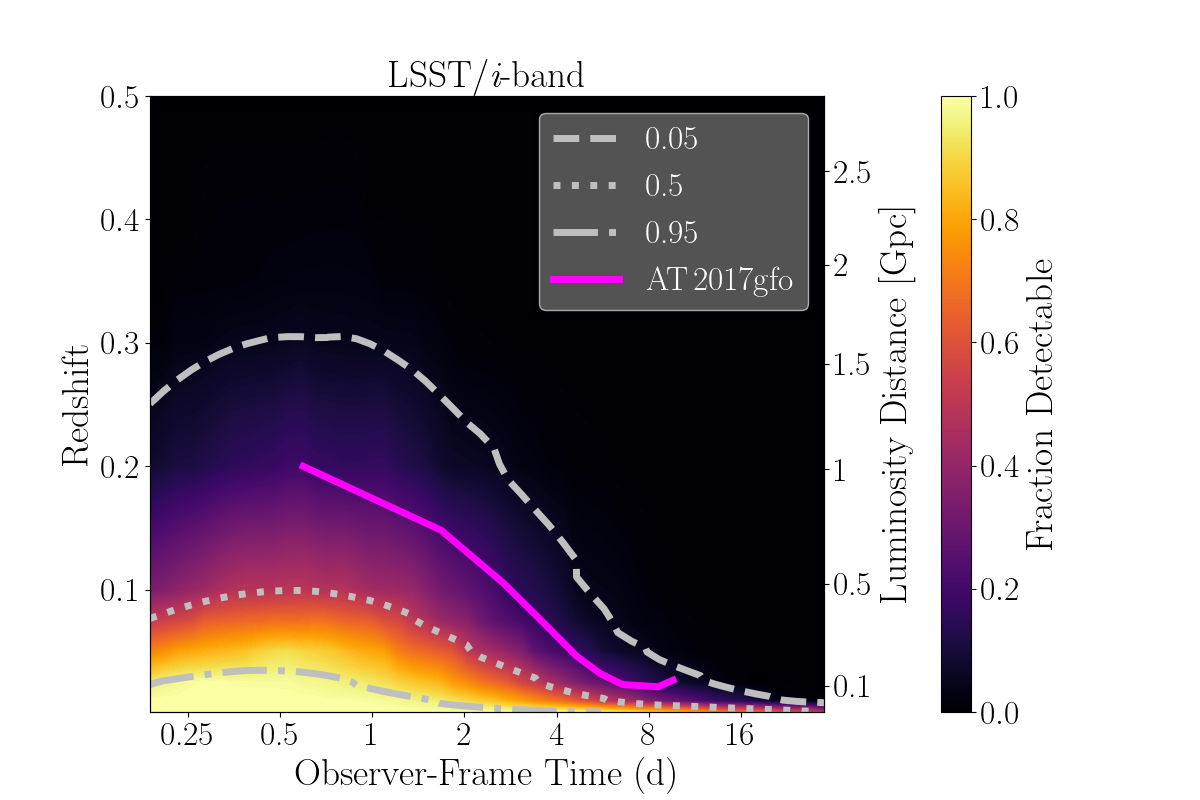}
\includegraphics[width=0.4\columnwidth, trim=20 0 100 0, clip]{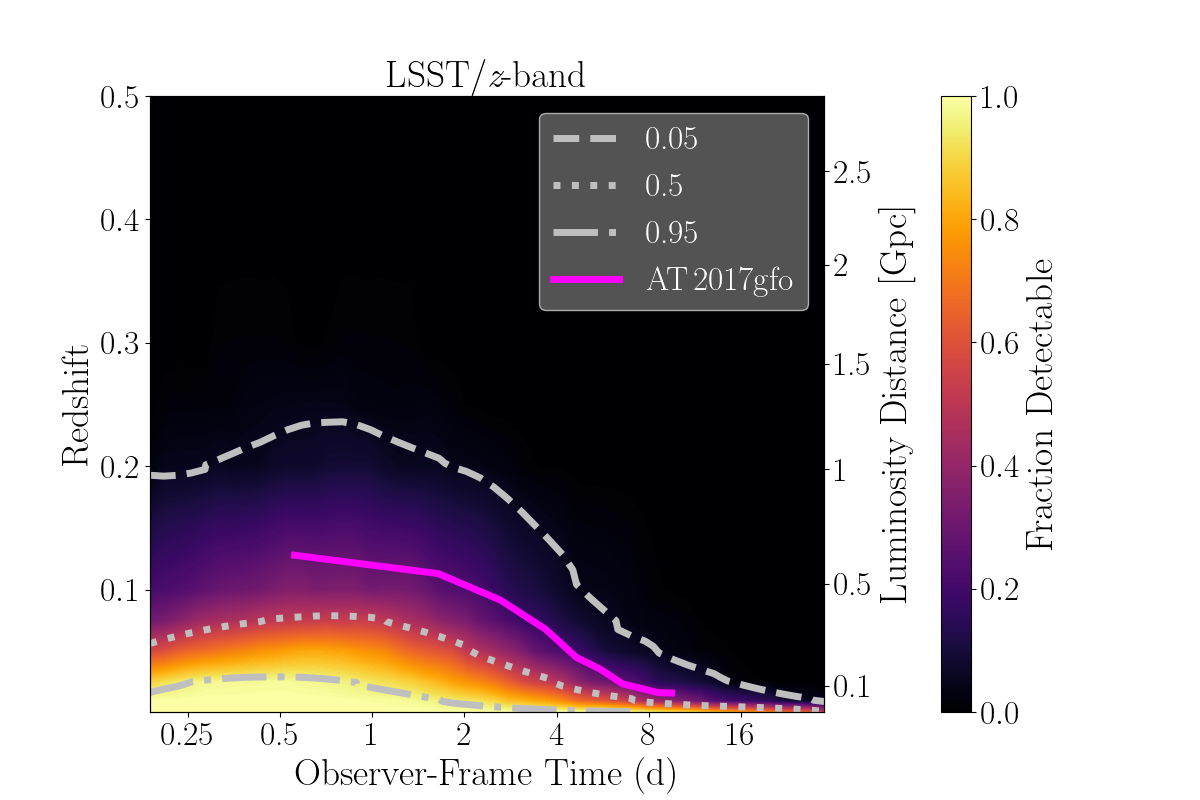}
\includegraphics[width=0.4\columnwidth, trim=20 0 100 0, clip]{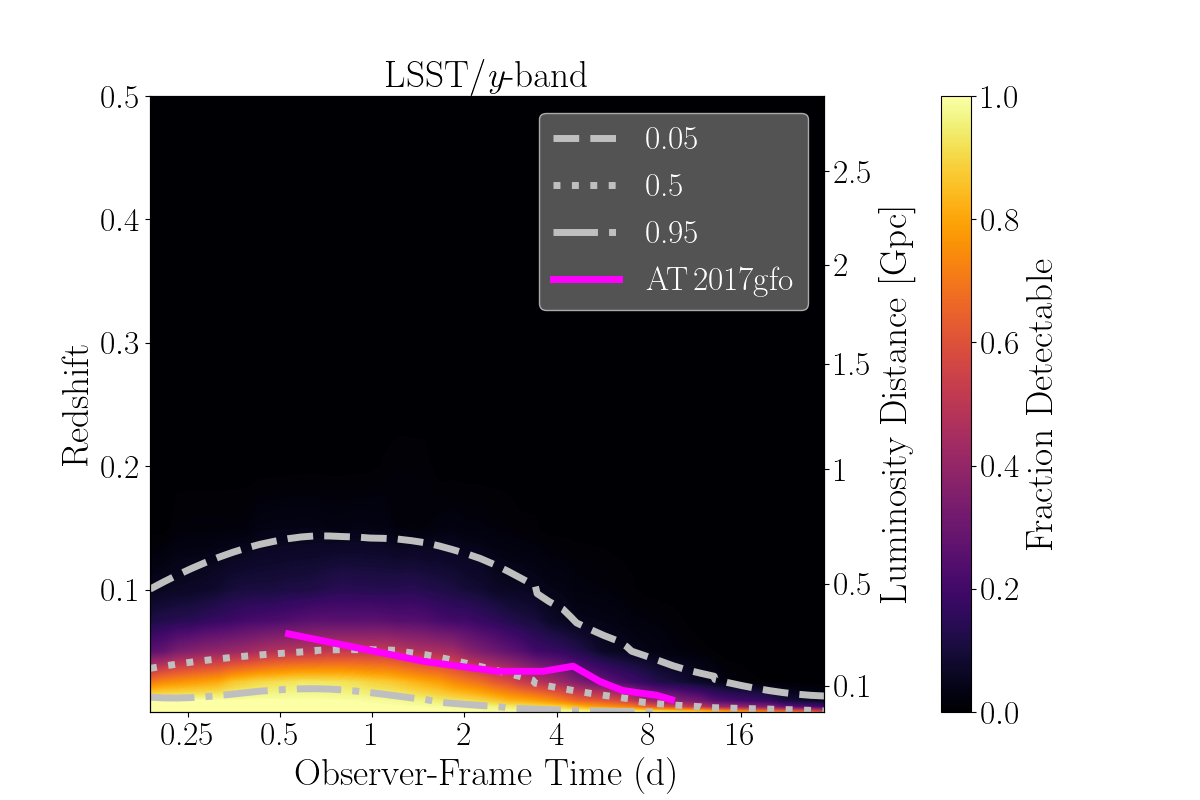}
\caption{Detectability contours for LSST (see Figure~\ref{fig: BlackGEM} caption).
}
\label{fig: LSST}
\end{figure*}

\begin{figure*}
\centering
\includegraphics[width=0.4\columnwidth, trim=20 0 100 00, clip]{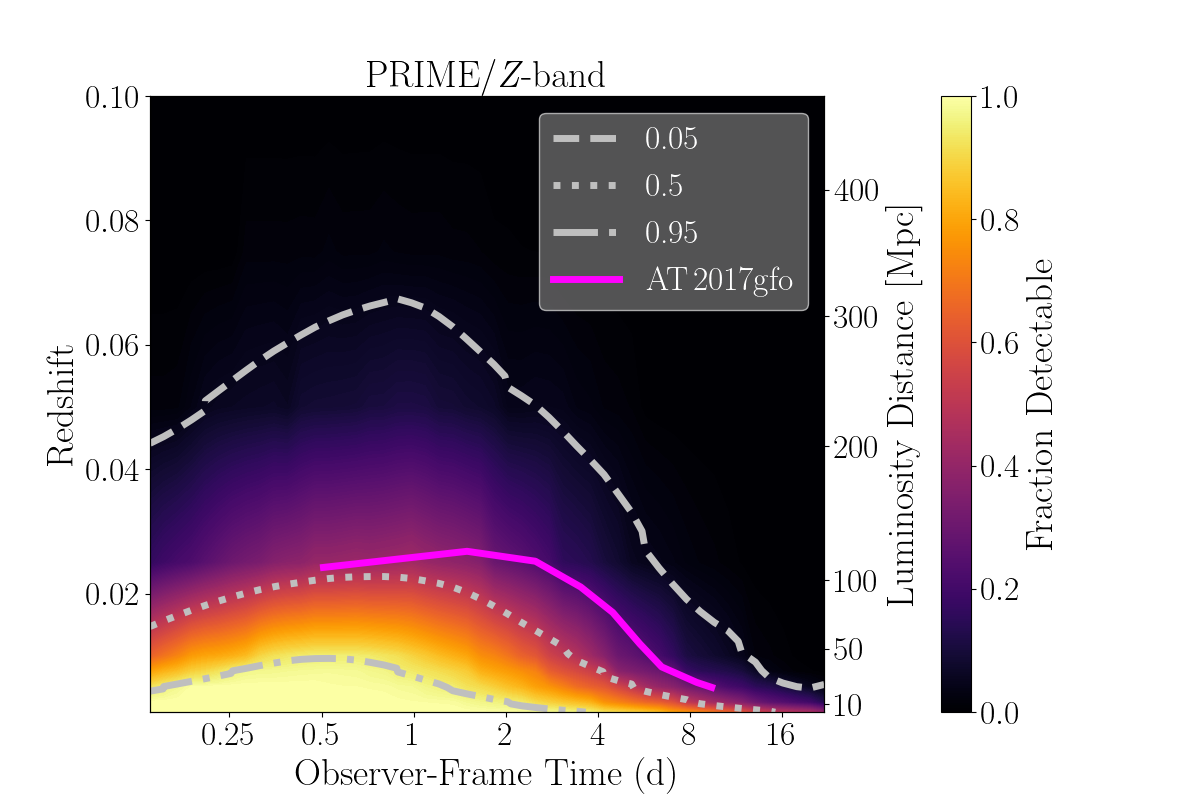}
\includegraphics[width=0.4\columnwidth, trim=20 0 100 00, clip]{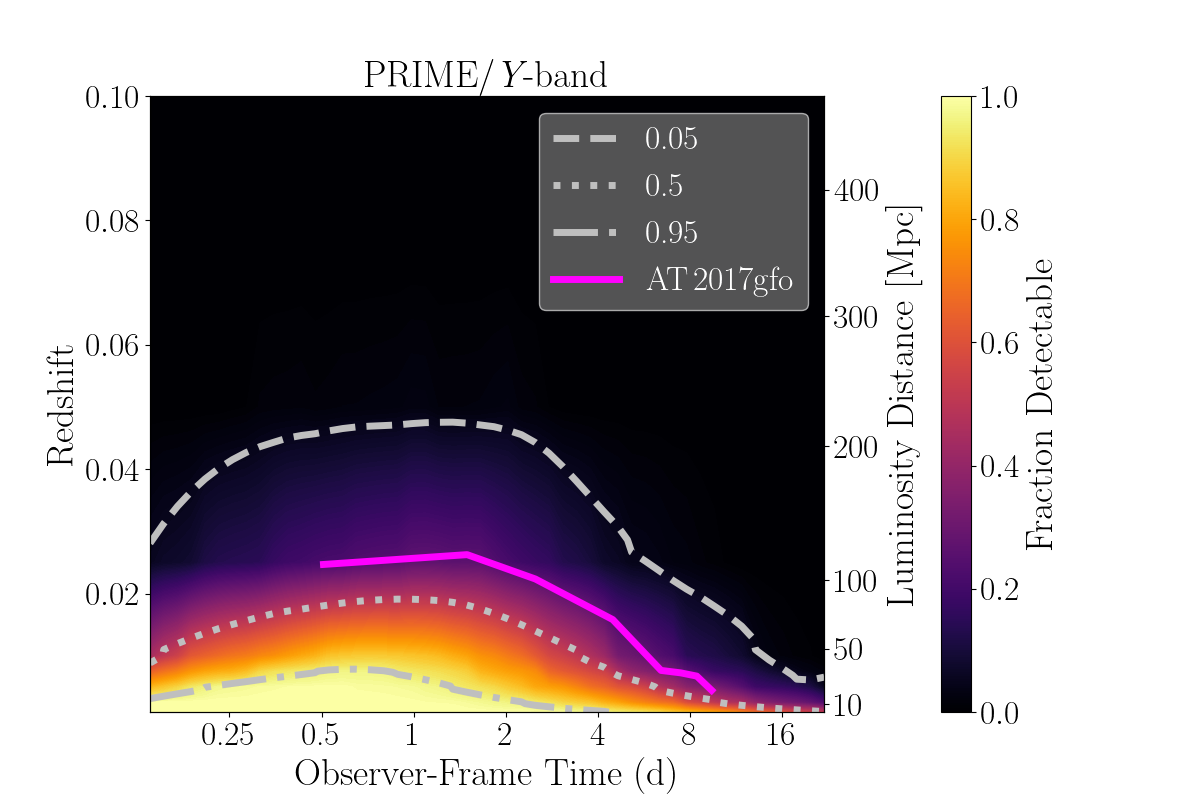}
\includegraphics[width=0.4\columnwidth, trim=20 0 100 00, clip]{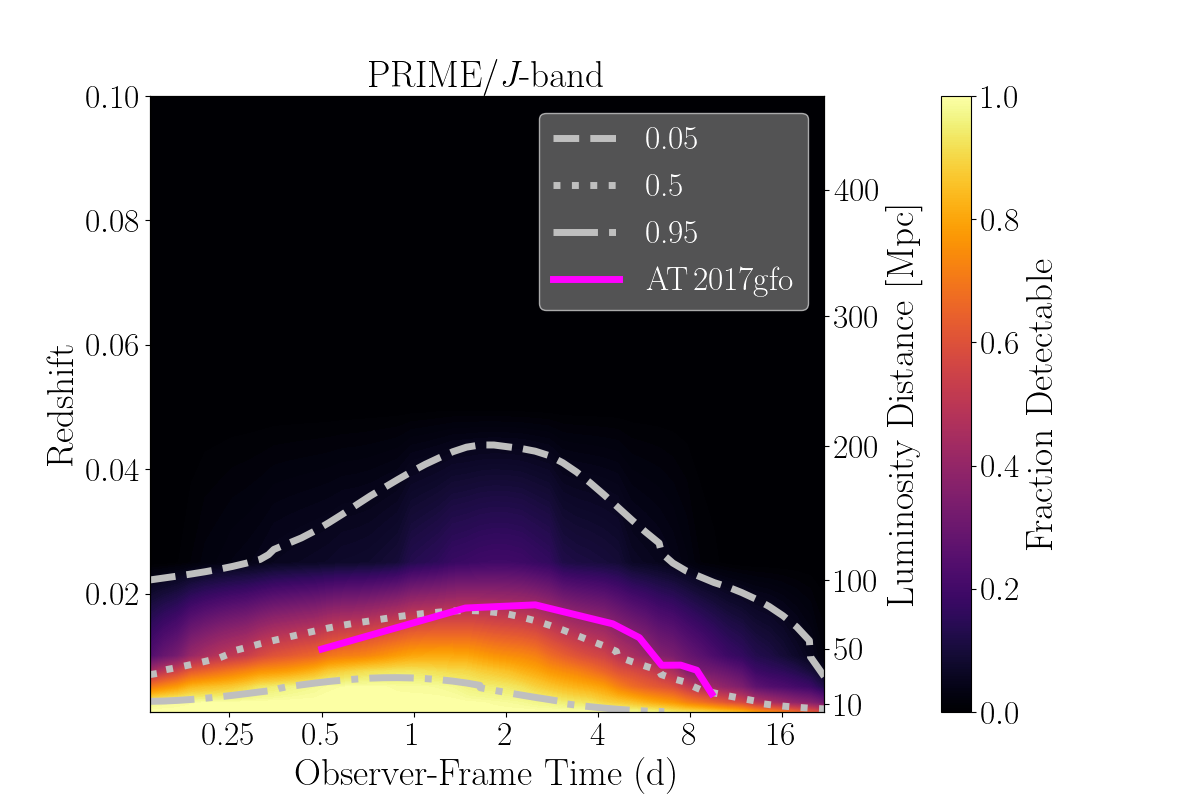}
\includegraphics[width=0.4\columnwidth, trim=20 0 100 00, clip]{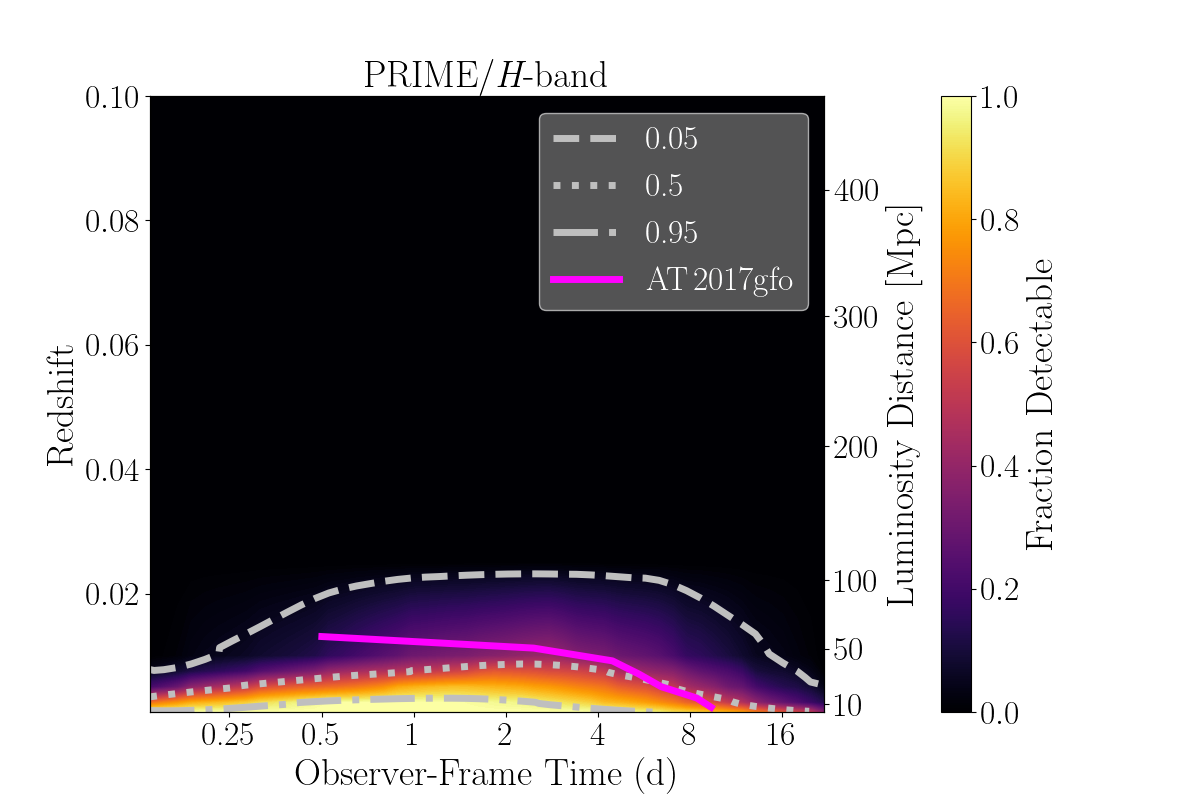}
\caption{Detectability contours for PRIME (see Figure~\ref{fig: BlackGEM} caption).
}
\label{fig: PRIME}
\end{figure*}

\begin{figure*}
\centering
\includegraphics[width=0.4\columnwidth, trim=20 0 100 00, clip]{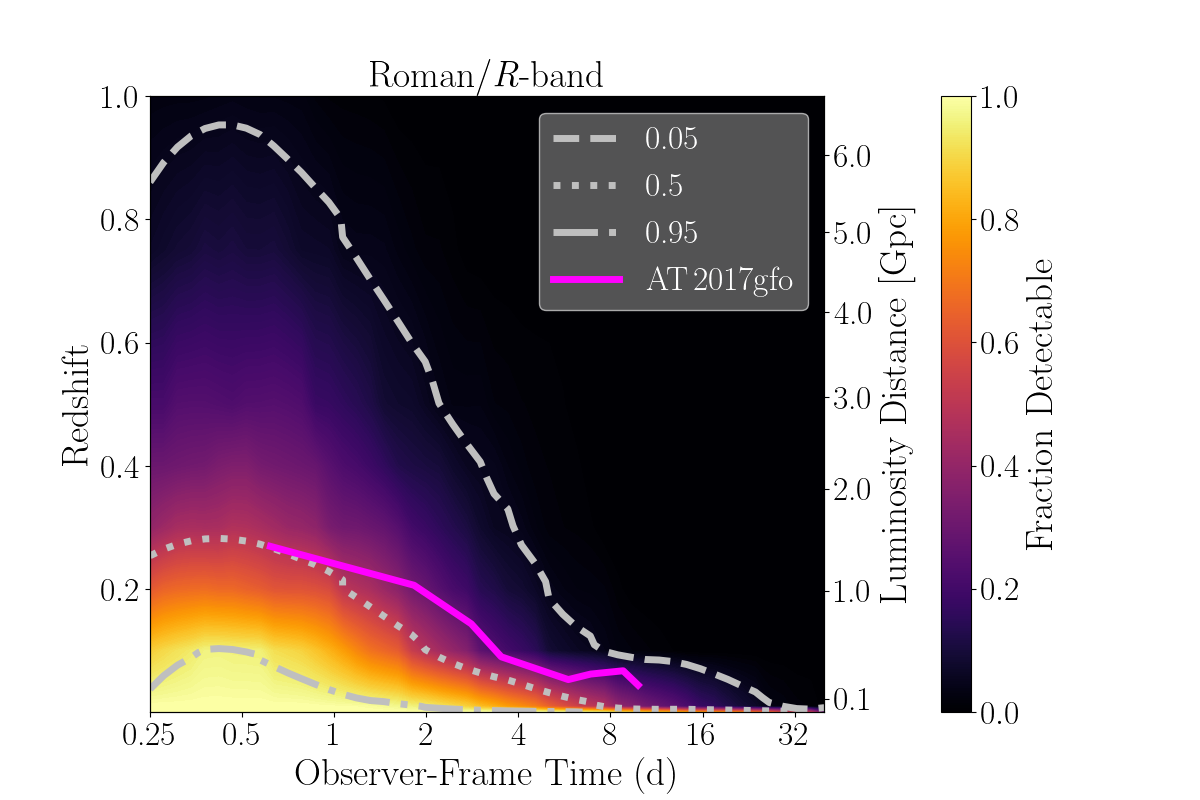}
\includegraphics[width=0.4\columnwidth, trim=20 0 100 00, clip]{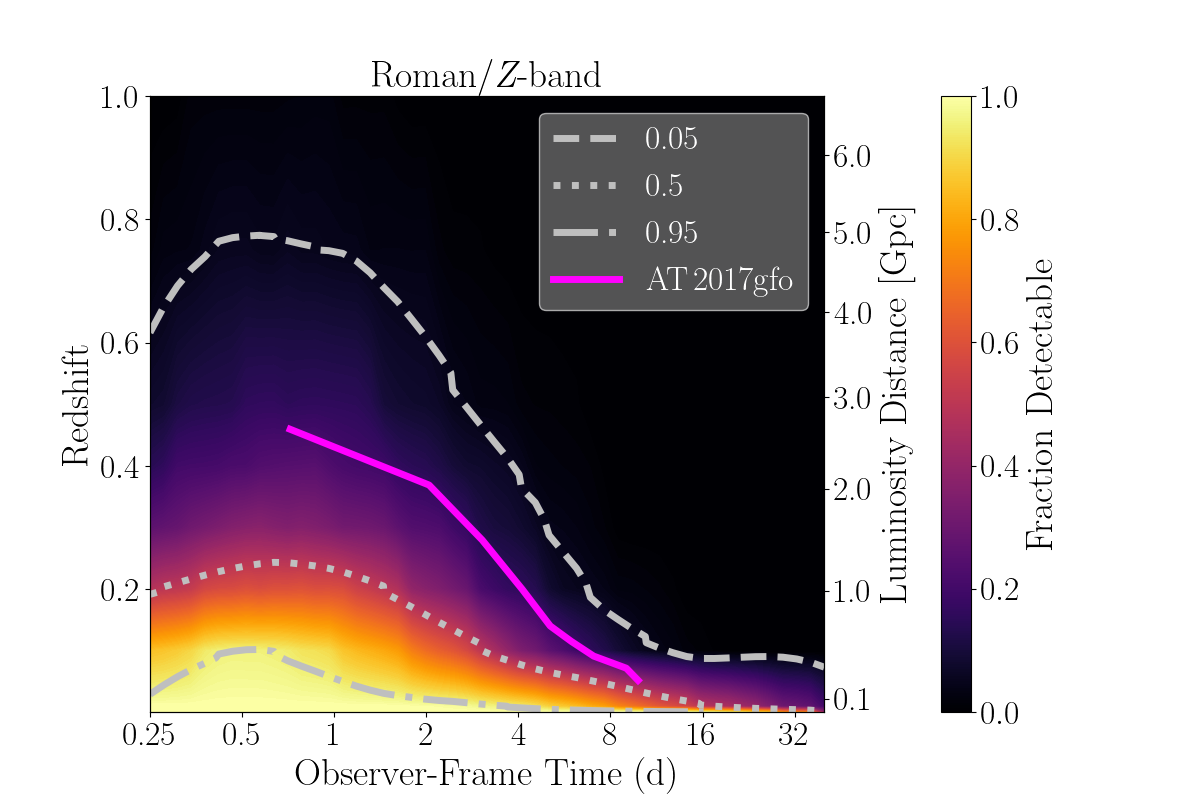}
\includegraphics[width=0.4\columnwidth, trim=20 0 100 00, clip]{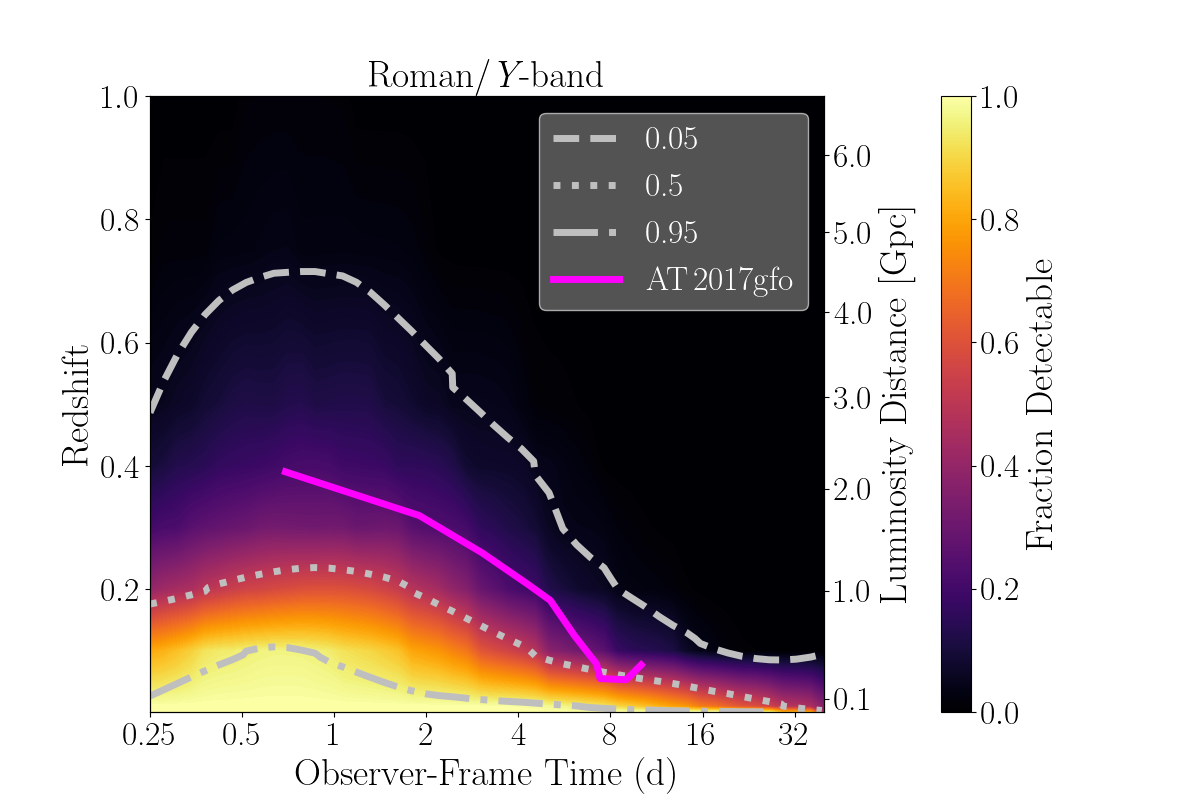}
\includegraphics[width=0.4\columnwidth, trim=20 0 100 00, clip]{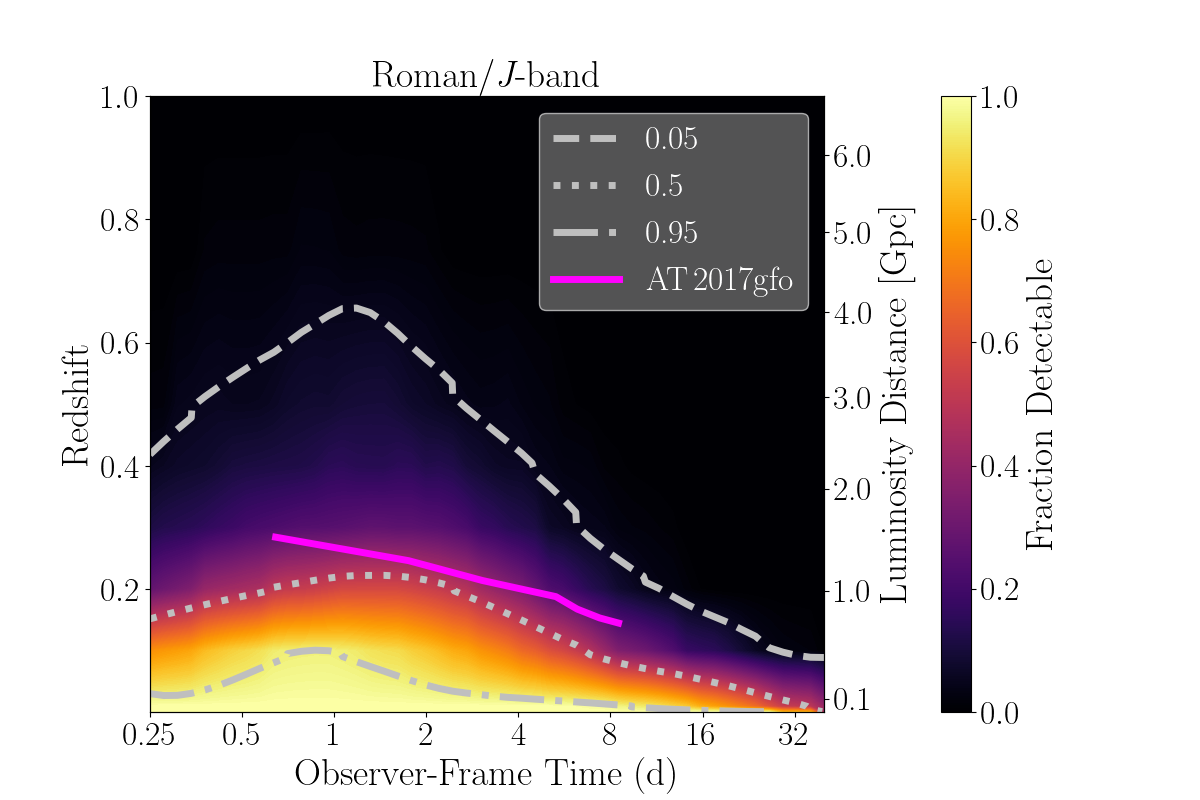}
\includegraphics[width=0.4\columnwidth, trim=20 0 100 00, clip]{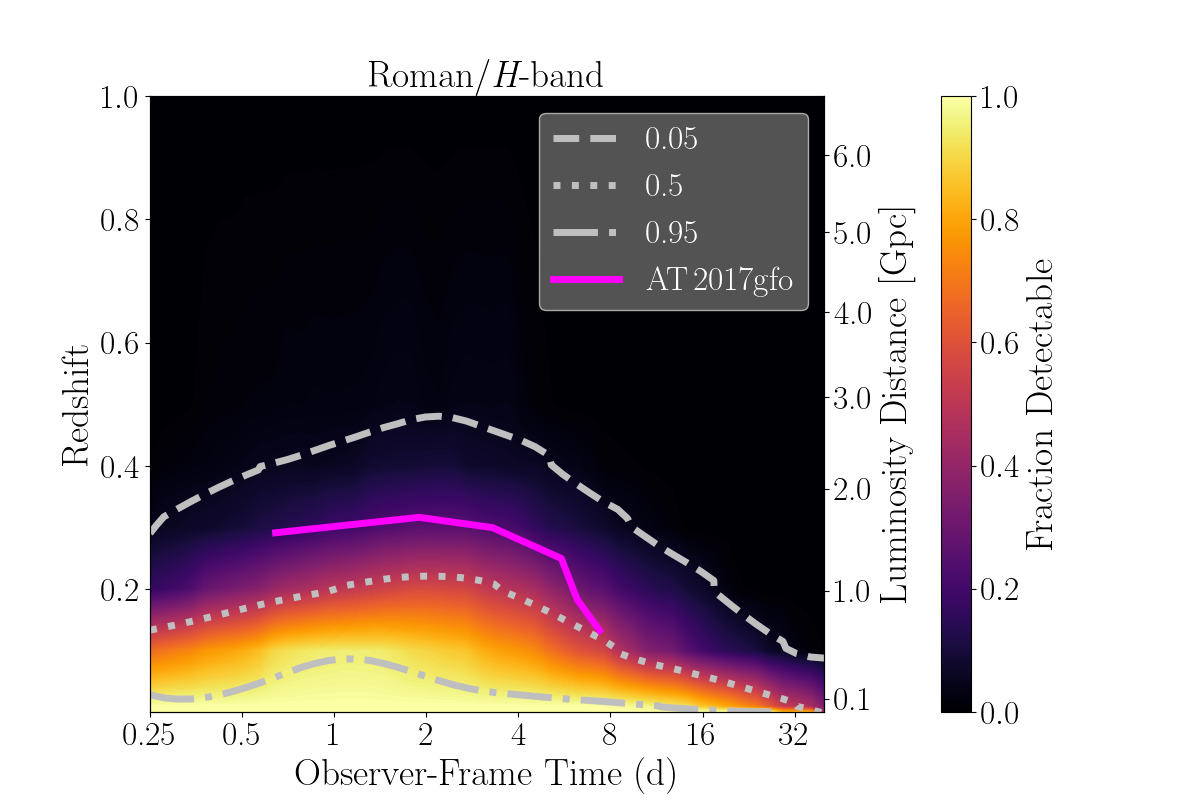}
\includegraphics[width=0.4\columnwidth, trim=20 0 100 00, clip]{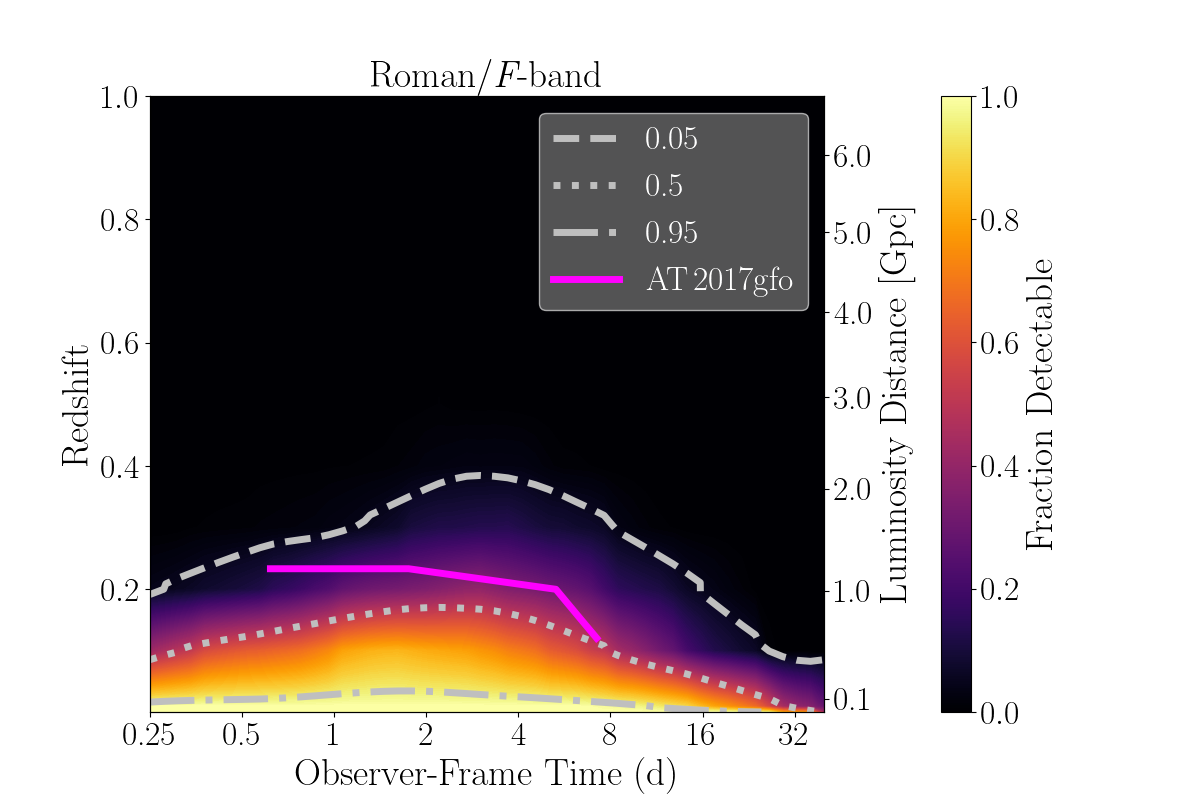}
\caption{Detectability contours for Roman (see Figure~\ref{fig: BlackGEM} caption).
}
\label{fig: Roman}
\end{figure*}

\begin{figure*}
\centering
\includegraphics[width=0.4\columnwidth, trim=20 0 100 00, clip]{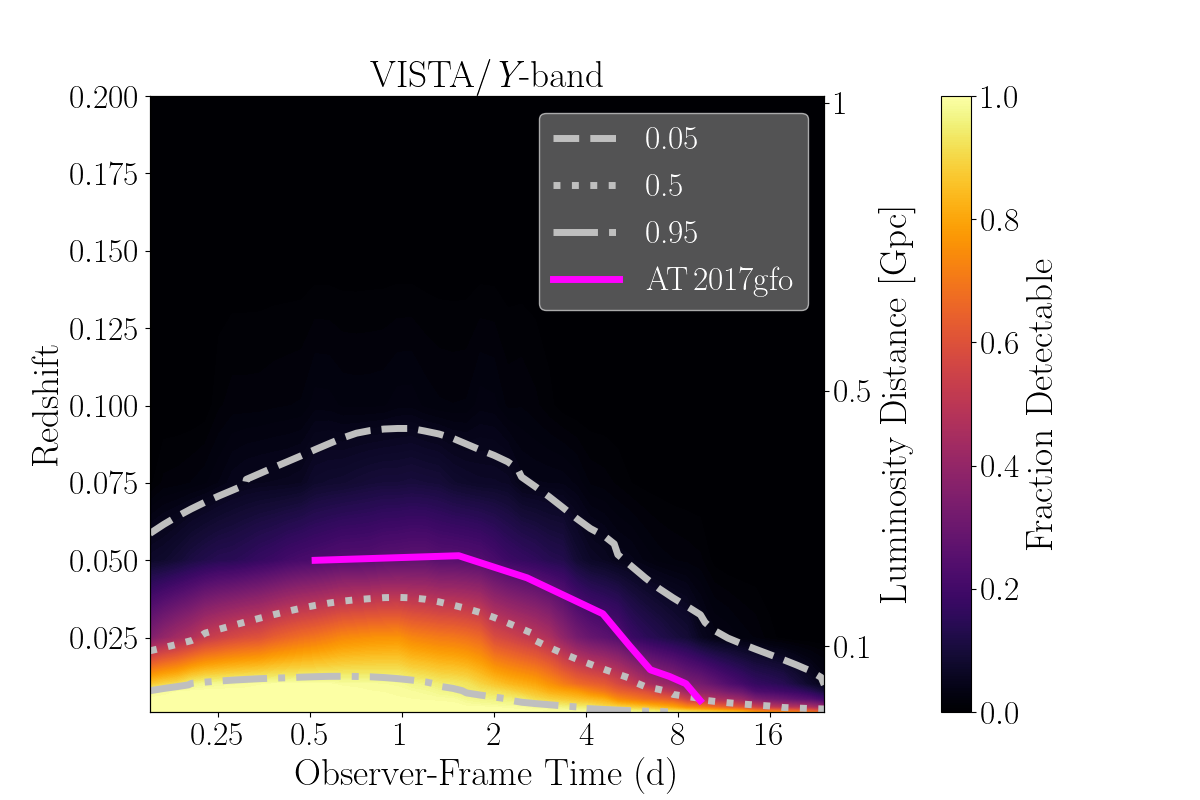}
\includegraphics[width=0.4\columnwidth, trim=20 0 100 00, clip]{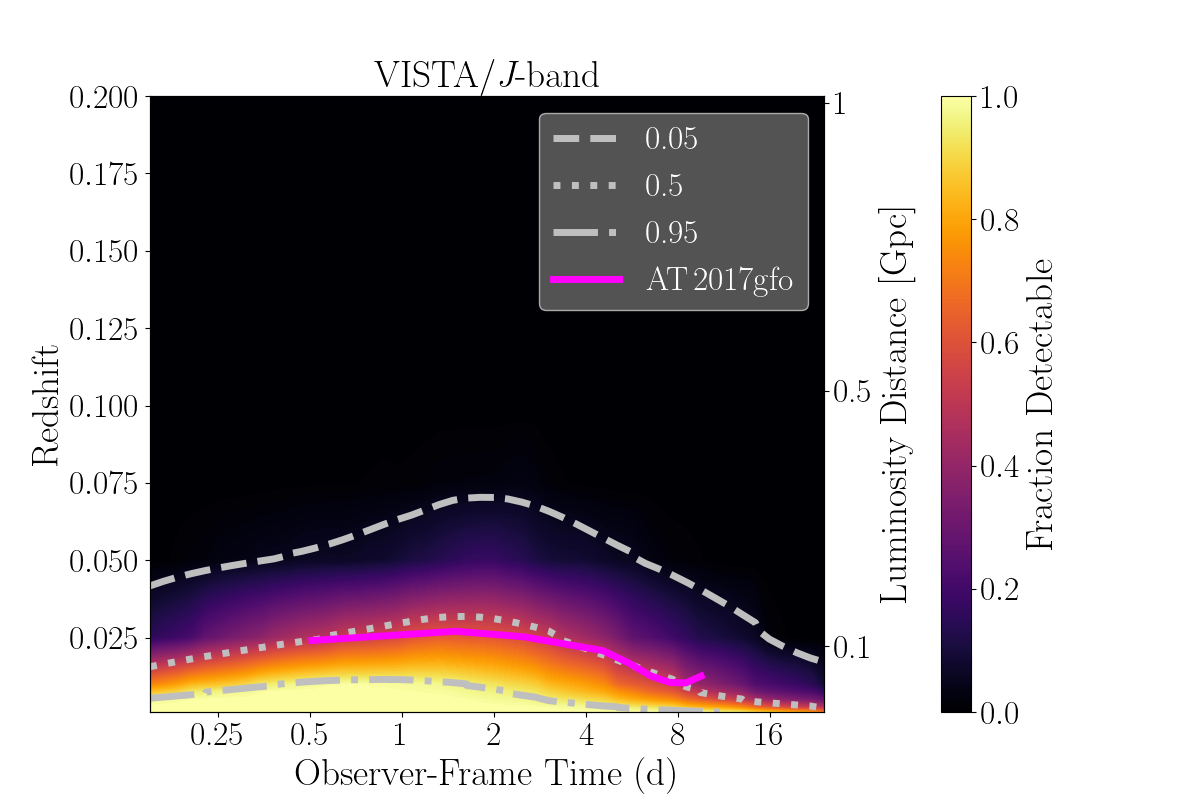}
\includegraphics[width=0.4\columnwidth, trim=20 0 100 00, clip]{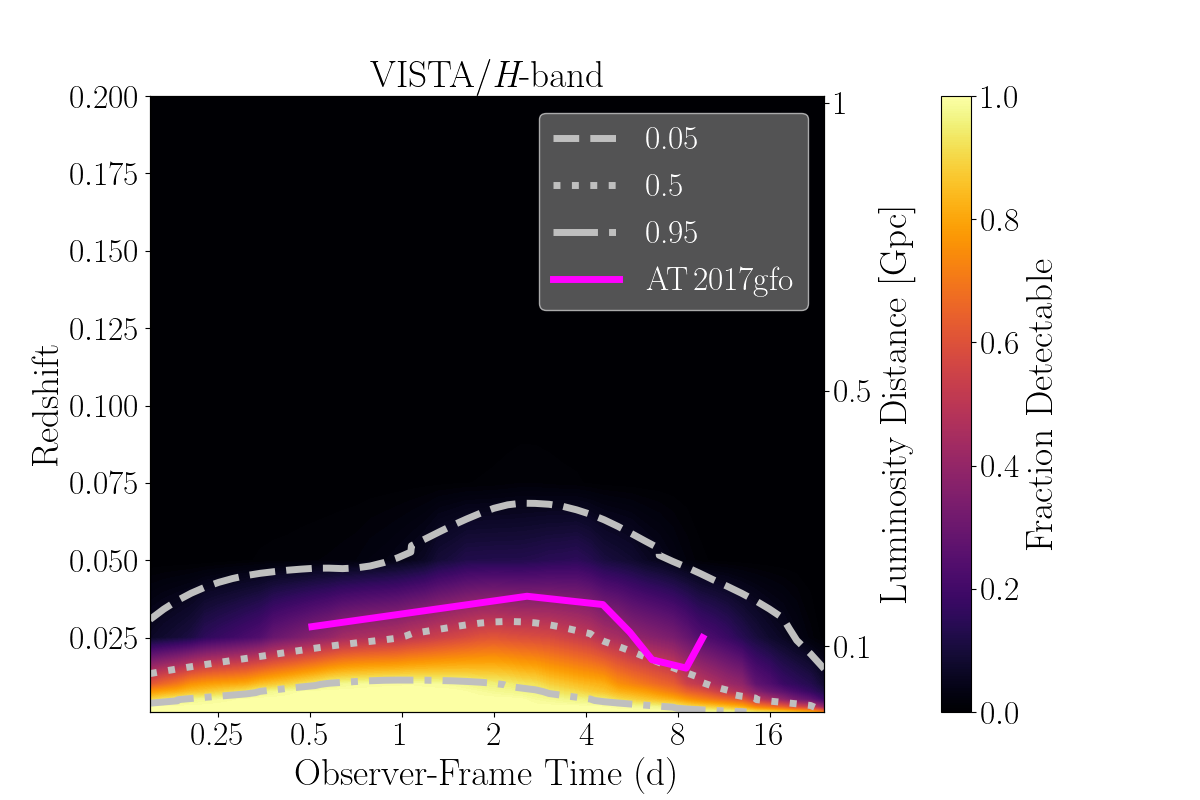}
\includegraphics[width=0.4\columnwidth, trim=20 0 100 00, clip]{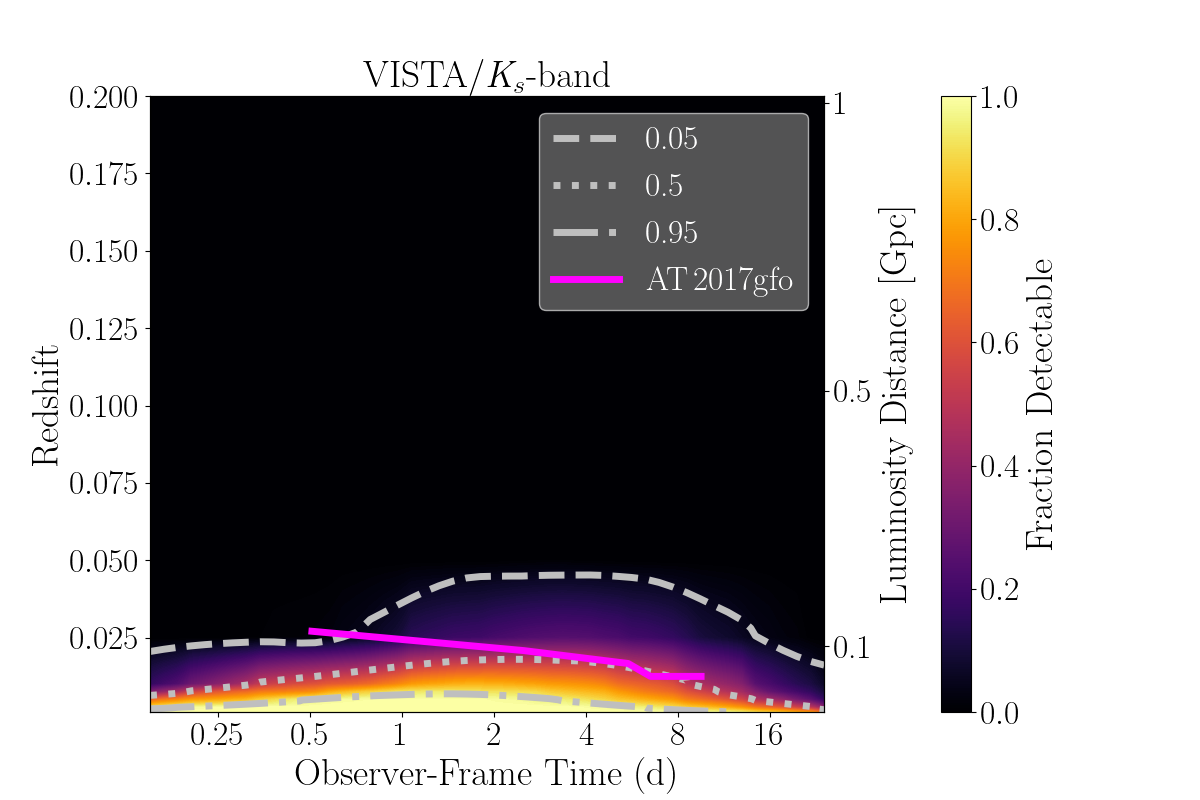}
\caption{Detectability contours for VISTA (see Figure~\ref{fig: BlackGEM} caption).
}
\label{fig: VISTA}
\end{figure*}

\begin{figure*}
\centering
\includegraphics[width=0.4\columnwidth, trim=20 0 100 00, clip]{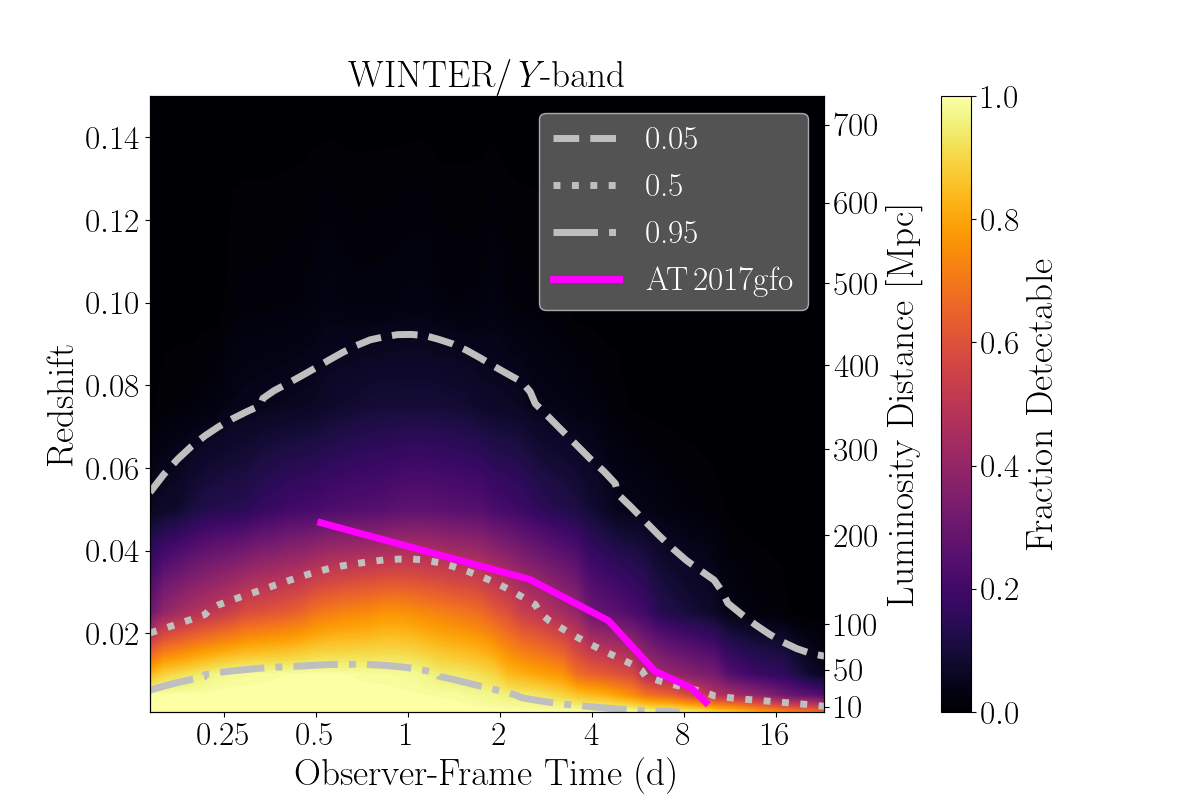}
\includegraphics[width=0.4\columnwidth, trim=20 0 100 00, clip]{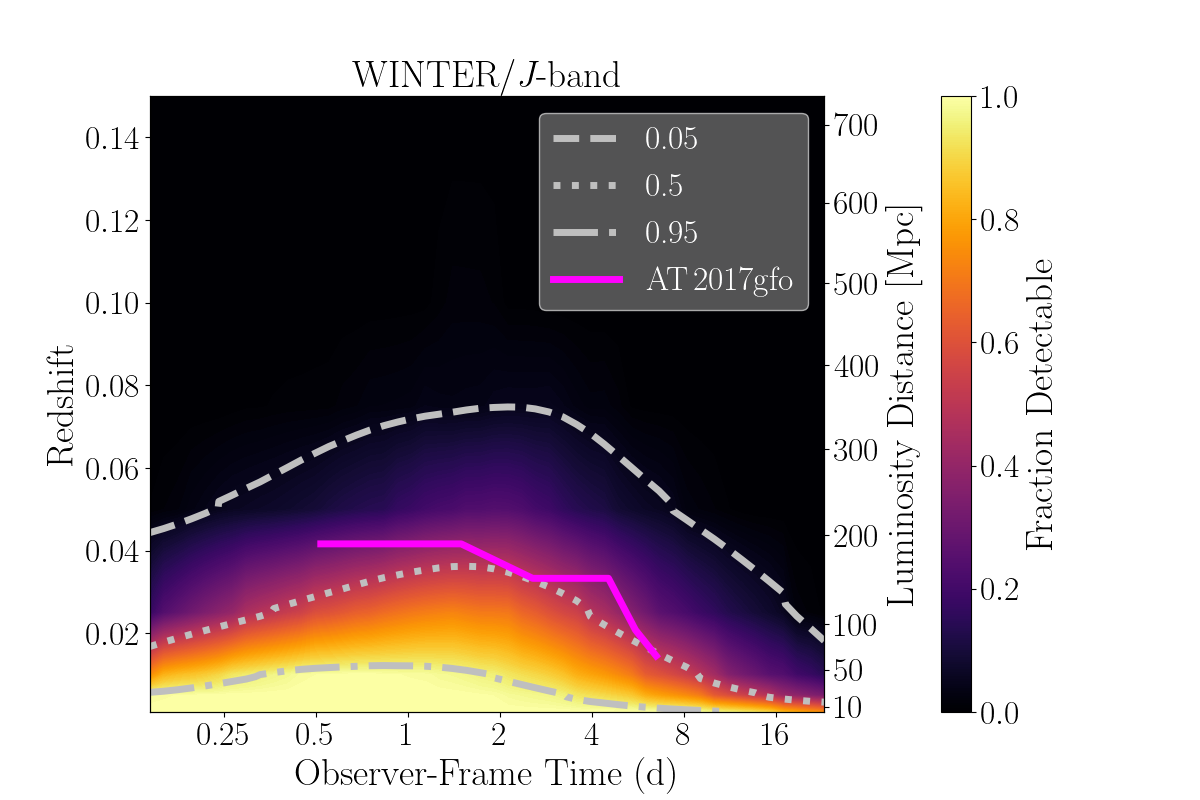}
\includegraphics[width=0.4\columnwidth, trim=20 0 100 00, clip]{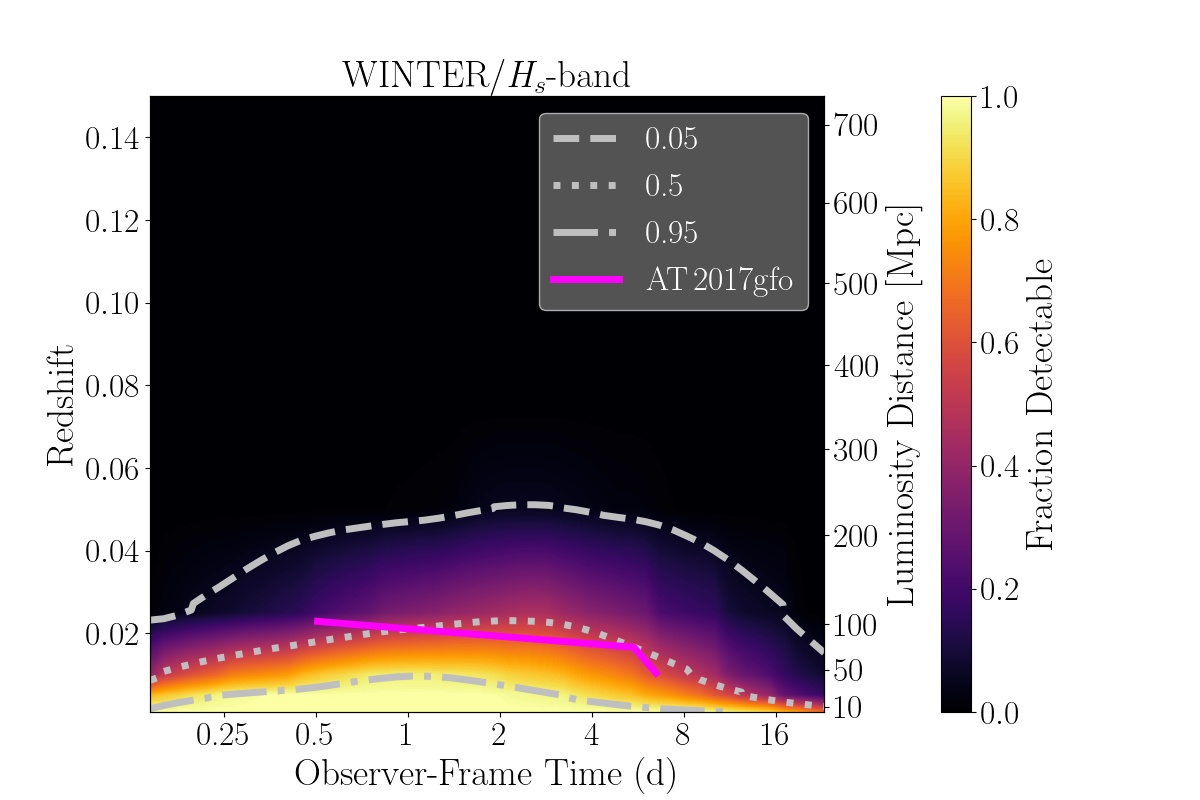}
\caption{Detectability contours for WINTER (see Figure~\ref{fig: BlackGEM} caption).
}
\label{fig: WINTER}
\end{figure*}

\begin{figure*}[h]
\centering
\includegraphics[width=0.4\columnwidth, trim=20 0 100 00, clip]{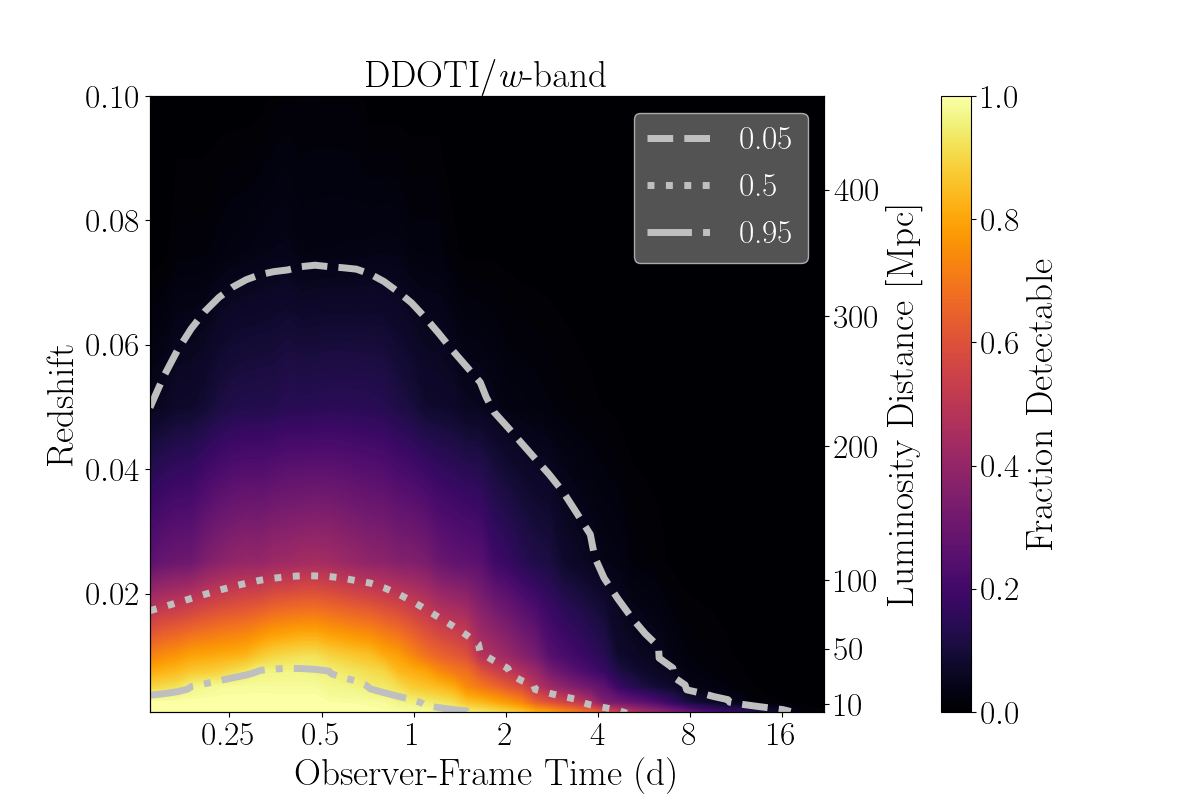}
\includegraphics[width=0.4\columnwidth, trim=20 0 100 00, clip]{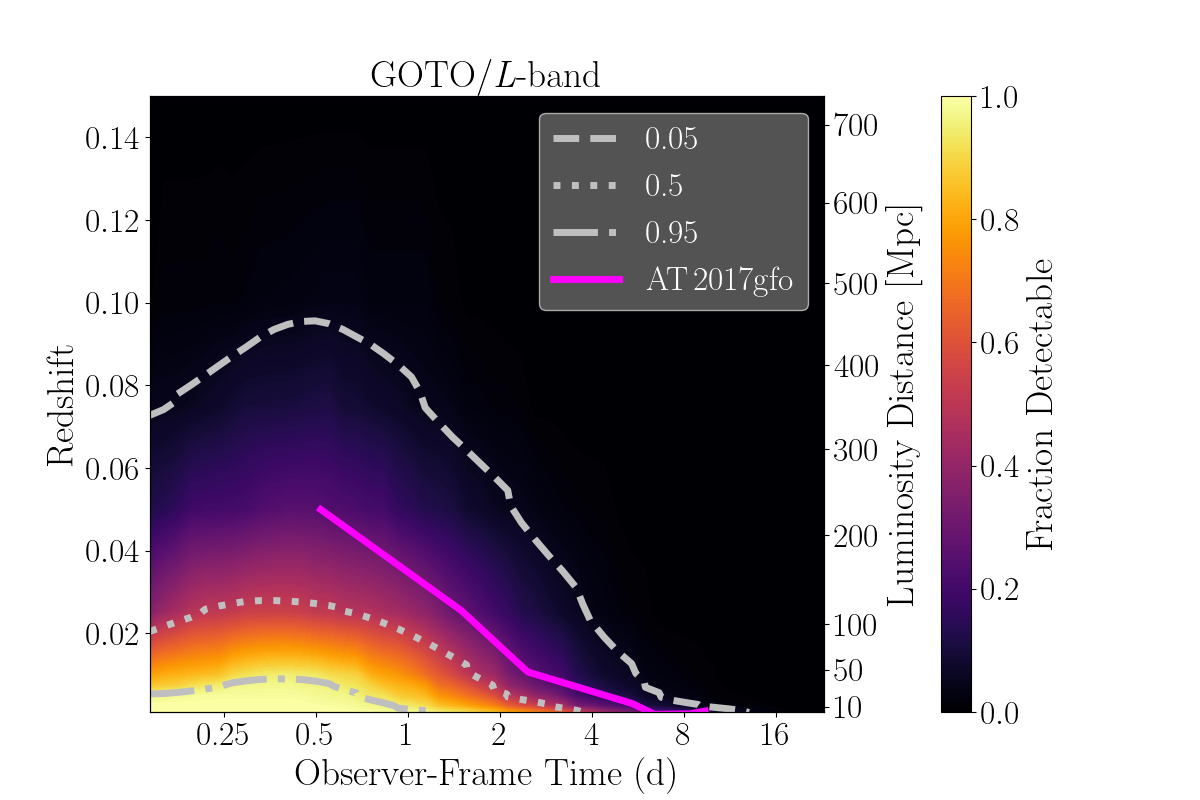}
\includegraphics[width=0.4\columnwidth, trim=20 0 100 00, clip]{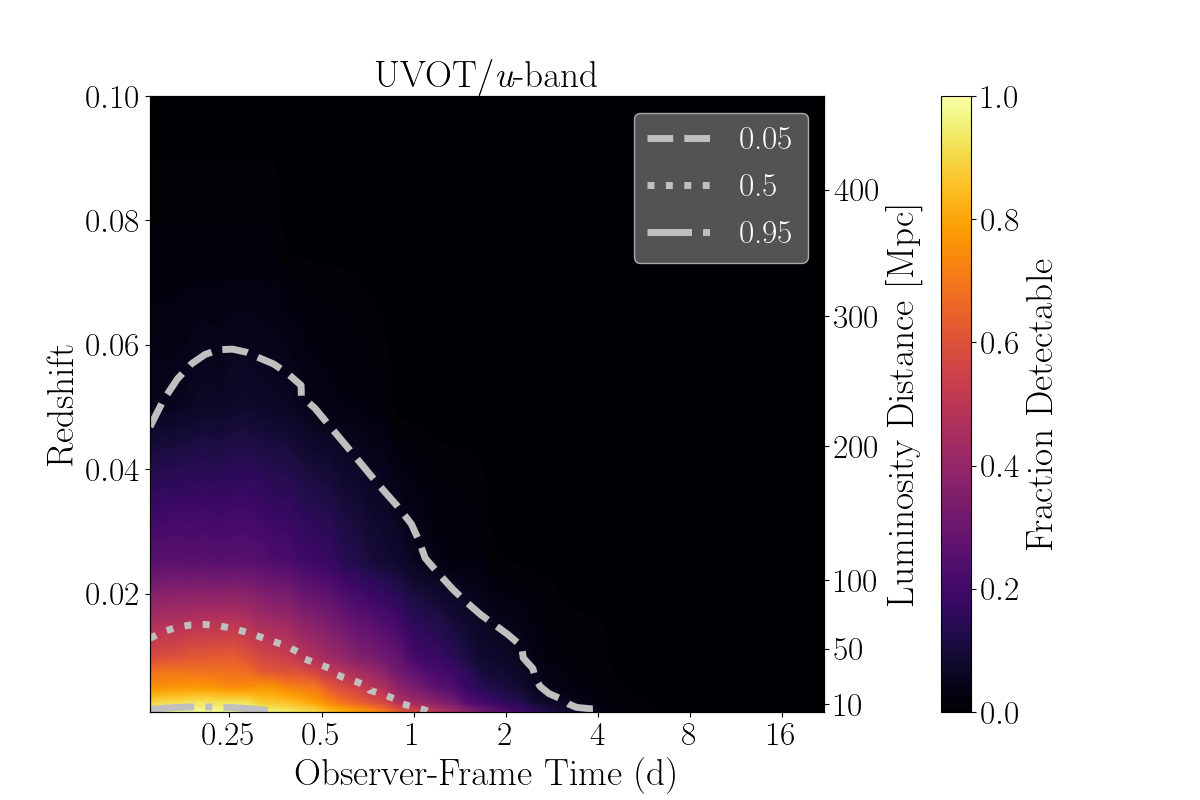}
\includegraphics[width=0.4\columnwidth, trim=20 0 100 00, clip]{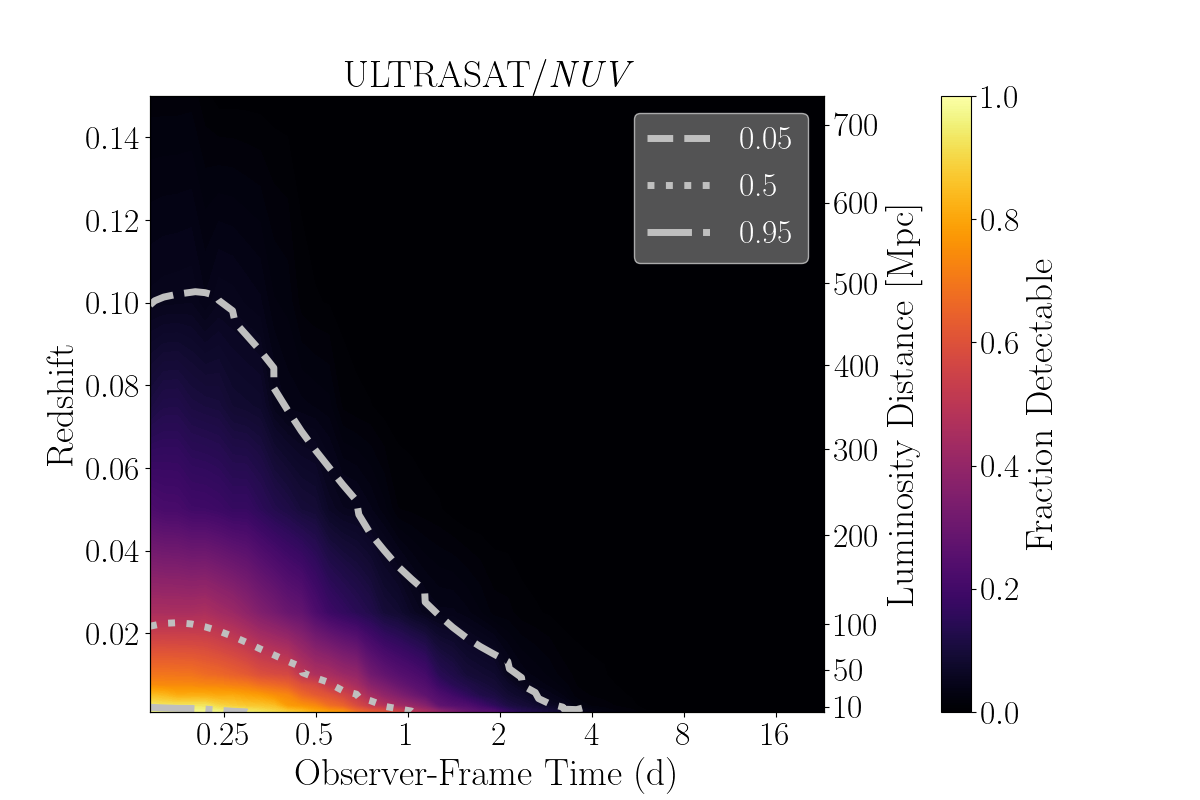}
\caption{Detectability contours for instruments with only one filter included in the study: DDOTI (\textit{top left}), GOTO (\textit{top right}), \textit{Swift}/UVOT (\textit{bottom left}), and ULTRASAT (\textit{bottom right}). See Figure~\ref{fig: BlackGEM} for more details. Note that the range of the vertical axis is not consistent in each panel.}
\end{figure*}

\end{document}